\documentclass[onecollarge,runningheads]{amsart}

\usepackage{caption}

\DeclareMathAlphabet{\mathpzc}{OT1}{pzc}{m}{it}
\usepackage[mathscr]{euscript}
\usepackage{amsmath} 
\usepackage{amsfonts}
\usepackage{amssymb}
\usepackage{ifthen} 
\usepackage{graphicx}
\usepackage{upgreek}
\usepackage{calligra}
\usepackage{calrsfs}
\usepackage{chngcntr}
\usepackage{cancel}
\usepackage{ifthen}
\usepackage{amsmath}
\usepackage{amsopn}
\usepackage{amsxtra}
\usepackage{amssymb}
\usepackage{amsbsy}
\usepackage{amscd}

\usepackage{mathtools}
\usepackage{amsmath,environ}
\NewEnviron{eqn}{
\begin{equation}\begin{split}
 \BODY
\end{split}\end{equation}
}
\NewEnviron{equn}{
\begin{equation*}\begin{split}
 \BODY
\end{split}\end{equation*}
}
\newcommand{\beqans}{\begin{subequations}\begin{eqnarray}}
\newcommand{\eeqans}[1]{\end{eqnarray}\label{#1}\end{subequations}}
\newcommand{\beqan}{\begin{eqnarray}}
\newcommand{\eeqan}{\end{eqnarray}}

\usepackage{xcolor}
\usepackage{color}

\usepackage{mathtools}
\usepackage{cases}

\usepackage[breaklinks]{hyperref}
\hypersetup{colorlinks, citecolor=blue, linkcolor=blue, backref=true}

\usepackage[doi=true,
isbn=false,url=true, bibstyle=authoryear, firstinits=true, 
eprint=false,natbib=true
]{biblatex}

\DeclareFieldFormat{bibentrysetcount}{\mkbibparens{\mknumalph{#1}}}
\DeclareFieldFormat{labelnumberwidth}{\mkbibbrackets{#1}}

\defbibenvironment{bibliography}{\list{\printtext[labelnumberwidth]{\printfield{prefixnumber}\printfield{labelnumber}}}{\setlength{\labelwidth}{\labelnumberwidth}\setlength{\leftmargin}{\labelwidth}\setlength{\labelsep}{\biblabelsep}\addtolength{\leftmargin}{\labelsep}\setlength{\itemsep}{\bibitemsep}\setlength{\parsep}{\bibparsep}}}{\endlist}{\item}

\usepackage[figuresright]{rotating}
\usepackage{float}
\usepackage[figuresright]{rotating}

\usepackage{mdframed}

\usepackage{graphicx}

\newcommand*{\bfrac}[2]{\genfrac{}{}{0pt}{}{\raisebox{-.3em}{\scriptsize$#1$}}{\raisebox{.4em}{\scriptsize$#2$}}}
\newcommand{\pd}[2]{\frac{\partial#1}{\partial#2}}
\newcommand{\od}[2]{\frac{d#1}{d#2}}

\DeclareMathOperator{\sLNsq}{\mathpzc{L}}
\DeclareMathOperator{\sLNtg}{\mathpzc{L}}
\newcommand{\srad}[1]{\tilde{#1}}
\DeclareMathOperator{\sr}{\mathpzc{r}}

\begin{document}

\title[Phonon transmission in partly-unzipped tubes]{Kinematically restricted phonon transmission in partly-unzipped tubes of square and triangular lattices}
\author{Basant Lal Sharma}\address{Department of Mechanical Engineering, Indian Institute of Technology Kanpur, Kanpur, U. P. 208016, India \\
Tel.: +91 512 2596173\\
Fax: +91 512 2597408\\
}
\email{bls@iitk.ac.in}
\date{Submitted: \today}
\maketitle
\begin{abstract}
{An analysis of the transmission of `scalar' phonons across partially unzipped square and triangular lattice tubes, assuming nearest neighbor interactions between particles, is presented. The phonon transport is assumed to involve the out-of-plane phonons in the unzipped portion and the radial phonons in the tubular portion. An exact expression of
reflectance and transmittance for the waves incident from either portions of the waveguide are provided explicitly, in terms of the Chebyshev polynomials, 
which leads to the provision of a simple expression for the ballistic conductance. }

\keywords{Keywords: Lattice strips\and Normal modes\and Chebyshev\and Wiener--Hopf\and Landauer-B\"{u}ttiker}

\subjclass{Primary 
74J05, Secondary 
81U30, 
37L60, 
74S20, 
74A50, 
74A60, 
39A14, 
74R15 
}
\end{abstract}

\setcounter{section}{-1}
\section{Introduction}
\label{intro}
The rise of nanostructures in technological applications has invigorated a slurry of questions concerning the nature of thermal transport \cite{Cahill1,Nazarovbook2009,balandin2011thermal}.
The reduced physical dimensions overcome the hurdles of the phonon mean free path, while new physical processes become crucial in the transport problem \cite{segal2003thermal,adam2009crossover}.
For instance, the thermal transport across nanojunctions connecting two semi-infinite leads (as thermal reservoirs) is often ballistic \cite{wang2008quantum,chiu2005ballistic,ciraci2001quantum,mingo2005carbon,Yamamoto2004}.
In general, the nanostructural regime admits transport that is defined in terms of reflection and transmission of waves, i.e., following the {\em Landauer viewpoint} \cite{Landauer1957}.
The problem of transmission of waves across a junction of two one dimensional lattices is similar to the classical problems in wave mechanics \cite{Brillouin,Miklowitz}, 
except that lattice waveguides are quasi-one dimensional. 
From a practical viewpoint, on the other hand, during the last decade, several independent experiments have demonstrated the unzipping of tubes into ribbons at nanoscale.
The unzipping of carbon nanotubes \cite{kosynkin2009longitudinal}
creates a junction between two different lattice waveguides namely a ribbon on the one side while a tube on the other.
The analyses involving waveguide junctions \cite{Hopkins2009,ZhangMingo2007} typically involves the nonequilibrium Green's function to study phonon thermal transport and energy transmission across atomic junctions \cite{Yamamoto2006noneq,wang2008quantum,Hopkins2009} paralleling its original electronic form.
\begin{figure}[ht]
\centering
{\includegraphics[width=\linewidth]{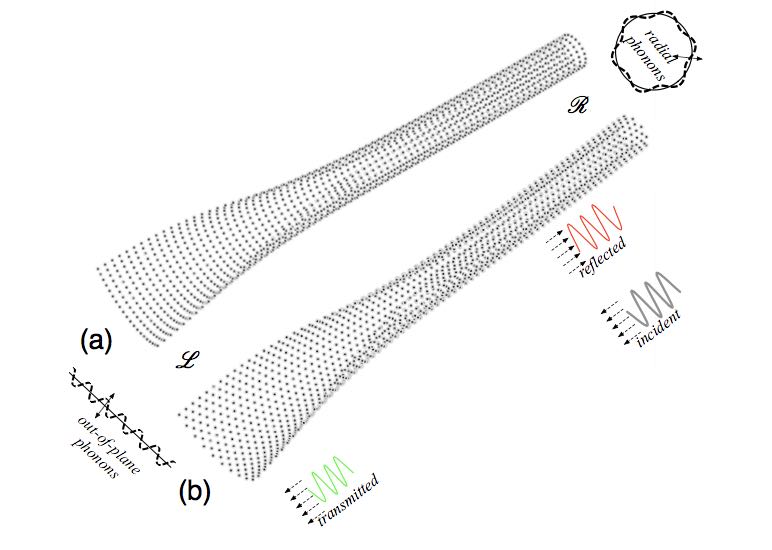}}
\caption{\footnotesize Schematic of a partially unzipped tube 
of square (a) and triangular (b) lattice structure, respectively.}
\label{sqtg_tube_phonons}
\end{figure}
\begin{figure*}[h]
\centering
{\includegraphics[width=\linewidth]{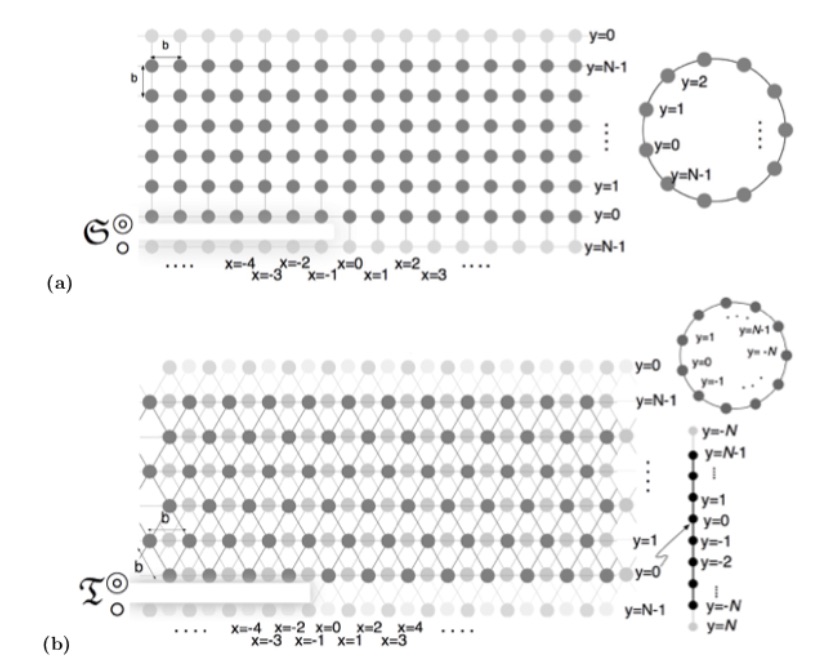}}
\caption{\footnotesize Partly-unzipped tube (${{\mathtt{N}}}=6$) of(a) square lattice 
and (b) triangular lattice. 
}
\label{latticestrip_sqtg_P_phonons}
\end{figure*}
In this paper, an analytical treatment to the mechanics of {phonons} is provided in an, arguably, simplified structure, taking the interpretation of phonons as lattice waves \cite{Maradudinbook} following the lines of bifurcated waveguides of square lattice structure \cite{Bls9s}. 
The problem discussed in this paper is shown schematically in Fig. \ref{sqtg_tube_phonons}(a) and (b) involving a partly unzipped tube of square and triangular lattices. 
Overall, the work can be seen as an extension of the recently present analysis for the honeycomb structures \cite{Bls5c_tube} and \cite{Bls5k_tube}; the former deals with the electronic counterpart of the problem and its results have been also reported in popular literature, for example \cite{Bls5c_tube_media}.

\section{Lattice model}
\subsection{Square lattice model}\label{latticemodel_sq} 

Consider a partly unzipped two-dimensional tube of square lattice, denoted by ${{\mathfrak{S}\hspace{-.4ex}}{\mathbin{\substack{{\circledcirc}\\{\circ}}}}},$ with ${\mathtt{N}}$ number of rows containing infinite number of particles with unit mass,
as shown schematically in Fig. \ref{sqtg_tube_phonons}(a). The `in-plane' nearest neighbors, with an equilibrium spacing ${\mathrm{b}}$, are connected by linearly elastic identical (massless) bonds with spring constant ${\mathrm{b}}^2$ (see \cite{Bls0} for the relevant scaling). 
Let ${\mathbb{Z}}$ denote the set of integers and ${\mathbb{Z}}_{m}^n$ (for $m\le n$) denote the set of integers $\{m, m+1, \dotsc, n-1, n\}$.
Let ${\mathbb{C}}$ denote the set of complex numbers.
Let ${\mathtt{u}}_{{\mathtt{x}}, {\mathtt{y}}}\in{\mathbb{C}}$ represent the `radial' displacement in the tubular part ${\mathcal{R}}$ and the `out-of-plane' displacement in the unzipped part ${\mathcal{L}}$;
with sites indexed by $({\mathtt{x}}, {\mathtt{y}})$ such that ${\mathtt{x}}\in{\mathbb{Z}}, {\mathtt{y}}\in{\mathbb{Z}}_0^{{\mathtt{N}}}$ as shown in Fig. \ref{latticestrip_sqtg_P_phonons}(a).
Suppose that
\begin{eqn}
{\mathtt{u}}^{i}_{{\mathtt{x}}, {\mathtt{y}}}(t)={{\mathrm{A}}}{{a}}_{({{{\kappa}}^{i}}){{\mathtt{y}}}}e^{-i{\upkappa}_x {\mathtt{x}}-i\omega t}, 
 ({\mathtt{x}}, {\mathtt{y}})\in{\mathfrak{S}\hspace{-.4ex}}{\mathbin{\substack{{\circledcirc}\\{\circ}}}},
 \label{uinc_sq}
\end{eqn}
represents a time harmonic {lattice wave} 
which is incident from the tubular part (the right side ${\mathcal{R}}$ of Fig. \ref{sqtg_tube_phonons}(a)). In other words, the wavenumber ${\upkappa}_x\in[-\pi, \pi]$ is such that the energy is transmitted from ${\mathcal{R}}$ towards ${\mathcal{L}}$.
In \eqref{uinc_sq}, $\omega$ denotes the frequency, ${{\mathrm{A}}}\in{\mathbb{C}}$ is constant amplitude, and ${{a}}_{({{{\kappa}}^{i}})}$ denotes the incident wave mode in the tubular portion of the structure. 

For convenience, let ${\upomega}$ be defined such that ${\upomega}{:=} {\mathrm{b}}\omega.$ By virtue of the fact that the equation of motion is satisfied by the incident wave mode 
\eqref{uinc_sq}, the triplet ${\upomega}, {\upkappa}_x, {{a}}_{({{{\kappa}}^{i}})}$ 
involves the mutually related entities \cite{Bls9}.
Following a traditional approach in diffraction theory, a vanishingly small amount of damping is introduced in the model. Consequently, 
${\upomega}={\upomega}_1+i{\upomega}_2, {\upomega}_2>0.$ Also in the remaining text, the explicit time dependence factor $e^{-i\omega t}$ is suppressed. The total displacement field ${\mathtt{u}}^{{t}}$, a sum of the incident wave field ${\mathtt{u}}^{i}$ and the scattered wave field ${\mathtt{u}}^{{s}}$, of an arbitrary particle in the lattice ${{\mathfrak{S}\hspace{-.4ex}}{\mathbin{\substack{{\circledcirc}\\{\circ}}}}}$ satisfies the discrete Helmholtz equation for all $({\mathtt{x}}, {\mathtt{y}})\in{\mathfrak{S}\hspace{-.4ex}}{\mathbin{\substack{{\circledcirc}\\{\circ}}}}$, {\em away} from unzipped half row of vertical bonds,
\begin{eqn}
\triangle{\mathtt{u}}^{{t}}_{{\mathtt{x}}, {\mathtt{y}}}+{\upomega}^2{\mathtt{u}}^{{t}}_{{\mathtt{x}}, {\mathtt{y}}}&=0 
\text{ with } {\mathtt{u}}_{{\mathtt{x}}, {\mathtt{y}}}^{{t}}={\mathtt{u}}_{{\mathtt{x}}, {\mathtt{y}}}^{i}+{\mathtt{u}}_{{\mathtt{x}}, {\mathtt{y}}},
\label{dHelmholtz_sq}
\end{eqn}
where $\triangle$ is the discrete Laplacian for square lattice, 
$\triangle{\mathtt{u}}_{{\mathtt{x}}, {\mathtt{y}}}={\mathtt{u}}_{{{\mathtt{x}}}+1, {{\mathtt{y}}}}+{\mathtt{u}}_{{{\mathtt{x}}}-1, {{\mathtt{y}}}}+{\mathtt{u}}_{{{\mathtt{x}}}, {{\mathtt{y}}}+1}+{\mathtt{u}}_{{{\mathtt{x}}}, {{\mathtt{y}}}-1}-4{\mathtt{u}}_{{{\mathtt{x}}}, {{\mathtt{y}}}}.$

\subsection{Triangular lattice model} 
\label{latticemodel_tg} 

Similar to the square lattice, ${\mathtt{N}}$ number of rows of particles with unit mass, are arranged in the form of one-dimensional lattices, but now these rows form a two-dimensional strip of triangular lattice ${{\mathfrak{T}\hspace{-.4ex}}{\mathbin{\substack{{\circledcirc}\\{\circ}}}}}$, as shown schematically in Fig. \ref{sqtg_tube_phonons}(b) and Fig. \ref{latticestrip_sqtg_P_phonons}(b).
The in-plane equilibrium spacing between the nearest neighbors in ${{\mathfrak{T}\hspace{-.4ex}}{\mathbin{\substack{{\circledcirc}\\{\circ}}}}}$ is ${\mathrm{b}}$ and 
they are connected by linearly elastic identical (massless) bonds with spring constant $2/3{\mathrm{b}}^2$ (see \cite{Bls4} for the relevant scaling). 
As a consequence of the geometric structure of the triangular lattice, a periodic boundary (associated with a singly rolled strip\footnote{\label{remk_tg_P} The case ${\mathtt{N}}=2{\mathrm{N}}+1$ for periodic boundary condition corresponds to a `doubly' rolled triangular lattice strip and appears to be less relevant, hence, its analysis is omitted in the paper.}) is possible only when ${\mathtt{N}}$ is an even integer, say $2{\mathrm{N}}$ (see bar and circle on the right side of Fig. \ref{latticestrip_sqtg_P_phonons}(b)). 
A rectangular coordinate system is used in this paper, i.e. a replica of ${{\mathfrak{T}\hspace{-.4ex}}{\mathbin{\substack{{\circledcirc}\\{\circ}}}}}$, 
denoted by ${{\mathfrak{T}\hspace{-.4ex}}{\mathbin{\substack{{\circledcirc}\\{\circ}}}}}{{^\mathrm{R}}}$, is juxtaposed with ${{\mathfrak{T}\hspace{-.4ex}}{\mathbin{\substack{{\circledcirc}\\{\circ}}}}}$ (see Fig. \ref{latticestrip_sqtg_P_phonons}(b)). 
The union of both lattice strips is a rectangular lattice strip (the original construction appeared in \cite{Bls4} and \cite{Bls9} for an infinite lattice), denoted by ${{\mathfrak{R}}{\mathbin{\substack{{\circledcirc}\\{\circ}}}}}$, with a period ${\mathrm{b}}/2$ horizontally and ${\sqrt{3}}{\mathrm{b}}/{2}$ vertically. Let ${\mathtt{u}}_{{\mathtt{x}}, {\mathtt{y}}}\in{\mathbb{C}}$ represent the displacement, similar to that for the square lattice model, at a site in ${{\mathfrak{R}}{\mathbin{\substack{{\circledcirc}\\{\circ}}}}}$ which is indexed by $({\mathtt{x}}, {\mathtt{y}})\in{\mathfrak{R}}{\mathbin{\substack{{\circledcirc}\\{\circ}}}}$. On the rectangular lattice strip ${{\mathfrak{R}}{\mathbin{\substack{{\circledcirc}\\{\circ}}}}}$, 
${\mathtt{u}}^{i}$ of the form \eqref{uinc_sq} is considered incident from the tubular part (right side ${\mathcal{R}}$ of the waveguide as shown in Fig. \ref{sqtg_tube_phonons}(b)). 
Without any loss of generality, it is assumed that 
\begin{eqn}
{{a}}_{({{{\kappa}}^{i}}){{\mathtt{y}}}}=-{{a}}_{({{{\kappa}}^{i}}){{\mathtt{N}}-{\mathtt{y}}-1}}, {\mathtt{y}}\in{\mathbb{Z}}_0^{{\mathtt{N}}},
\label{odd_tg}\end{eqn}
as the definition of the incident wave mode on the replica ${{\mathfrak{T}\hspace{-.4ex}}{\mathbin{\substack{{\circledcirc}\\{\circ}}}}}{{^\mathrm{R}}}$ in terms of that on ${{\mathfrak{T}\hspace{-.4ex}}{\mathbin{\substack{{\circledcirc}\\{\circ}}}}}$.
Above choice imposes an odd reflection symmetry (so it is a `manufactured' symmetry
 \cite{Bls4}) on ${{\mathfrak{R}}{\mathbin{\substack{{\circledcirc}\\{\circ}}}}}$. 
Note that the incident wave mode does not possess any symmetry in ${\mathfrak{T}\hspace{-.4ex}}{\mathbin{\substack{{\circledcirc}\\{\circ}}}}$ (since ${\mathtt{N}}$ is even,
see Footnote \ref{remk_tg_P}, the notion itself does not arise). 
The triplet ${\upomega}, {\upkappa}_x, {{a}}_{({{{\kappa}}^{i}})}$ correspond to the tubular part for this choice of incidence direction \cite{Bls9}. 
The total displacement field ${\mathtt{u}}^{{t}}$ of an arbitrary particle in the lattice ${{\mathfrak{R}}{\mathbin{\substack{{\circledcirc}\\{\circ}}}}}$ (and, therefore, in triangular lattices ${{\mathfrak{T}}}$ or ${{\mathfrak{T}}}{{^\mathrm{R}}}$) satisfies the discrete Helmholtz equation (compare with \eqref{dHelmholtz_sq})
\begin{eqn}
&\triangle{\mathtt{u}}^{{t}}_{{\mathtt{x}}, {\mathtt{y}}}+\frac{3}{2}{\upomega}^2{\mathtt{u}}^{{t}}_{{\mathtt{x}}, {\mathtt{y}}}=0,
({\mathtt{x}}, {\mathtt{y}})\in{\mathfrak{T}\hspace{-.4ex}}{\mathbin{\substack{{\circledcirc}\\{\circ}}}}, 
\label{dHelmholtz_tg}
\end{eqn}
\text{where }
$\triangle{\mathtt{u}}_{{\mathtt{x}}, {\mathtt{y}}}={\mathtt{u}}_{{{\mathtt{x}}}+2, {{\mathtt{y}}}}{}+{\mathtt{u}}_{{{\mathtt{x}}}-2, {{\mathtt{y}}}}{}+{\mathtt{u}}_{{{\mathtt{x}}}-1, {{\mathtt{y}}}+1}{}+{\mathtt{u}}_{{{\mathtt{x}}}-1, {{\mathtt{y}}}+1}{}+{\mathtt{u}}_{{{\mathtt{x}}}+1, {{\mathtt{y}}}-1}{}+{\mathtt{u}}_{{{\mathtt{x}}}-1, {{\mathtt{y}}}-1}{}-6{\mathtt{u}}_{{{\mathtt{x}}}, {{\mathtt{y}}}}{}$
{\em away} from unzipped half row of slant bonds. 

\section{Wiener--Hopf formulation}
\label{crack_sq}

The boundaries are such that ${{\mathtt{y}}}=-1$ and ${{\mathtt{y}}}={{\mathtt{N}}}-1$ are identified as same (shown schematically in Fig. \ref{latticestrip_sqtg_P_phonons}), while the unzipped portion lies along the broken bonds between ${\mathtt{y}}=0$ and ${\mathtt{y}}=-1\simeq {{\mathtt{N}}}-1$. 

\subsection{Square lattice tube} 
\label{crack_sq_P}
The equation of motion at ${\mathtt{y}}=0,$ ${\mathtt{y}}={\mathtt{N}}-1\simeq-1$, is
\begin{eqn}
{\upomega}^2{\mathtt{u}}^{{t}}_{{{\mathtt{x}}}, {{\mathtt{y}}}}&
={\mathtt{u}}^{{t}}_{{{\mathtt{x}}}+1, {{\mathtt{y}}}}+{\mathtt{u}}^{{t}}_{{{\mathtt{x}}}-1, {{\mathtt{y}}}}+{\mathtt{u}}^{{t}}_{{{\mathtt{x}}}, {{\mathtt{y}}}\mp1\pm{\mathtt{N}}}{\mathit{H}}({\mathtt{x}})+{\mathtt{u}}^{{t}}_{{{\mathtt{x}}}, {{\mathtt{y}}}\pm1}
-(3+{\mathit{H}}({\mathtt{x}})){\mathtt{u}}^{{t}}_{{{\mathtt{x}}}, {{\mathtt{y}}}}, {\mathtt{x}}\in{\mathbb{Z}},
\label{bc_sq_P}
\end{eqn}
where the letter ${{\mathit{H}}}$ stands for the Heaviside function: ${{\mathit{H}}}(m)=0, m<0$ and ${{\mathit{H}}}(m)=1, m\ge0$.
Suppose the discrete Fourier transform of $\{{\mathtt{u}}_m\}_{m\in{\mathbb{Z}}}$ is denoted by ${\mathtt{u}}^F$ (see Appendix 
\ref{sqtg_recall}). In this paper, the symbol ${{z}}$ is exclusively used throughout as a complex variable for the discrete Fourier transform.
Equation of motion \eqref{bc_sq_P} on rows ${\mathtt{y}}=0$ and ${\mathtt{y}}=-1$ (accounting for the unzipped portion for ${\mathtt{x}}<0$) imply that
\begin{eqn}
-2({\mathtt{u}}_{0;+}-{\mathtt{u}}_{-1;+})=2({{\mathtt{u}}_{-1}^{i}}^F-{{\mathtt{u}}_{0}^{i}}^F)_-
+({{\mathpzc{h}}}^2+1) ({\mathtt{u}}^F_0-{\mathtt{u}}^F_{-1})-{\mathtt{u}}^F_1+{\mathtt{u}}^F_{-2}.
\label{u0un1diff}
\end{eqn}
Using the identification ${\mathtt{y}}={\mathtt{N}}-1\simeq-1$ and the general solution as described in \ref{sqtg_recall}, it follows that 
${\mathtt{u}}^F_{{\mathtt{y}}}={\mathtt{u}}^F_{0}(\frac{1-{\lambda}^{-{\mathtt{N}}}}{{\lambda}^{{\mathtt{N}}}-{\lambda}^{-{\mathtt{N}}}}{\lambda}^{{\mathtt{y}}}-\frac{1-{\lambda}^{{\mathtt{N}}}}{{\lambda}^{{\mathtt{N}}}-{\lambda}^{-{\mathtt{N}}}}{\lambda}^{-{\mathtt{y}}}), {\mathtt{y}}\in{\mathbb{Z}}_0^{{\mathtt{N}}-1}.$
Introducing 
\begin{eqn}
{\mathrm{v}}_{{\mathtt{x}}}={\mathtt{u}}_{{\mathtt{x}}, 0}-{\mathtt{u}}_{{\mathtt{x}}, -1}, {\mathrm{v}}^{i}_{{\mathtt{x}}}={\mathtt{u}}^{i}_{{\mathtt{x}}, 0}-{\mathtt{u}}^{i}_{{\mathtt{x}}, -1},
\label{vspring_sq}\end{eqn}
with ${\mathrm{v}}^F={\mathtt{u}}^F_{0}-{\mathtt{u}}^F_{-1},$
it is found that ${\mathtt{u}}^F_{0}={{\mathrm{v}}}^F{{\mathtt{U}}_{{\mathtt{N}}-1}}/({{\mathtt{U}}_{{\mathtt{N}}-1}-{\mathtt{U}}_{{\mathtt{N}}-2}-1}),$
as well as in the context of \eqref{u0un1diff}, $-{\mathtt{u}}^F_{1}+{\mathtt{u}}^F_{-2}\simeq$$-{\mathtt{u}}^F_{1}+{\mathtt{u}}^F_{{\mathtt{N}}-2}=$
${{\mathrm{v}}}^F({1+{\mathtt{U}}_{{\mathtt{N}}-2}-{\mathtt{U}}_{{\mathtt{N}}-3}-2{\vartheta}})/({-{\mathtt{U}}_{{\mathtt{N}}-1}+{\mathtt{U}}_{{\mathtt{N}}-2}+1}),$ where using \eqref{q2_sq}, ${{\vartheta}}={\frac{1}{2}}(4-{{z}}-{{z}}^{-1}-{\upomega}^2)$.\footnote{\label{Chebvar}Throughout the paper, ${\vartheta}$, which itself is a function of ${z}$, appears as an argument in the definition of Chebyshev polynomials \cite{Chebyshev00,MasonHand,Bls9} (see also Appendix of \cite{Bls9});
it is defined by \begin{eqn}
{{\vartheta}}{:=}{\frac{1}{2}}{\mathpzc{Q}}.
\label{chebx_sq}
\end{eqn}}
Finally, using \eqref{annAALk_sq}, for ${{z}}\in{{\mathscr{A}}}$\footnote{\label{annulus}${\mathscr{A}}$ is an annulus containing the unit circle ${\mathbb{T}}$ in the complex plane that makes the Wiener--Hopf problem
well-defined.}
\begin{eqn}
{{\mathrm{v}}}_+({{z}})+\sLNsq_{{}}({{z}}){{\mathrm{v}}}_-({{z}})&=(1-\sLNsq({{z}})){{\mathrm{A}}}{{{\mathrm{v}}}_{({{{\kappa}}^{i}})}}\delta_{D-}({{z}}{{z}}_{{P}}^{-1}),
\label{WH_sq_P}
\text{with }
{{{\mathrm{v}}}_{({{{\kappa}}^{i}})}}{:=}{{a}}_{({{{\kappa}}^{i}})0}-{{a}}_{({{{\kappa}}^{i}}){{\mathtt{N}}}-1},
\end{eqn}
is obtained as a Wiener--Hopf equation, using \eqref{zPdef_sq}, with the kernel $\sLNsq$ given by $\sLNsq_{{}}={2({\vartheta}-1){\mathtt{U}}_{{\mathtt{N}}-1}}/({{\mathtt{U}}_{{\mathtt{N}}}-{\mathtt{U}}_{{\mathtt{N}}-2}-2})$, i.e.,
\begin{eqn}
\sLNsq_{{}}
&=\frac{{\mathpzc{H}}{\mathtt{U}}_{{\mathtt{N}}-1}}{2{\mathtt{T}}_{{\mathtt{N}}}-2}{=:}\frac{{\mathscr{N}}}{{\mathscr{D}}}
=\begin{cases}
\dfrac{\prod\nolimits_{j=1}^{{\mathrm{N}}}({{\mathpzc{h}}}^2+4\sin^2\frac{(j-{\frac{1}{2}})\pi}{{2{\mathrm{N}}}})}{\sr^{2}\prod\nolimits_{j=1}^{{\mathrm{N}}-1}({{\mathpzc{h}}}^2+4\sin^2\frac{j\pi}{2{\mathrm{N}}})}&\text{ if }{\mathtt{N}}=2{\mathrm{N}}\\
\dfrac{\prod\nolimits_{j=1}^{{\mathrm{N}}}({{\mathpzc{h}}}^2+4\sin^2\frac{(j-{\frac{1}{2}})\pi}{2{\mathrm{N}}+1})}{\prod\nolimits_{j=1}^{{\mathrm{N}}}({{\mathpzc{h}}}^2+4\sin^2\frac{j\pi}{2{\mathrm{N}}+1})}&\text{ if }{\mathtt{N}}=2{\mathrm{N}}+1
\end{cases},
\label{Lk_sq_P}
\end{eqn}

Notice that the denominator of $\sLNsq_{{}}$ \eqref{Lk_sq_P} contains factors involving the dispersion relation for a periodic strip (all modes in the lattice tube of width ${{\mathtt{N}}}$) while the numerator contains those due to dispersion relation for waves in a strip corresponding to unzipped portion of width ${\mathtt{N}}$ \cite{Bls9}.
Based on the intuition gathered from the analysis of bifurcated lattice waveguides \cite{Bls9s}, it can be anticipated that the kernel is ${\sLNsq}_{{}}={\prod\nolimits_{j=1}^{{\mathtt{N}}}({{\mathpzc{h}}}^2+4\sin^2{\frac{1}{2}}\frac{j-1}{{\mathtt{N}}}\pi)}/{\prod\nolimits_{j=1}^{{\mathtt{N}}}({{\mathpzc{h}}}^2+4\sin^2\frac{j}{{\mathtt{N}}}\pi)}$; indeed, this expression reduces to \eqref{Lk_sq_P} as stated
in \ref{appLk_sq_P}.
Additionally, using the even/odd mode based factorizations given by \cite{Bls9}, from \eqref{Lk_sq_P}, 
\begin{subequations}
\begin{numcases}{{\sLNsq}_{{}}=}
\frac{2{\mathtt{T}}_{{\mathrm{N}}}}{2({\vartheta}+1){\mathtt{U}}_{{\mathrm{N}}-1}}, & \text{ for } ${\mathtt{N}}=2{\mathrm{N}}$ 
\label{Lk_sq_P_case1}\\
\frac{{\mathtt{V}}_{{\mathrm{N}}}}{{\mathtt{W}}_{{\mathrm{N}}}}, & \text{ for } ${\mathtt{N}}=2{\mathrm{N}}+1$.
\label{Lk_sq_P_case2} 
\end{numcases}
\end{subequations}

This completes the Wiener--Hopf formulation for ${\mathfrak{S}\hspace{-.4ex}}{\mathbin{\substack{{\circledcirc}\\{\circ}}}}$ assuming incidence of wave modes from the tubular portion.
As ${\mathtt{N}}\to\infty$, it is easily seen that ${\sLNsq}_{{}}$ \eqref{Lk_sq_P} approaches the kernel for a single mode III crack, i.e. ${{\mathpzc{h}}}/{{\sr}}$, in infinite square lattice \cite{Bls0}. 

\subsection{Triangular lattice tube}
\label{crack_tg_P}
In terms of a more convenient labelling shown in Fig. \ref{latticestrip_sqtg_P_phonons}(b), it is required that ${\mathtt{u}}^F_{{\mathrm{N}}}={\mathtt{u}}^F_{-{\mathrm{N}}}$. Due to the manufactured symmetry on ${\mathfrak{R}}{\mathbin{\substack{{\circledcirc}\\{\circ}}}}$, i.e., odd symmetry \eqref{odd_tg} in the choice of the incident wave mode \eqref{uinc_sq}, 
using same calculation as that described in \ref{sqtg_recall} except that ${\mathpzc{Q}}$ is given \eqref{q2_tg},
it is found that
$${\mathtt{u}}^F_{{\mathtt{y}}}={\mathtt{u}}^F_0\frac{(1+{\lambda}^{-1}){{\lambda}}^{{\mathtt{y}}}-(1+{\lambda}){{\lambda}}^{2{{\mathrm{N}}}-2-{{\mathtt{y}}}}}{(1+{\lambda}^{-1})-(1+{\lambda}){{\lambda}}^{2{{\mathrm{N}}}-2}};$$
in particular, ${\mathtt{u}}_{1}^F={\mathtt{u}}^F_0{{\mathtt{W}}}_{{\mathrm{N}}-2}/{{\mathtt{W}}}_{{\mathrm{N}}-1}$.
Thus, ${\mathtt{u}}_{0;+}$ and ${\mathtt{u}}_{0;-}$ satisfy the Wiener--Hopf equation (recall Footnote \ref{annulus} and \eqref{annAALk_tg})
\begin{subequations}
\begin{eqn}
{\mathtt{u}}_{0;+}({{z}})+{\sLNtg}_{{}}({{z}}){\mathtt{u}}_{0;-}({{z}})={\frac{1}{2}}\frac{1-{\sLNtg}_{{}}({{z}})}{1+{2}({{z}}+{{z}}^{-1})^{-1}}
(q^F({{z}})-(1+{{z}}){\mathtt{u}}_{-1, 0}), {{z}}\in{{\mathscr{A}}}_{{}}, \label{WHKeq}\\
\end{eqn}
\begin{eqn}
\text{ where }
{\sLNtg}_{{}}({{z}}){:=}1-\frac{1+2({{z}}+{{z}}^{-1})^{-1}}{1+{\mathpzc{Q}}-{\mathpzc{V}}}, 
\label{Lkz}
\end{eqn}
\begin{eqn}
{\mathpzc{Q}}({z})=\frac{6-({z}^2+{z}^{-2})-\frac{3}{2}{\upomega}^2}{({z}+{z}^{-1})}.
\label{q2_tg}
\end{eqn}
In \eqref{WHKeq} (also see the details in \eqref{qF_plusinc}),
\begin{eqn}
q^F({{z}})
&={{\mathrm{A}}}{{a}}_{({{{\kappa}}^{i}}){0}}(1+{{z}})(1+{{\,\gimel\,}}{{z}}_{{P}}^{-1})\delta_{D-}({{\,\gimel\,}}{{z}}{{z}}_{{P}}^{-1}),
\label{qF}
\end{eqn}
\end{subequations}
where {${\,\gimel\,}={\pm}$ with} the upper sign $+$ (resp. lower sign $-$) refers to the case when the incident wave mode possesses {odd} (resp. {even}) symmetry on ${{\mathfrak{R}}{\mathbin{\substack{{\circledcirc}\\{\circ}}}}}$, i.e., in addition to \eqref{odd_tg}, it is assumed that
\begin{eqn}
{{a}}_{({{{\kappa}}^{i}}){{\mathtt{y}}}}={{\,\gimel\,}}{{a}}_{({{{\kappa}}^{i}}){-{\mathtt{y}}-1}}, {\mathtt{y}}\in{\mathbb{Z}}_0^{{\mathrm{N}}}.
\label{evenoddsymm}
\end{eqn}
The details of aforementioned `even' and `odd' modes has been provided by \cite{Bls9}.

The Wiener--Hopf 
kernel \eqref{Lkz} can be simplified to (compare with \eqref{Lk_sq_P_case1})
\begin{eqn}
{\sLNtg}_{{}}({{z}})
&=\frac{1}{{\frac{1}{2}}({z}+{z}^{-1})}\frac{{\frac{1}{2}}({z}+{z}^{-1}){{\mathtt{W}}}_{{\mathrm{N}}}-{{\mathtt{W}}}_{{\mathrm{N}}-1}}{2({\vartheta}+1){{\mathtt{U}}}_{{\mathrm{N}}-1}}{=:}\frac{{\mathscr{N}}}{{\mathscr{D}}},
\label{Lk_tg_P}
\end{eqn}
where ${\vartheta}$, the argument of the Chebyshev polynomials, is given by \eqref{chebx_sq} using the definition of ${\mathpzc{Q}}$ for triangular lattice given by \eqref{q2_tg}, i.e. ${\vartheta}=({3-{\frac{1}{2}}({z}^2+{z}^{-2})-\frac{3}{4}{\upomega}^2})/{({z}+{z}^{-1})}$.

Notice that the denominator ${\mathscr{D}}$ of ${\sLNtg}_{{}}$ (in \eqref{Lk_tg_P}) 
contains dispersion relations for the odd wave modes in the tubular portion of the rectangular lattice structure \cite{Bls9}. On the other hand, the numerator ${\mathscr{N}}$ of ${\sLNtg}_{{}}$ contains dispersion relations for the wave modes in the unzipped portion albeit in the form of rectangular lattice strip \cite{Bls9}. 
This is a surprising osbervation in view of the allowance of both even and odd symmetries in the incident wave mode which are subject to scattering by the edge of unzipped portion. In order to unearth a subtlety behind this, a key role is played by a well known property of Chebyshev polynomials of second kind, namely ${{\mathtt{U}}}_{n}(-{\vartheta})=(-1)^n{{\mathtt{U}}}_{n}({\vartheta})$, and that (using \eqref{chebx_sq}) ${\vartheta}(-{z})=-{\vartheta}({z})$. Indeed, ${\mathscr{N}}(-{z})=-{\frac{1}{2}}({z}+{z}^{-1})({{\mathtt{U}}}_{{\mathrm{N}}}({\vartheta}(-{z}))+{{\mathtt{U}}}_{{\mathrm{N}}-1}({\vartheta}(-{z})))-({{\mathtt{U}}}_{{\mathrm{N}}-1}({\vartheta}(-{z}))+{{\mathtt{U}}}_{{\mathrm{N}}-2}({\vartheta}(-{z})))=(-1)^{{\mathrm{N}}}({\frac{1}{2}}({z}+{z}^{-1}){{\mathtt{V}}}_{{\mathrm{N}}}({z})-{{\mathtt{V}}}_{{\mathrm{N}}-1}({z}))$ and ${\mathscr{D}}(-{z})=-{\frac{1}{2}}({z}+{z}^{-1})(1+{\vartheta}(-{z})){{\mathtt{U}}}_{{\mathrm{N}}-1}({\vartheta}(-{z}))=(-1)^{{\mathrm{N}}}{\frac{1}{2}}({z}+{z}^{-1})(1-{\vartheta}({z})){{\mathtt{U}}}_{{\mathrm{N}}-1}({\vartheta}({z}))$ which correspond to dispersion relations for the {\em even} wave modes in the tubular and unzipped portions (of the rectangular lattice structure) \cite{Bls9}.

\subsection{Incidence from unzipped portion} 
\label{unzippedinc}
In contrast to \S\ref{crack_sq_P} and \S\ref{crack_tg_P} discussed so far in this paper, consider the case when a wave mode is incident from the unzipped portion. 
The scattering of such a wave (Fig. \ref{sqtg_tube_phonons}) occurs due to the `new' bonds placed between the `through-crack'. It is assumed that the incident wave has the form \eqref{uinc_sq}, 
with ${\upkappa}_x$ such that the energy is carried towards the tubular side ${\mathcal{R}}$,
while ${{a}}_{({{\kappa}}^{i})}$ is the corresponding normal mode in the unzipped portion ${\mathcal{L}}$.

\begin{itemize}
\item 
For the square lattice structrue, using \eqref{bc_sq_P}, after arriving at an equation similar to \eqref{u0un1diff}, the counter part of \eqref{WH_sq_P} is found to be (using the definitions in \eqref{zPdef_sq}) 
\begin{eqn}
{{\mathrm{v}}}_+({{z}})+\sLNsq_{{}}({{z}}){{\mathrm{v}}}_-({{z}})&=-(1-\sLNsq({{z}})){{\mathrm{A}}}{{{\mathrm{v}}}_{({{{\kappa}}^{i}})}}\delta_{D+}({{z}}{{z}}_{{P}}^{-1}),
\label{WH_sq_P_altinc}
\end{eqn}
\item
For the triangular lattice structure, note that the odd reflection (manufactured) symmetry continues to hold. 
The scattered displacement ${\mathtt{u}}$ at ${\mathtt{y}}=0$ satisfies
\begin{eqn}
-\frac{3}{2}{\upomega}^2{\mathtt{u}}_{{{\mathtt{x}}}, 0}=({\mathtt{u}}^{i}_{{{\mathtt{x}}}, 0}-{\mathtt{u}}^{i}_{{{\mathtt{x}}}+1, -1}){\mathit{H}}(-{{\mathtt{x}}}-2)
+({\mathtt{u}}^{i}_{{{\mathtt{x}}}, 0}-{\mathtt{u}}^{i}_{{{\mathtt{x}}}-1, -1}){\mathit{H}}(-{{\mathtt{x}}}-1)\\
-({\mathtt{u}}_{{{\mathtt{x}}}, 0}-{\mathtt{u}}_{{{\mathtt{x}}}+1, -1}){\mathit{H}}({{\mathtt{x}}}+1)-({\mathtt{u}}_{{{\mathtt{x}}}, 0}-{\mathtt{u}}_{{{\mathtt{x}}}-1, -1}){\mathit{H}}({{\mathtt{x}}})\\
+({\mathtt{u}}_{{{\mathtt{x}}}+2, 0}+{\mathtt{u}}_{{{\mathtt{x}}}-2, 0}+{\mathtt{u}}_{{{\mathtt{x}}}+1, 1}+{\mathtt{u}}_{{{\mathtt{x}}}-1, 1}-4{\mathtt{u}}_{{{\mathtt{x}}}, 0}), {\mathtt{x}}\in{\mathbb{Z}}.
\label{y0_tg_P}
\end{eqn}
Applying discrete Fourier transform on \eqref{y0_tg_P}, it is found that
\begin{eqn}
2{\mathtt{u}}_{0;+}+{{z}}^{-1}{\mathtt{u}}_{0;+}+(1+{{z}}){\mathtt{u}}_{-1, 0}+{{z}} {\mathtt{u}}_{0;+}
=q^F({{z}})+({{z}}^2+{{z}}^{-2}-4+\frac{3}{2}{\upomega}^2){\mathtt{u}}_{0}^F+({{z}}+{{z}}^{-1}){\mathtt{u}}_{1}^F,
\end{eqn}
where (in contrast to \eqref{qF})
\begin{eqn}
q^F({{z}})=-{{\mathrm{A}}}{{a}}_{({{{\kappa}}^{i}}){0}}(1+{{z}})(1+{{\,\gimel\,}}{{z}}_{{P}}^{-1})\delta_{D+}({{\,\gimel\,}}{{z}}{{z}}_{{P}}^{-1}).
\label{qF_altinc}
\end{eqn}
The relevant details regarding derivation of $q^F$ \eqref{qF_altinc} are provided in \eqref{qF_minusinc}.
This yields Wiener--Hopf equation \eqref{WHKeq} except that $q^F$ of \eqref{qF} is replaced by \eqref{qF_altinc}.
\end{itemize}

\section{Exact Solution of Wiener--Hopf equation}
\label{exact_sol_WH}
The multiplicative factors for the Wiener--Hopf kernels \eqref{Lk_sq_P}, and \eqref{Lk_tg_P}, can be constructed easily (see Appendix 
\ref{appWHfac} for the details). 
Using $\sLNsq_{{}}=\sLNsq_{{}+}\sLNsq_{{}-}$, Wiener--Hopf equations \eqref{WH_sq_P} and \eqref{WH_sq_P_altinc} (resp. \eqref{WHKeq}), can be expressed as ${{\sLNsq}_{{}+}^{-1}({{z}})}{{{\mathrm{v}}}_{+}({{z}})}+{\sLNsq}_{{}-}({{z}}){{\mathrm{v}}}_-({{z}})
={{\mathpzc{C}}}({{z}}), {{z}}\in{{\mathscr{A}}}$ (resp. ${{\sLNtg}_{{}+}^{-1}({{z}})}{{\mathtt{u}}_{+}({{z}})}+{\sLNtg}_{{}-}({{z}}){\mathtt{u}}_{-}({{z}})=({1+{{z}}^{-1}})^{-1}{{\mathpzc{C}}}({{z}})$).
Analogous to the case of infinite lattice \cite{Bls0, Bls4}\footnote{\eqref{Cz} and \eqref{CpmK_sq} can be compared with (2.28b) and (2.29) of \cite{Bls0}}, an additive factorization, ${{\mathpzc{C}}}={{\mathpzc{C}}}_{+}({{z}})+{{\mathpzc{C}}}_{-}({{z}}),$ is constructed by elementary means \cite{Noble}.
It is easy to see that ${{\mathpzc{C}}}_+({{z}})$ and ${{\mathpzc{C}}}_-({{z}})$ are analytic at ${{z}}\in{\mathbb{C}}$ with $|{{z}}|>\max\{{{\mathit{R}}}_+, {{\mathit{R}}}_{L_{{}}}\},$ $|{{z}}|<\min\{{{\mathit{R}}}_-, {{\mathit{R}}}_{L_{{}}}^{-1}\}$, respectively (${{\mathit{R}}}_\pm$ are given \eqref{annAu_sq} in Appendix 
\ref{sqtg_recall}). It is understood as an implicit notation that ${{\mathfrak{s}}}$ denotes either case of incidence.

\subsection{Square lattice structure}
\label{crack_sq_WHsol}
Using \eqref{WH_sq_P} for ${{\mathfrak{s}}}={{\mathcal{R}}}$ and \eqref{WH_sq_P_altinc} for ${{\mathfrak{s}}}={{\mathcal{L}}}$,
\begin{subequations}
\begin{eqn}
{{\mathpzc{C}}}({{z}})&=(\frac{1}{{\sLNsq}_{{}+}({{z}})}-{\sLNsq}_{{}-}({{z}})){{\mathrm{A}}}{{{\mathrm{v}}}_{({{{\kappa}}^{i}})}}
(\delta_{D-}({{z}} {{z}}_{{P}}^{-1})\delta_{{{\mathfrak{s}}},{{\mathcal{R}}}}-\delta_{D+}({{z}} {{z}}_{{P}}^{-1})\delta_{{{\mathfrak{s}}},{{\mathcal{L}}}}), 
 {{z}}\in{{\mathscr{A}}}, \label{Cz},
\end{eqn}
leading to
\begin{eqn}
{{\mathpzc{C}}}_\pm({{z}})&=\mp{{\mathrm{A}}}{{{\mathrm{v}}}_{({{{\kappa}}^{i}})}}(\frac{1}{{\sLNsq}_{{}+}({{z}}_{{P}})}-{\sLNsq}_{{}\pm}^{\mp1}({{z}}))\delta_{D-}({{z}} {{z}}_{{P}}^{-1})\delta_{{{\mathfrak{s}}},{{\mathcal{R}}}}
\pm{{\mathrm{A}}}{{{\mathrm{v}}}_{({{{\kappa}}^{i}})}}({\sLNsq}_{{}-}({{z}}_{{P}})-{\sLNsq}_{{}\pm}^{\mp1}({{z}}))\delta_{D+}({{z}} {{z}}_{{P}}^{-1})\delta_{{{\mathfrak{s}}},{{\mathcal{R}}}}, {{z}}\in{\mathscr{A}}.
\label{CpmK_sq}
\end{eqn}
\end{subequations}
Note that ${{{\mathrm{v}}}_{({{{\kappa}}^{i}})}}$ denotes the expression provided in \eqref{WH_sq_P}. Invoking the Liouville's theorem \cite{Noble,Bls0}, the solution of the discrete Wiener--Hopf equation is obtained (see \ref{Liouville_argument}).
In terms of the one-sided discrete Fourier transform,
\begin{eqn}
{{\mathtt{v}}}_{\pm}({{z}})={{\mathpzc{C}}}_{\pm}({{z}}){{{\mathpzc{L}}_{{}}}}_{\pm}^{\pm1}({{z}}), {{z}}\in{\mathbb{C}}, |{{z}}|\gtrless\bfrac{\max}{\min}\{{{\mathit{R}}}_{\pm}, {{\mathit{R}}}_{L_{{}}}^{\pm1}\}.
\label{vpm_sq}
\end{eqn}
and the complex function ${\mathtt{v}}^F={{\mathtt{v}}}_{+}+{{\mathtt{v}}}_{-}$ (for ${{z}}\in{{\mathscr{A}}}$) is 
\begin{eqn}
{{\mathrm{v}}}^F({{z}})&={{\mathrm{A}}}{\mathtt{C}}_0\frac{{{z}} {{\mathpzc{K}}}({{z}})}{{{z}}-{{z}}_{{P}}}, 
{{\mathpzc{K}}}({{z}}){:=}\frac{1-{\sLNsq}_{{}}({{z}})}{{\sLNsq}_{{}-}({{z}})}, 
{\mathtt{C}}_0{:=}-{{{\mathrm{v}}}_{({{{\kappa}}^{i}})}}({\sLNsq}_{{}+}^{-1}({{z}}_{{P}})\delta_{{{\mathfrak{s}}},{{\mathcal{R}}}}+{\sLNsq}_{{}-}({{z}}_{{P}})\delta_{{{\mathfrak{s}}},{{\mathcal{L}}}})\in{\mathbb{C}}.
\label{vzsol}
\end{eqn}

\subsection{Triangular lattice structure}
Using \eqref{qF} for ${{\mathfrak{s}}}={{\mathcal{R}}}$ and \eqref{qF_altinc} for ${{\mathfrak{s}}}={{\mathcal{L}}}$, in the context of the first paragraph of this section \S\ref{exact_sol_WH},\footnote{\label{symmetryaspect}The odd symmetry of the incident wave on ${{\mathfrak{T}\hspace{-.4ex}}{\mathbin{\substack{{\circledcirc}\\{\circ}}}}}$ is assumed (i.e., $-$ sign in \eqref{evenoddsymm}) since the case corresponding to the even symmetry (i.e., $+$ sign in \eqref{evenoddsymm}) can be obtained simply by the substitution ${{z}}_{{P}}{\mapsto}-{{z}}_{{P}}.$}
\begin{subequations}
\begin{eqn}
{{\mathpzc{C}}}({{z}})&=\frac{1-{\sLNtg}_{{}}({{z}})}{{\sLNtg}_{{}+}({{z}})}({{\mathrm{A}}}{{a}}_{({{{\kappa}}^{i}}){0}}(1+{{z}}_{{P}}^{-1})
(\delta_{D-}({{z}}{{z}}_{{P}}^{-1})\delta_{{{\mathfrak{s}}},{{\mathcal{R}}}}-\delta_{D+}({{z}}{{z}}_{{P}}^{-1})\delta_{{{\mathfrak{s}}},{{\mathcal{L}}}})-{\mathtt{u}}_{-1, 0}),
\label{Cz_tg}
\end{eqn}
which leads to the additive factors
\begin{eqn}
&{{\mathpzc{C}}}_\pm({{z}})=\mp{\mathtt{u}}_{-1, 0}{\sLNtg}_{{}\pm}({{z}})^{\mp1}\pm{{l}_{{}}}_{-0}{\mathtt{u}}_{-1, 0}
\pm{{\mathrm{A}}}{{a}}_{({{{\kappa}}^{i}}){0}}(1+{{z}}_{{P}}^{-1})\big({\sLNtg}_{{}\pm}({{z}})^{\mp1}-\frac{1}{{\sLNtg}_{{}+}({{z}}_{{P}})}\big)\delta_{D-}({{z}}{{z}}_{{P}}^{-1})\delta_{{{\mathfrak{s}}},{{\mathcal{R}}}}\\
&\mp{{\mathrm{A}}}{{a}}_{({{{\kappa}}^{i}}){0}}(1+{{z}}_{{P}}^{-1})\big({\sLNtg}_{{}\pm}({{z}})^{\mp1}-{\sLNtg}_{{}-}({{z}}_{{P}})\big)\delta_{D+}({{z}}{{z}}_{{P}}^{-1})\delta_{{{\mathfrak{s}}},{{\mathcal{L}}}},
\label{CpmK_tg}
\end{eqn}
\end{subequations}
where ${l_{{}}}_{-0}=\lim_{{{z}}\to0}{{\sLNtg}_{{}}}_-({{z}})$. 
As a consequence of the standard application of the Liouville's theorem \cite{Noble}, following the results obtained by \cite{Bls4}, \eqref{WHKeq} yields 
\begin{eqn}
{\mathtt{u}}_{{0};\pm}({{z}})&=\frac{{\mathpzc{L}}^{\pm1}_{{}\pm}({{z}})}{1+{{z}}^{-1}}({\mathpzc{C}}_\pm({{z}})\mp{{l}_{{}}}_{-0}{\mathtt{u}}_{-1, 0}).
\label{uzfull_tg}
\end{eqn}
The solution of the discrete Wiener--Hopf equation \eqref{WHKeq}, in the form of a discrete Fourier transform, is written as 
\begin{eqn}
{\mathtt{u}}_{{0}}^F({{z}})={\mathtt{u}}_{{0};+}({{z}})+{\mathtt{u}}_{{0};-}({{z}})={{\mathrm{A}}}{\mathtt{C}}_{0}\frac{{{z}}{{\mathpzc{K}}}({{z}})}{{{z}}-{{z}}_{{P}}}, {{z}}\in{{\mathscr{A}}}_{{}}, \label{uzsolK_tg}
\text{with }
\end{eqn}
\beqans
\hspace{-.4in}
{\mathtt{C}}_{0}{:=}-{{a}}_{({{{\kappa}}^{i}}){0}}(1+{{z}}_{{P}}^{-1})
({\sLNtg}_{{}+}^{-1}({{z}}_{{P}})\delta_{{{\mathfrak{s}}},{{\mathcal{R}}}}+{\sLNtg}_{{}-}({{z}}_{{P}})\delta_{{{\mathfrak{s}}},{{\mathcal{L}}}})\in{\mathbb{C}}, 
\label{C0}\\
\hspace{-.4in}\text{ and }
{{\mathpzc{K}}}({{z}}){:=}\frac{1}{1+{{z}}^{-1}}\frac{1-{\sLNtg}_{{}}({{z}})}{{\sLNtg}_{{}-}({{z}})}, 
{{z}}\in{{\mathscr{A}}}_{{}}.\label{KK}
\eeqans{C012}
Let the scattered displacement in one type of the slant bonds be given by
\begin{eqn}
{{\mathrm{v}}}_{{{\mathtt{x}}}}{:=}{\mathtt{u}}_{{{\mathtt{x}}}, {0}}-{\mathtt{u}}_{{{\mathtt{x}}}-1, {-1}}. 
\label{vspring_tg}
\end{eqn}
Note that $\sum\nolimits_{{\mathtt{x}}\in{\mathbb{Z}}}({\mathtt{u}}_{{\mathtt{x}}, 0}-{\mathtt{u}}_{{\mathtt{x}}-1, -1}){{z}}^{-{\mathtt{x}}}
=(1+{{z}}^{-1}){\mathtt{u}}_{{0}}^F({{z}}),$ 
and therefore, using 
\eqref{vspring_tg}, 
\begin{eqn}
(1+{{z}}^{-1}){\mathtt{u}}_{{0};\pm}({{z}})\pm{\mathtt{u}}_{-1, 0}={{\mathrm{v}}}_{\pm}.
\label{vpm_tg}
\end{eqn}
Note that the Fourier transform of the second type of slant bondlength is given by $({\mathtt{u}}_{{{\mathtt{x}}}, {0}}-{\mathtt{u}}_{{{\mathtt{x}}}+1, {-1}})^F({{z}})={z}{{\mathrm{v}}}^F({{z}})$.

\begin{figure}[h]
\centering
{}{{\includegraphics[width=\linewidth]{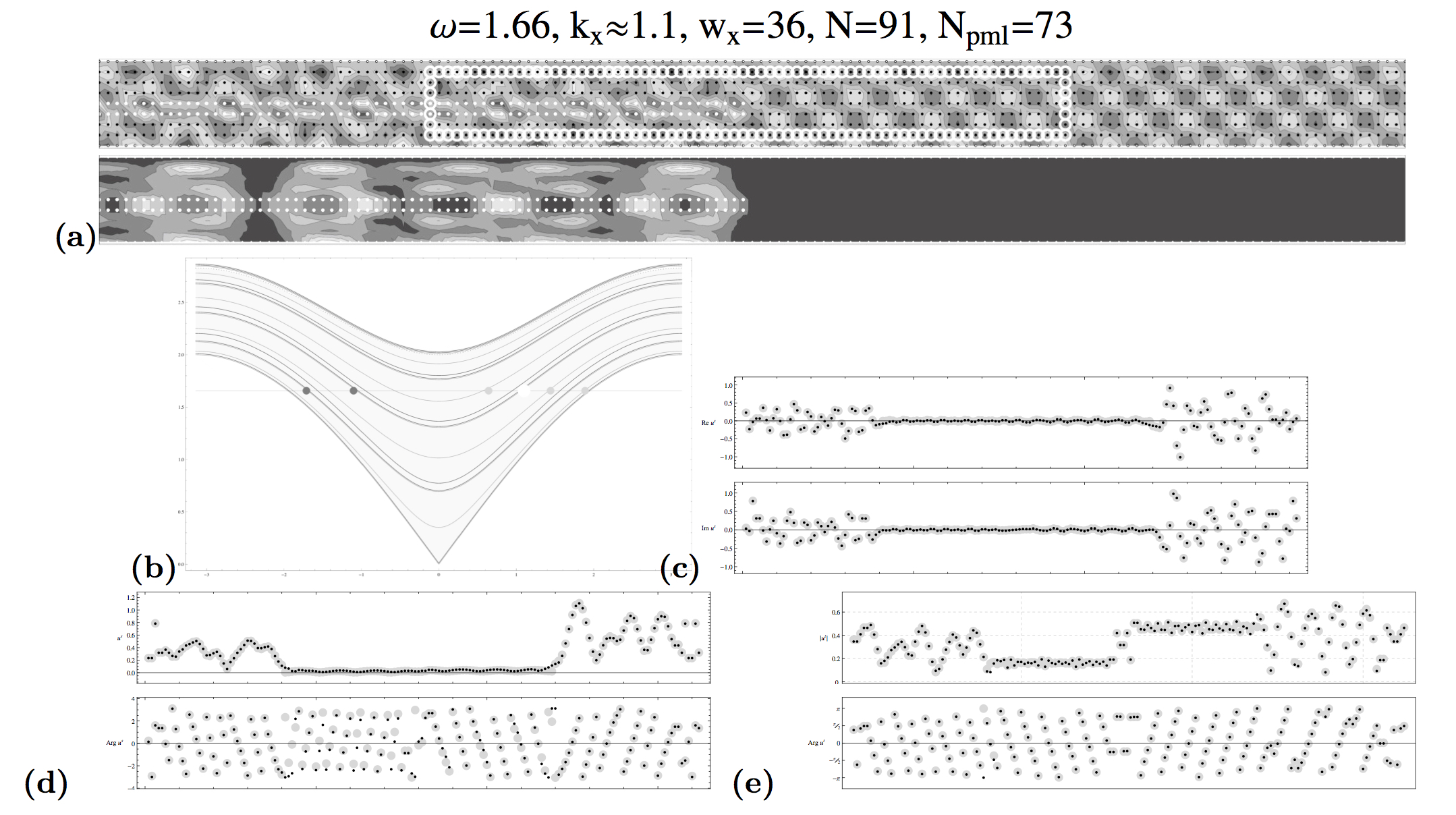}}}
\caption{\footnotesize (a) $\Re{\mathtt{u}}^{{t}}$ (top), $|{\mathtt{u}}^{{s}}|$ (bottom) for 
square lattice structure with ${\mathtt{N}}=9$ and incidence from the tubular side. (b) Dispersion curves for the wave modes ahead (blue) and behind (black) the edge of unzipped portion. (c) Comparison of $\Re{\mathtt{u}}^{{s}}$ and $\Im{\mathtt{u}}^{{s}}$ between the far-field approximation and the numerical solution on a finite grid (a). (d) (resp. (e)) provide the plots of $|{\mathtt{u}}^{{s}}|$ and $\arg{\mathtt{u}}^{{s}}$ (resp. $|{\mathtt{u}}^{{t}}|$ and $\arg{\mathtt{u}}^{{t}}$). The horizontal axis corresponds to the integral labels for the lattice sites forming discrete rectangles of white bubbles shown in (a). }
\label{crackSol_sq_P}\end{figure}
\begin{figure}[h]
\centering
{}{{\includegraphics[width=\linewidth]{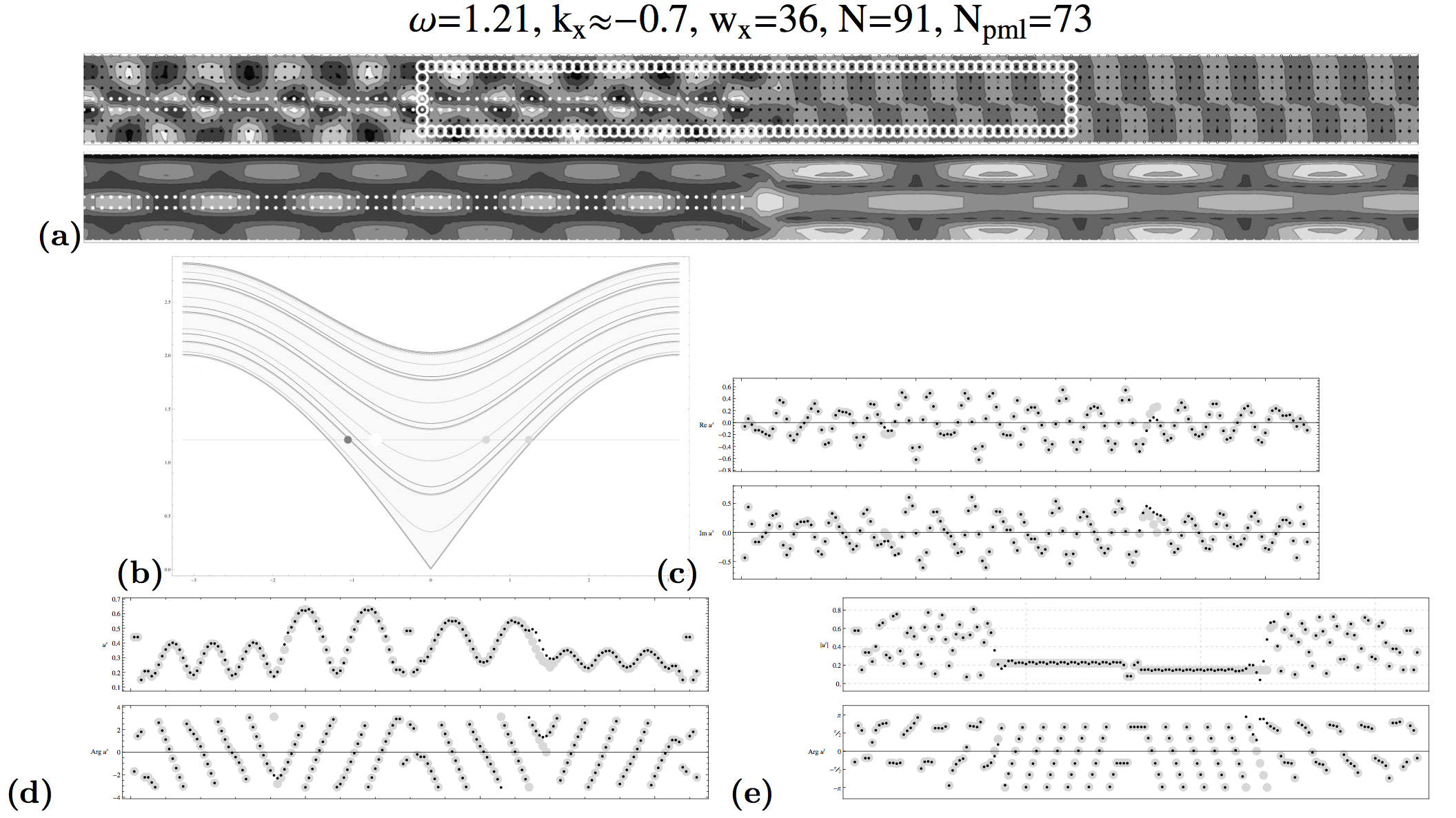}}}
\caption{\footnotesize Illustration for 
square lattice structure with details same as those in Fig. \ref{crack_sq_P} except for the incidence from the unzipped side. }
\label{crackSol_sq_P_altinc}\end{figure}

\section{Scattering matrix}
Under the limit ${\upomega}_2=\Im{\upomega}\to0$ and $|{\mathtt{x}}|\to\infty$, it is easy to see that 
the far-field can be determined (suitably) in terms of the propagating waves associated with the two different portions of the lattice strip. 
Suppose that the symbol ${\mathbb{T}}$ denotes the unit circle (as a counterclockwise contour if applicable) in complex plane. 
Based on the discussion in the paragraphs following equations \eqref{Lk_sq_P} and \eqref{Lk_tg_P}, the sets of ${z}$ corresponding to outgoing (away from edge of unzipped portion) waves are
\begin{eqn}
\hspace{-.1in}{{\mathcal{Z}}}^+_{{{\mathcal{R}}}}=\{{z}\in{\mathbb{T}} \big| {{\mathscr{D}}_+({{z}})=0}\}, {{\mathcal{Z}}}^{-}_{{{\mathcal{L}}}}=\{{z}\in{\mathbb{T}} \big| {{\mathscr{N}}_-({{z}})=0}\}. 
\label{Zer_sq_k}
\end{eqn}
Correspondingly, the sets of ${z}_{{P}}$ for the incoming (towards the edge of unzipped portion) waves are 
\begin{eqn}
\hspace{-.1in}{{\mathcal{Z}}}^{-}_{{{\mathcal{R}}}}=\{{z}\in{\mathbb{T}} \big| {{\mathscr{D}}_-({{z}})=0}\}, 
{{\mathcal{Z}}}^+_{{{\mathcal{L}}}}=\{{z}\in{\mathbb{T}} \big| {{\mathscr{N}}_+({{z}})=0}\}.
\label{Zer_sq_k_inc}
\end{eqn}
Note that $\#{{\mathcal{Z}}}^+_{{{\mathcal{R}}}}=\#{{\mathcal{Z}}}^-_{{{\mathcal{R}}}}$, etc, except at non-generic points corresponding to zero group velocity \cite{Brillouin}.
Suppose that the elements of ${{\mathcal{Z}}}^+_{{{\mathcal{R}}}}$ (resp. ${{\mathcal{Z}}}^-_{{{\mathcal{R}}}}$) are indexed\footnote{For the purpose of the ordering, one possible and easy choice is to associate the index with the dispersion curves labelled according to increasing values of ${\upomega}$ at zero wave number.} by ${\mathsf{a}}$ (resp. ${\srad{{\mathsf{a}}}}$) with a range $1\dotsc N^{{{\mathcal{R}}}}=\#{{\mathcal{Z}}}^+_{{{\mathcal{R}}}}$, while elements of ${{\mathcal{Z}}}^-_{{{\mathcal{L}}}}$ (resp. ${{\mathcal{Z}}}^+_{{{\mathcal{L}}}}$) are indexed by ${\mathsf{b}}$ (resp. ${\srad{{\mathsf{b}}}}$) ranging from $1$ to $N^{{{\mathcal{L}}}}=\#{{\mathcal{Z}}}^-_{{{\mathcal{L}}}}.$ Thus, \eqref{Zer_sq_k} and \eqref{Zer_sq_k_inc} can be written as, respectively,
${{\mathcal{Z}}}^+_{{{\mathcal{R}}}}=\{{z}_{{\mathsf{a}}}\}_{{\mathsf{a}}=1}^{N^{{{\mathcal{R}}}}}, {{\mathcal{Z}}}^{-}_{{{\mathcal{L}}}}=\{{z}_{{\mathsf{b}}}\}_{{\mathsf{b}}=1}^{N^{{{\mathcal{L}}}}},$
and
${{\mathcal{Z}}}^{-}_{{{\mathcal{R}}}}=\{{z}_{\srad{{\mathsf{a}}}}\}_{\srad{{\mathsf{a}}}=1}^{N^{{{\mathcal{R}}}}}, 
{{\mathcal{Z}}}^+_{{{\mathcal{L}}}}=\{{z}_{\srad{{\mathsf{b}}}}\}_{\srad{{\mathsf{b}}}=1}^{N^{{{\mathcal{L}}}}}.$

\begin{figure}[h]
\centering
{{}{\raisebox{-0.05\height}{\includegraphics[width=\linewidth]{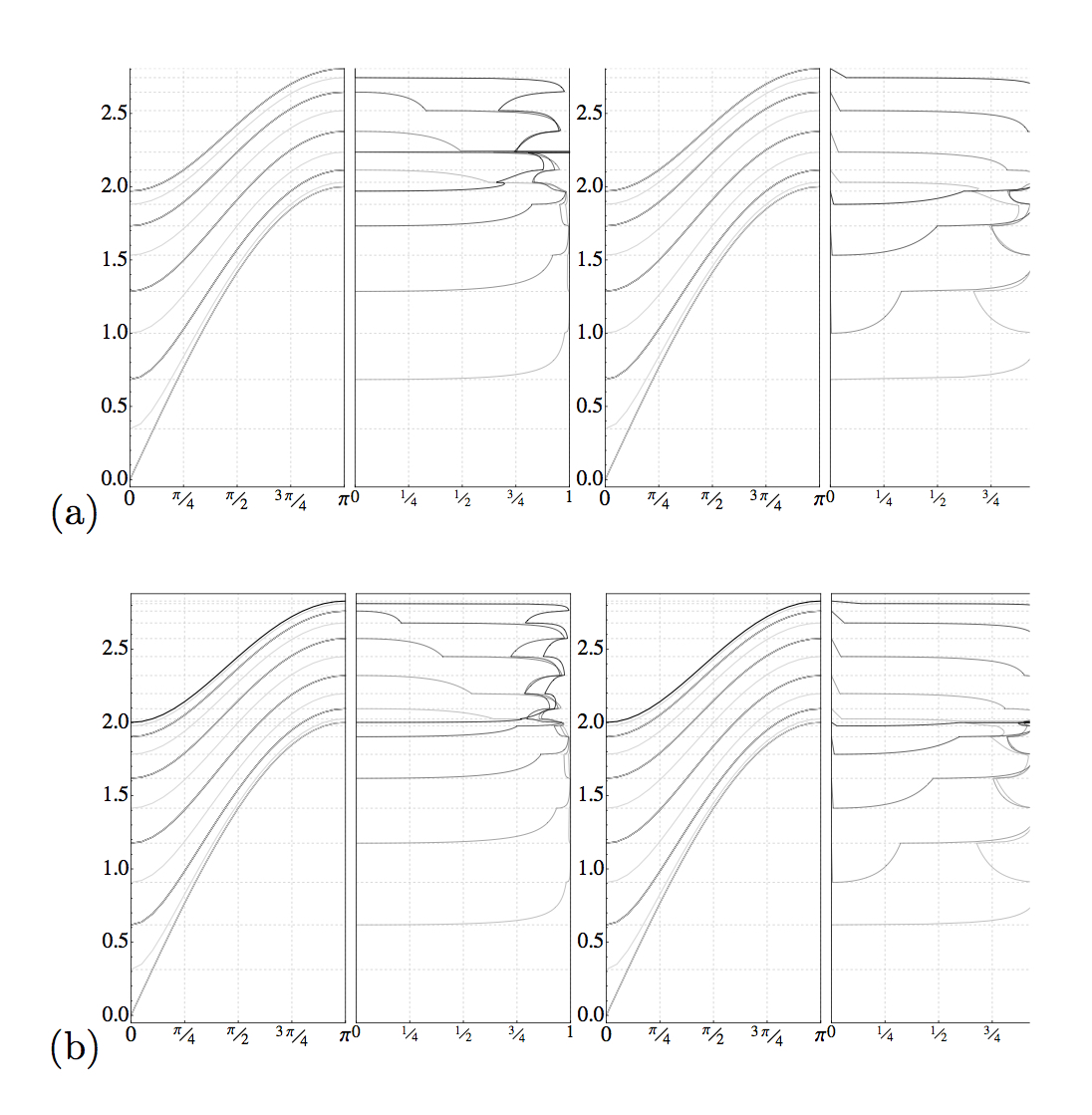}}}}
\caption{Transmittance ${\mathscr{T}}$ for incidence from tube (left) and the unzipped portion (right), for ${\mathtt{N}}=9$ (top) and ${\mathtt{N}}=10$ (bottom). The darker curves for transmittance ${\mathscr{T}}$ correspond to incident wave with wavenumber lying on the higher dispersion curve. The vertical (frequency) axis demonstrates the correspondence between dispersion curves for the two sides of the partly unzipped lattice tube
and critical values attained in transmittance.}\label{latticestrip_Transmittance_sq}\end{figure}

\begin{figure}[h]\centering
{{}{\raisebox{-0.05\height}{\includegraphics[width=\linewidth]{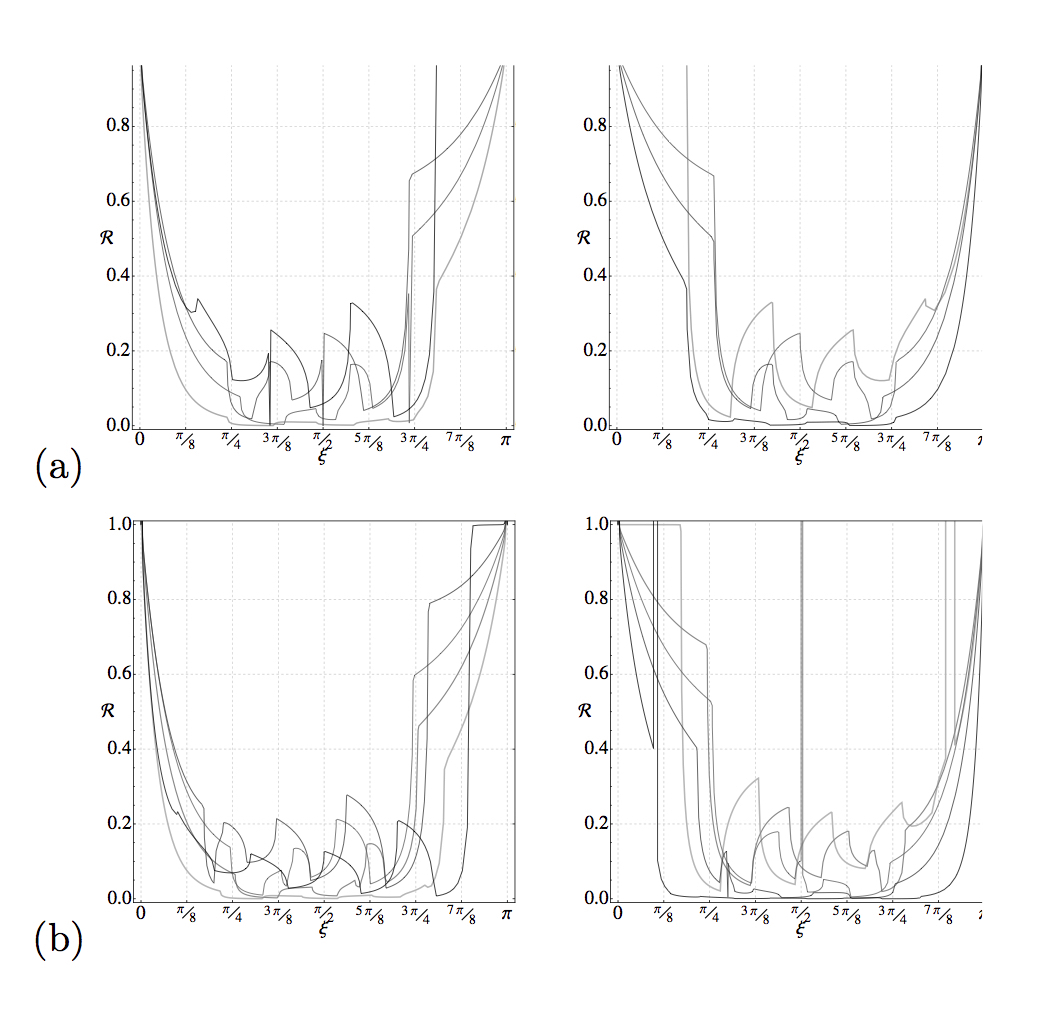}}}}
\caption{Reflectance ${\mathscr{R}}$ (left: incidence from tube, right: incidence from the lattice strip) for square lattice structure, (a) ${\mathtt{N}}=9,$ (b) ${\mathtt{N}}=10$. 
The critical values attained in reflectance ${\mathscr{R}}$ correspond to the critical values of the dispersion relation. }\label{latticestrip_Reflectance_sq}\end{figure}

\subsection{Square lattice structure}
\label{reftrans_sq}
The asymptotic expression for the bond lengths \eqref{vspring_sq} between the rows at ${\mathtt{y}}=0$ and ${\mathtt{y}}={\mathtt{N}}-1\simeq-1$, i.e., ${\mathtt{v}}$, can be obtained by analyzing \eqref{vpm_sq} with
${\mathtt{v}}_{{\mathtt{x}}}=\frac{1}{2\pi i}\oint_{{\mathbb{T}}} {\mathtt{v}}_\pm({{z}}){{z}}^{{\mathtt{x}}-1}d{{z}},$ 
as ${\mathtt{x}}\to\pm\infty$.
Indeed, by applying residue calculus, 
using \eqref{CpmK_sq} and \eqref{Zer_sq_k}, with ${{{\mathrm{v}}}_{(\srad{{\mathsf{a}}})}}={{{\mathrm{v}}}_{({{{\kappa}}^{i}})}}$,
\begin{eqn}
{\mathtt{v}}_{{\mathtt{x}}}&\sim{{\mathrm{A}}}{{{\mathrm{v}}}_{(\srad{{\mathsf{a}}})}}\frac{{\mathscr{D}}_+({{z}}_{\srad{{\mathsf{a}}}})}{{\mathscr{N}}_+({{z}}_{\srad{{\mathsf{a}}}})}\sum\limits_{{\mathsf{a}}=1}^{N^{{{\mathcal{R}}}}}\frac{{\mathscr{N}}_+({{z}}_{{\mathsf{a}}})}{{\mathscr{D}}'_+({{z}}_{{\mathsf{a}}})}\frac{{{z}}_{{\mathsf{a}}}^{{\mathtt{x}}}}{{{z}}_{{\mathsf{a}}}-{{z}}_{\srad{{\mathsf{a}}}}}, {\mathtt{x}}\to+\infty,\\
{\mathtt{v}}_{{\mathtt{x}}}&\sim{{\mathrm{A}}}{{{\mathrm{v}}}_{(\srad{{\mathsf{a}}})}}(-{{z}}_{\srad{{\mathsf{a}}}}^{{\mathtt{x}}}+\frac{{\mathscr{D}}_+({{z}}_{\srad{{\mathsf{a}}}})}{{\mathscr{N}}_+({{z}}_{\srad{{\mathsf{a}}}})}\sum\limits_{{\mathsf{b}}=1}^{N^{{{\mathcal{L}}}}}\frac{{\mathscr{D}}_-({{z}}_{{\mathsf{b}}})}{{\mathscr{N}}'_-({{z}}_{{\mathsf{b}}})}\frac{{{z}}_{{\mathsf{b}}}^{{\mathtt{x}}}}{{{z}}_{{\mathsf{b}}}-{{z}}_{\srad{{\mathsf{a}}}}}), {\mathtt{x}}\to-\infty.
\label{vm0asym_sq}
\end{eqn}
The asymptotic expression for the corresponding total field is given by ${\mathtt{v}}^{{t}}_{{\mathtt{x}}}={\mathtt{v}}^{i}_{{\mathtt{x}}}+{\mathtt{v}}_{{\mathtt{x}}}$. 
The eigenmodes for a square lattice waveguide with periodic or free boundary are well known (see also \cite{Bls9} for a systematic catalogue). 
Thus, it is required that ahead and behind the edge of unzipped portion
\begin{eqn}
{\mathtt{u}}_{{\mathtt{x}}, {\mathtt{y}}}\sim{{\mathrm{A}}}\sum\nolimits_{{\mathsf{a}}=1}^{N^{{{\mathcal{R}}}}}{{\mathrm{A}}}_{{{\mathsf{a}}}}{{a}}_{({{\mathsf{a}}}){{\mathtt{y}}}}{{z}}_{{\mathsf{a}}}^{{\mathtt{x}}}, {\mathtt{x}}\to+\infty,
{\mathtt{u}}_{{\mathtt{x}}, {\mathtt{y}}}\sim-{{\mathrm{A}}}{{a}}_{({\srad{{\mathsf{a}}}}){{\mathtt{y}}}}{{z}}_{\srad{{\mathsf{a}}}}^{{\mathtt{x}}}+{{\mathrm{A}}}\sum\nolimits_{{\mathsf{b}}=1}^{N^{{\mathcal{L}}}}{{\mathrm{A}}}_{{{\mathsf{b}}}}{{a}}_{({{\mathsf{b}}}){{\mathtt{y}}}}{{z}}_{{\mathsf{b}}}^{{\mathtt{x}}}, {\mathtt{x}}\to-\infty.
\label{modeexpand_sq}
\end{eqn}
Comparing \eqref{vm0asym_sq} with the expression of ${\mathtt{v}}$ (${\mathtt{v}}_{{\mathtt{x}}}={\mathtt{u}}_{{\mathtt{x}}, 0}-{\mathtt{u}}_{{\mathtt{x}}, -1}$) based on \eqref{modeexpand_sq}, the coefficients ${{\mathrm{A}}}_{{{\mathsf{a}}}}$, ${{\mathrm{A}}}_{{{\mathsf{b}}}}$ can be found.
Thus, the total displacement field is given by
\begin{eqn}
{\mathtt{u}}^{{t}}_{{\mathtt{x}}, {\mathtt{y}}}\sim{{\mathrm{A}}}{{a}}_{(\srad{{\mathsf{a}}}){{\mathtt{y}}}}{{z}}_{\srad{{\mathsf{a}}}}^{{\mathtt{x}}}+{{\mathrm{A}}}\frac{{\mathscr{D}}_+({{z}}_{\srad{{\mathsf{a}}}})}{{\mathscr{N}}_+({{z}}_{\srad{{\mathsf{a}}}})}\sum\limits_{{\mathsf{a}}=1}^{N^{{{\mathcal{R}}}}}\frac{{{{\mathrm{v}}}_{(\srad{{\mathsf{a}}})}}}{{{\mathrm{v}}}_{({{\mathsf{a}}})}}\frac{{{a}}_{({{\mathsf{a}}}){{\mathtt{y}}}}{{z}}_{{\mathsf{a}}}^{{\mathtt{x}}}}{{{z}}_{{\mathsf{a}}}-{{z}}_{\srad{{\mathsf{a}}}}} \frac{{\mathscr{N}}_+({{z}}_{{\mathsf{a}}})}{{\mathscr{D}}'_+({{z}}_{{\mathsf{a}}})},
{\mathtt{u}}^{{t}}_{{\mathtt{x}}, {\mathtt{y}}}\sim{{\mathrm{A}}}\frac{{\mathscr{D}}_+({{z}}_{\srad{{\mathsf{a}}}})}{{\mathscr{N}}_+({{z}}_{\srad{{\mathsf{a}}}})}\sum\limits_{{\mathsf{b}}=1}^{N^{{{\mathcal{L}}}}}\frac{{{{\mathrm{v}}}_{(\srad{{\mathsf{a}}})}}}{{{\mathrm{v}}}_{({{\mathsf{b}}})}}\frac{{{a}}_{({{\mathsf{b}}}){{\mathtt{y}}}}{{z}}_{{\mathsf{b}}}^{{\mathtt{x}}}}{{{z}}_{{\mathsf{b}}}-{{z}}_{\srad{{\mathsf{a}}}}} \frac{{\mathscr{D}}_-({{z}}_{{\mathsf{b}}})}{{\mathscr{N}}'_-({{z}}_{{\mathsf{b}}})},
\label{farfield_k_sq}
\end{eqn}
as ${\mathtt{x}}\to+\infty$ and ${\mathtt{x}}\to-\infty$, respectively, where ${{{\mathrm{v}}}_{(\srad{{\mathsf{a}}})}}={{a}}_{(\srad{{\mathsf{a}}}){0}}-{{a}}_{(\srad{{\mathsf{a}}}){{\mathtt{N}}-1}}$, ${{{\mathrm{v}}}_{({{\mathsf{a}}})}}={{a}}_{({{\mathsf{a}}}){0}}-{{a}}_{({{\mathsf{a}}}){{\mathtt{N}}-1}}$ and ${{{\mathrm{v}}}_{({{\mathsf{b}}})}}={{a}}_{({{\mathsf{b}}}){0}}-{{a}}_{({{\mathsf{b}}}){{\mathtt{N}}-1}}$. 
For the wave incidence from the unizipped portion (corresponding to index $\srad{{\mathsf{b}}}$), an expression similar to \eqref{farfield_k_sq} can be also obtained though the details are omitted.
The part (a) of Fig. \ref{crackSol_sq_P} and Fig. \ref{crackSol_sq_P_altinc} provides an illustration of the total and scattered displacement due to the partially unzipped tube ${\mathfrak{S}\hspace{-.4ex}}{\mathbin{\substack{{\circledcirc}\\{\circ}}}}$. The dispersion curves for the portion of the waveguide ahead and behind the unzipped portion are shown in part (b) of the same figures; the black dots represent the wave numbers in the region ${\mathcal{R}}$ while gray dots represent those in ${\mathcal{L}}$, white dot represents the incident wave. These illustrations of the numerical solution are based on the scheme summarized in an Appendix of \cite{Bls9s}.
A comparison between the far-field approximation based on the analytical solution, i.e. \eqref{farfield_k_sq}, and the numerical solution on a finite grid is shown in the same figures in (c), (d), and (e), where the displacement of particles located at lattice sites forming a discrete rectangle (shown as white bubbles in top of part (a)) with two sides aligned with `boundary' of the strips is plotted. 
Note that the far field displacement, for any of the lattice waveguides, is dominated by the wave modes in the pass band allowed by the corresponding dispersion curves as indicated by green dots for the transmitted wave modes and red dots for the reflected wave modes (also see Fig. \ref{sqtg_tube_phonons}).

Suppose that the symbol ${{\upxi}}$ denotes the real wave number in horizontal direction, while ${\upeta}$ denotes its real or complex equivalent along the vertical direction. 
The reflectance (resp. transmittance) is obtained by taking the ratio of the total energy flux in outgoing wavemodes ahead of (resp. behind) the edge of unzipped portion vs the energy flux carried by the incident wave mode \cite{Brillouin}. With ${\upeta}$ independent of ${\upxi}$ for a wave mode in the square lattice tube/lattice strip, ${\upomega}^2=4(\sin^2{\frac{1}{2}}{\upxi}+\sin^2{\frac{1}{2}}{\upeta}), {\upxi}\in [-\pi, \pi].$ Hence, the group velocity \cite{Brillouin} of the propagating wave with wave number ${\upxi}$ is
${\mathtt{V}_g}({\upxi}){:=}\pd{{\upomega}}{{\upxi}}={\upomega}^{-1}\sin{\upxi}$. The energy flux, for instance, for the incident wave mode is given by
\begin{eqn}
{\mathscr{E}^{i}}&=\sum\limits_{{\mathtt{y}}\in{\mathbb{Z}}_0^{{\mathtt{N}}-1}}|{{\mathrm{A}}}|^2|{{a}}_{(\srad{{\mathsf{a}}}){{\mathtt{y}}}}|^2|{\mathtt{V}_g}({\upxi}_{\srad{{\mathsf{a}}}})|={\upomega}^{-1}|{{\mathrm{A}}}|^2\sin|{\upxi}_{\srad{{\mathsf{a}}}}|,
\label{energyflux_inc}
\end{eqn}
where ${\upxi}_{\srad{{\mathsf{a}}}}={\upkappa}_{{\mathtt{x}}}>0$ when the wave is incident from the tubular portion while it is $<0$ when the wave is incident from the unzipped lattice strip. The expression for the reflectance and the transmittance (using the far-field expansion \eqref{farfield_k_sq} for a wave incident from the tubular portion) is given by.
\beqans
\hspace{-.45in}
{\mathscr{R}}_{{{\mathcal{L}}}\leftarrow{{\mathcal{R}}}}&=&\frac{-{{\mathtt{V}_g}}({\upxi}_{\srad{{\mathsf{a}}}})^{-1}}{|{\sLNsq}_{{}+}({{z}}_{\srad{{\mathsf{a}}}})|^2}\sum\limits_{{\mathsf{a}}=1}^{N^{{{\mathcal{R}}}}}\bigg|\frac{{{{\mathrm{v}}}_{(\srad{{\mathsf{a}}})}}}{{{\mathrm{v}}}_{+({{\mathsf{a}}})}}\bigg|^2\frac{{{\mathtt{V}_g}}({\upxi})}{|{{z}}_{{\mathsf{a}}}-{{z}}_{\srad{{\mathsf{a}}}}|^2} \bigg|\frac{{\mathscr{N}}_+({{z}}_{{\mathsf{a}}})}{{\mathscr{D}}'_+({{z}}_{{\mathsf{a}}})}\bigg|^2, \label{R_sq_P_eq1}\\
\hspace{-.45in}{\mathscr{T}}_{{{\mathcal{L}}}\leftarrow{{\mathcal{R}}}}&=&\frac{{{\mathtt{V}_g}}({\upxi}_{\srad{{\mathsf{a}}}})^{-1}}{|{\sLNsq}_{{}+}({{z}}_{\srad{{\mathsf{a}}}})|^2}\sum\limits_{{\mathsf{b}}=1}^{N^{{{\mathcal{L}}}}}\bigg|\frac{{{{\mathrm{v}}}_{(\srad{{\mathsf{a}}})}}}{{{\mathrm{v}}}_{-({{\mathsf{b}}})}}\bigg|^2\frac{{{\mathtt{V}_g}}({\upxi})}{|{{z}}_{{\mathsf{b}}}-{{z}}_{\srad{{\mathsf{a}}}}|^2} \bigg|\frac{{\mathscr{D}}_-({{z}}_{{\mathsf{b}}})}{{\mathscr{N}}'_-({{z}}_{{\mathsf{b}}})}\bigg|^2.\label{T_sq_P_eq1}
\eeqans{RT_sq}
In \eqref{R_sq_P_eq1}, the minus sign appears
because the group velocity of incident and reflected waves are opposite in sign.

In the tubular portion (of period ${\mathtt{N}}$), i.e. ${\mathfrak{S}{{\circledcirc}}}$, it is easy to see that
${{{\mathrm{v}}}_{({{\mathsf{a}}})}}=\sqrt{1/{\mathtt{N}}}2i\sin{\frac{1}{2}}{{\upeta}_{{\kappa}}}e^{i({\mathrm{N}}+{\frac{1}{2}}){{\upeta}_{{\kappa}}}}$ with ${\upeta}_{{\kappa}}=2{\kappa}\pi/{\mathtt{N}}$ for a suitable ${\kappa}$
 \cite{Bls9}. 
And for the unzipped portion, ${{{\mathrm{v}}}_{({{\mathsf{b}}})}}=2/\sqrt{{\mathtt{N}}}\cos{{\frac{1}{2}}}{{\upeta}}$ for an appropriate ${\upeta}$ \cite{Bls9}.
After various manipulations, analagous to those applied for the bifurcated waveguides of square lattice \cite{Bls9s}, it is found that the reflectance ${\mathscr{R}}_{{{\mathcal{L}}}\leftarrow{{\mathcal{R}}}}$ and transmittance ${\mathscr{T}}_{{{\mathcal{L}}}\leftarrow{{\mathcal{R}}}}$ are given by the following general expressions for the wave incidence from the tubular portion (corresponding to index $\srad{{\mathsf{a}}}$):
\begin{subequations}
\begin{eqn}
\hspace{-.25in}{\mathscr{R}}_{{{\mathcal{L}}}\leftarrow{{\mathcal{R}}}}={\mathtt{C}}_{RT}\sum\nolimits_{{\mathsf{a}}=1}^{N^{{{\mathcal{R}}}}}\frac{\overline{{\mathscr{D}}_-({{z}}_{{\mathsf{a}}}){\mathscr{N}}_+({{z}}_{{\mathsf{a}}})}}{{\mathscr{N}}_-({{z}}_{{\mathsf{a}}}){\mathscr{D}}'_+({{z}}_{{\mathsf{a}}})}\frac{{{z}}_{\srad{{\mathsf{a}}}}}{({{z}}_{{\mathsf{a}}}-{{z}}_{\srad{{\mathsf{a}}}})^2},
{\mathscr{T}}_{{{\mathcal{L}}}\leftarrow{{\mathcal{R}}}}={\mathtt{C}}_{RT}\sum\nolimits_{{\mathsf{b}}=1}^{N^{{{\mathcal{L}}}}}\frac{\overline{{\mathscr{D}}_-({{z}}_{{\mathsf{b}}}){\mathscr{N}}_+({{z}}_{{\mathsf{b}}})}}{{\mathscr{N}}'_-({{z}}_{{\mathsf{b}}}){\mathscr{D}}_+({{z}}_{{\mathsf{b}}})}\frac{{{z}}_{\srad{{\mathsf{a}}}}}{({{z}}_{{\mathsf{b}}}-{{z}}_{\srad{{\mathsf{a}}}})^2},
\label{RefTrans_sq_P}
\end{eqn}
\begin{eqn}
\text{where }
{\mathtt{C}}_{RT}&=\frac{{{z}}_{\srad{{\mathsf{a}}}}{\mathscr{N}}_-({{z}}_{\srad{{\mathsf{a}}}}){\mathscr{D}}_+({{z}}_{\srad{{\mathsf{a}}}})}{\overline{{\mathscr{D}}'_-({{z}}_{\srad{{\mathsf{a}}}})}\overline{{\mathscr{N}}_+({{z}}_{\srad{{\mathsf{a}}}})}}.
\label{CRexp_sq}
\end{eqn}
For the wave incidence from the unizipped portion (corresponding to index $\srad{{\mathsf{b}}}$), it is found that the transmittance ${\mathscr{T}}_{{{\mathcal{L}}}\to{{\mathcal{R}}}}$ and reflectance ${\mathscr{R}}_{{{\mathcal{L}}}\to{{\mathcal{R}}}}$ are respectively given by \eqref{RefTrans_sq_P}, i.e. interchange the expressions for incidence from the tube, with
\begin{eqn}
{\mathtt{C}}_{RT}&=\frac{{{z}}_{\srad{{\mathsf{b}}}}{\mathscr{N}}_-({{z}}_{\srad{{\mathsf{b}}}}){\mathscr{D}}_+({{z}}_{\srad{{\mathsf{b}}}})}{\overline{{\mathscr{N}}'_+({{z}}_{\srad{{\mathsf{b}}}})}\overline{{\mathscr{D}}_-({{z}}_{\srad{{\mathsf{b}}}})}}.
\label{CRexp_sq_altinc}
\end{eqn}
\end{subequations}
The expression \eqref{RefTrans_sq_P} for transmittance ${\mathscr{T}}_{{{\mathcal{L}}}\leftarrow{{\mathcal{R}}}}$ and its counterpart ${\mathscr{T}}_{{{\mathcal{L}}}\to{{\mathcal{R}}}}$ is plotted in Fig. \ref{latticestrip_Transmittance_sq} with ${\upomega}$ as a variable on horizontal axis, while the expression \eqref{RefTrans_sq_P} for the corresponding reflectance ${\mathscr{R}}_{{{\mathcal{L}}}\leftarrow{{\mathcal{R}}}}$ and its counterpart ${\mathscr{R}}_{{{\mathcal{L}}}\to{{\mathcal{R}}}}$ is plotted in Fig. \ref{latticestrip_Reflectance_sq} with ${\upxi}$ as a variable on horizontal axis.

\begin{figure}[h]
\centering
{\includegraphics[width=\linewidth]{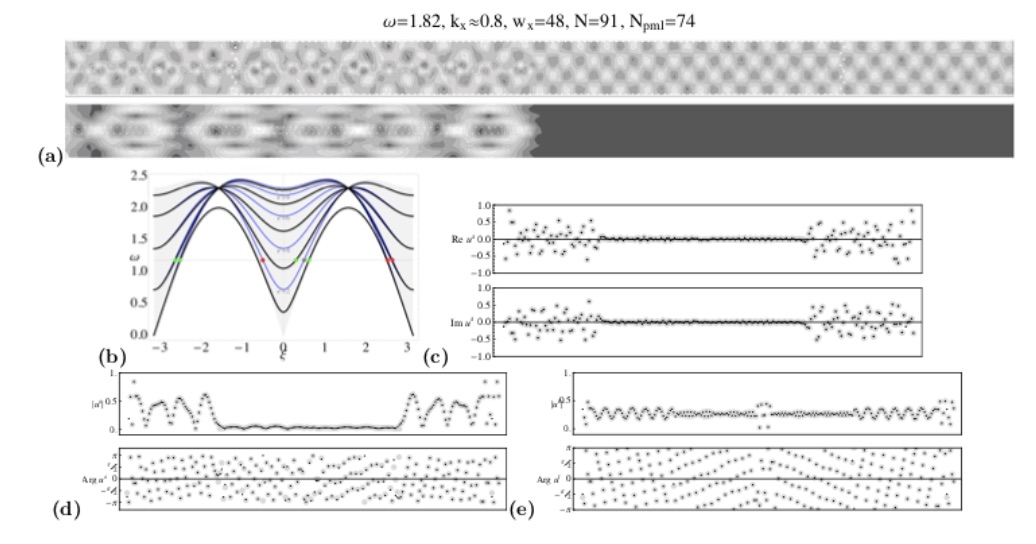}}
\caption{\footnotesize (a) $\Re{\mathtt{u}}^{{t}}$ (top), $|{\mathtt{u}}^{{s}}|$ (bottom) for 
triangular lattice structure with ${\mathrm{N}}=5$ and incidence from the tubular side. (b) Dispersion curves for the wave modes ahead (blue), with odd symmetry, and behind (black) the edge of unzipped portion. (c) Comparison of $\Re{\mathtt{u}}^{{s}}$ and $\Im{\mathtt{u}}^{{s}}$ between the far-field approximation and the numerical solution on a finite grid (a). (d) (resp. (e)) provide the plots of $|{\mathtt{u}}^{{s}}|$ and $\arg{\mathtt{u}}^{{s}}$ (resp. $|{\mathtt{u}}^{{t}}|$ and $\arg{\mathtt{u}}^{{t}}$). The horizontal axis corresponds to the integral labels for the lattice sites forming discrete rectangles of white bubbles shown in (a). {N in the plot label refers to N$_{\text{grid}}$.}}
\label{crackSol_tg_P}\end{figure}

\begin{figure}[h]
\centering
{\includegraphics[width=\linewidth]{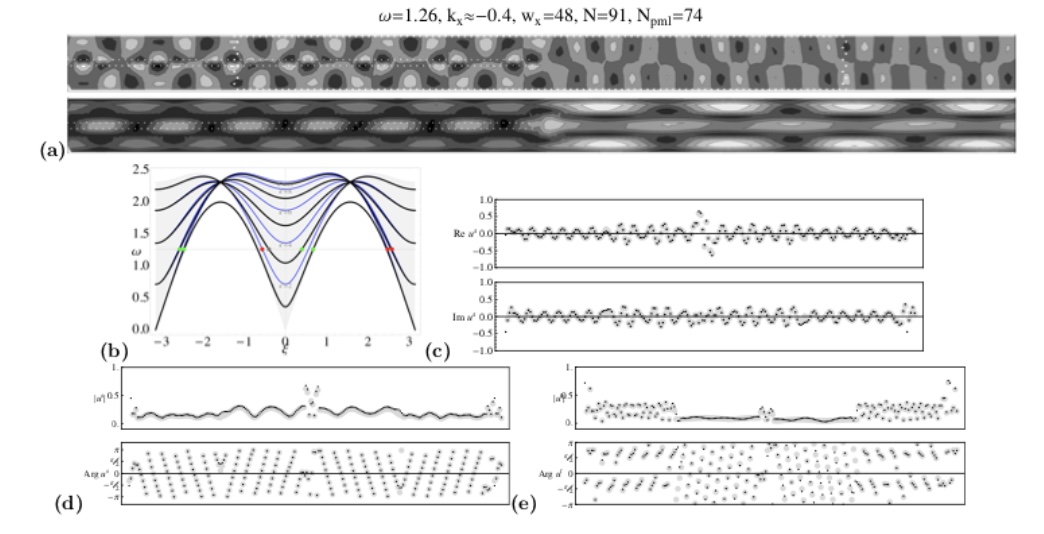}}
\caption{\footnotesize Illustration for triangular lattice structure with details same as those in Fig. \ref{crack_tg_P} except for the incidence from the unzipped side. }
\label{crackSol_tg_P_altinc}\end{figure}

\subsection{Triangular lattice structure}
\label{reftrans_tg}

Taking the limit ${\upomega}_2\to0^+$ and considering $|{\mathtt{x}}|\to\infty$, 
using \eqref{CpmK_tg}, \eqref{uzfull_tg}, and 
\eqref{vspring_tg}, further simplification of \eqref{vpm_tg} can be carried out for the incidence from tubular portion (corresponding to index $\srad{{\mathsf{a}}}$).
As a result the following asymptotic expression holds,
\begin{eqn}
{\mathtt{v}}_{{\mathtt{x}}}
&\sim{{\mathrm{A}}}{{a}}_{(\srad{{\mathsf{a}}}){0}}({1+{{z}}_{\srad{{\mathsf{a}}}}^{-1}})\frac{{\mathscr{D}}_+({{z}}_{\srad{{\mathsf{a}}}})}{{\mathscr{N}}_+({{z}}_{\srad{{\mathsf{a}}}})}\sum\nolimits_{{\mathsf{a}}=1}^{N^{{{\mathcal{R}}}}}
\frac{1}{{{z}}_{{\mathsf{a}}}-{{z}}_{\srad{{\mathsf{a}}}}}\frac{{\mathscr{N}}_+({{z}}_{{\mathsf{a}}})}{{\mathscr{D}}'_+({{z}}_{{\mathsf{a}}})}{{{z}}_{{\mathsf{a}}}^{{\mathtt{x}}}}, \\
{\mathtt{v}}_{{\mathtt{x}}}&\sim{{\mathrm{A}}}{{a}}_{(\srad{{\mathsf{a}}}){0}}(1+{z}_{\srad{{\mathsf{a}}}}^{-1})(-{z}_{\srad{{\mathsf{a}}}}^{{\mathtt{x}}}
+\frac{{\mathscr{D}}_+({{z}}_{\srad{{\mathsf{a}}}})}{{\mathscr{N}}_+({{z}}_{\srad{{\mathsf{a}}}})}\sum\nolimits_{{\mathsf{b}}=1}^{N^{{{\mathcal{L}}}}}
\frac{1}{{{z}}_{{\mathsf{b}}}-{{z}}_{\srad{{\mathsf{a}}}}}\frac{{\mathscr{D}}_-({{z}}_{{\mathsf{b}}})}{{\mathscr{N}}'_-({{z}}_{{\mathsf{b}}})}{{z}}_{{\mathsf{b}}}^{{\mathtt{x}}}), 
\label{vm0asym_tg}
\end{eqn}
as ${\mathtt{x}}\to\pm\infty,$ respectively,
where the values of ${z}$, corresponding to the outgoing wave modes, that appear in the summands are elements defined in \eqref{Zer_sq_k} but borrowing the kernel suited for the triangular lattice.

For the case of incidence from the unzipped portion of the waveguide (corresponding to index $\srad{{\mathsf{b}}}$), analysis similar to above 
leads to the following asymptotic expression (on the lattice ${\mathfrak{T}\hspace{-.4ex}}{\mathbin{\substack{{\circledcirc}\\{\circ}}}}$ where ${\mathtt{x}}\in2{\mathbb{Z}}$ at ${\mathtt{y}}=0$), as ${\mathtt{x}}\to+\infty$ and ${\mathtt{x}}\to-\infty$, respectively, 
\begin{eqn}
{\mathtt{v}}_{{\mathtt{x}}}&\sim-{{\mathrm{A}}}{{a}}_{(\srad{{\mathsf{b}}}){0}}{z}_{\srad{{\mathsf{a}}}}^{{\mathtt{x}}}
+{{\mathrm{A}}}{{a}}_{(\srad{{\mathsf{b}}}){0}}({1+{{z}}_{\srad{{\mathsf{a}}}}^{-1}})\frac{{\mathscr{N}}_-({{z}}_{\srad{{\mathsf{a}}}})}{{\mathscr{D}}_-({{z}}_{\srad{{\mathsf{a}}}})}
\sum\limits_{{\mathsf{a}}=1}^{N^{{{\mathcal{R}}}}}\frac{1}{{{z}}_{{\mathsf{a}}}-{{z}}_{\srad{{\mathsf{a}}}}}\frac{{\mathscr{N}}_+({{z}}_{{\mathsf{a}}})}{{\mathscr{D}}'_+({{z}}_{{\mathsf{a}}})}{{z}}_{{\mathsf{a}}}^{{\mathtt{x}}},\\
{\mathtt{v}}_{{\mathtt{x}}}&\sim{{\mathrm{A}}}{{a}}_{(\srad{{\mathsf{b}}}){0}}(1+{z}_{\srad{{\mathsf{a}}}}^{-1})\frac{{\mathscr{N}}_-({{z}}_{\srad{{\mathsf{a}}}})}{{\mathscr{D}}_-({{z}}_{\srad{{\mathsf{a}}}})}\sum\limits_{{\mathsf{b}}=1}^{N^{{{\mathcal{L}}}}}\frac{1}{{{z}}_{{\mathsf{b}}}-{{z}}_{\srad{{\mathsf{a}}}}}\frac{{\mathscr{D}}_-({{z}}_{{\mathsf{b}}})}{{\mathscr{N}}'_-({{z}}_{{\mathsf{b}}})}{{z}}_{{\mathsf{b}}}^{{\mathtt{x}}}.
\label{vm0asym_tg_altinc}
\end{eqn}
Same as explained above for the case of square lattice, for example \eqref{modeexpand_sq}, the total displacement in the far-field can be also, directly, represented in terms of the wave modes.
Comparing \eqref{vm0asym_tg} and \eqref{vm0asym_tg_altinc} with the expression of ${\mathtt{v}}$ \eqref{vspring_tg} (i.e. ${\mathtt{v}}_{{\mathtt{x}}}={\mathtt{u}}_{{{\mathtt{x}}}, {0}}-{\mathtt{u}}_{{{\mathtt{x}}}-1, {-1}}$) obtained by substitution of the normal mode expansion, the coefficients ${{\mathrm{A}}}_{{{\mathsf{a}}}}, {{\mathrm{A}}}_{{{\mathsf{b}}}}$ can be found. 
Further, on the lattice ${\mathfrak{T}\hspace{-.4ex}}{\mathbin{\substack{{\circledcirc}\\{\circ}}}}$ where ${\mathtt{x}}\in2{\mathbb{Z}}$ at ${\mathtt{y}}=0$, 
the total displacement field in the far-field is asymptotically given by (with ${{a}}_{({{{\kappa}}^{i}})}={{a}}_{(\srad{{\mathsf{a}}})}\delta_{{{\mathfrak{s}}}, {{\mathcal{R}}}}+{{a}}_{(\srad{{\mathsf{b}}})}\delta_{{{\mathfrak{s}}}, {{\mathcal{R}}}}$)\footnote{In view of Footnote \ref{symmetryaspect}, ${{z}}_{\srad{{\mathsf{a}}}}$ and ${{z}}_{\srad{{\mathsf{b}}}}$ can be replaced by $-{{z}}_{\srad{{\mathsf{a}}}}$ and $-{{z}}_{\srad{{\mathsf{b}}}}$, respectivey, for the case of even symmetry.}
{
\begin{eqn}
{\mathtt{u}}^{{t}}_{{\mathtt{x}}, {\mathtt{y}}}&\sim{\mathtt{u}}_{{\mathtt{x}}, {\mathtt{y}}}^{i}\delta_{{{\mathfrak{s}}}, {{\mathcal{R}}}}+{{\mathrm{A}}}(\frac{{\mathscr{D}}_+({{z}}_{\srad{{\mathsf{a}}}})}{{\mathscr{N}}_+({{z}}_{\srad{{\mathsf{a}}}})}\delta_{{{\mathfrak{s}}}, {{\mathcal{R}}}}+\frac{{\mathscr{N}}_-({{z}}_{\srad{{\mathsf{b}}}})}{{\mathscr{D}}_-({{z}}_{\srad{{\mathsf{b}}}})}\delta_{{{\mathfrak{s}}}, {{\mathcal{L}}}})\sum\nolimits_{{\mathsf{a}}=1}^{N^{{{\mathcal{R}}}}}\frac{{{a}}_{({\srad{{\mathsf{a}}},\srad{{\mathsf{b}}}}){0}}(1+{{z}}_{\srad{{\mathsf{a}}},\srad{{\mathsf{b}}}}^{-1})}{{{a}}_{({{\mathsf{a}}})0}(1+{{z}}_{{\mathsf{a}}}^{-1})}\frac{1}{{{z}}_{{\mathsf{a}}}-{{z}}_{\srad{{\mathsf{a}}},\srad{{\mathsf{b}}}}} \frac{{\mathscr{N}}_+({{z}}_{{\mathsf{a}}})}{{\mathscr{D}}'_+({{z}}_{{\mathsf{a}}})}{{a}}_{({{\mathsf{a}}}){{\mathtt{y}}}}{{z}}_{{\mathsf{a}}}^{{\mathtt{x}}},\\
{\mathtt{u}}^{{t}}_{{\mathtt{x}}, {\mathtt{y}}}&\sim{\mathtt{u}}_{{\mathtt{x}}, {\mathtt{y}}}^{i}\delta_{{{\mathfrak{s}}}, {{\mathcal{L}}}}+{{\mathrm{A}}}(\frac{{\mathscr{D}}_+({{z}}_{\srad{{\mathsf{a}}}})}{{\mathscr{N}}_+({{z}}_{\srad{{\mathsf{a}}}})}\delta_{{{\mathfrak{s}}}, {{\mathcal{R}}}}+\frac{{\mathscr{N}}_-({{z}}_{\srad{{\mathsf{b}}}})}{{\mathscr{D}}_-({{z}}_{\srad{{\mathsf{b}}}})}\delta_{{{\mathfrak{s}}}, {{\mathcal{L}}}})\sum\nolimits_{{\mathsf{b}}=1}^{N^{{{\mathcal{L}}}}}\frac{{{a}}_{({\srad{{\mathsf{a}}},\srad{{\mathsf{b}}}}){0}}(1+{{z}}_{\srad{{\mathsf{a}}},\srad{{\mathsf{b}}}}^{-1})}{{{a}}_{({{\mathsf{b}}})0}(1+{{z}}_{{\mathsf{b}}}^{-1})}\frac{1}{{{z}}_{{\mathsf{b}}}-{{z}}_{\srad{{\mathsf{a}}},\srad{{\mathsf{b}}}}} \frac{{\mathscr{D}}_-({{z}}_{{\mathsf{b}}})}{{\mathscr{N}}'_-({{z}}_{{\mathsf{b}}})}{{a}}_{({{\mathsf{a}}}){{\mathtt{y}}}}{{z}}_{{\mathsf{b}}}^{{\mathtt{x}}}, 
\label{farfield_k_tg}
\end{eqn}
}
as ${\mathtt{x}}\to+\infty$ and ${\mathtt{x}}\to-\infty$, respectively, where ${{\mathfrak{s}}}={{\mathcal{R}}}$ (resp. ${{\mathfrak{s}}}={{\mathcal{L}}}$) for incidence from the tubular 
(resp. unzipped) portion ahead (resp. behind) the edge of unzipped portion. 
The part (a) of Fig. \ref{crackSol_tg_P} and Fig. \ref{crackSol_tg_P_altinc} provides an illustration of the total and scattered displacement due to the partially unzipped tube ${\mathfrak{T}\hspace{-.4ex}}{\mathbin{\substack{{\circledcirc}\\{\circ}}}}$. 
The numerical solution of the wave propagation problem is based on the scheme summarized in an Appendix of \cite{Bls9s}.
These illustrations of the numerical solution of the wave propagation problem are based on the scheme summarized in an Appendix of \cite{Bls4} and its 
analogue for a triangular lattice tube. 
The dispersion curves for the portion of the waveguide ahead and behind the unzipped portion are shown in part (b); the red dots represent the wave numbers in ${\mathcal{R}}$ while green dots represent those in ${\mathcal{L}}$, the incident wave number is denoted by the big gray dot.
A comparison between the far-field approximation based on the analytical solution, i.e. \eqref{farfield_k_tg}, and the numerical solution on a finite grid is shown in the same figures in (c), (d), and (e), where the displacement of particles located at lattice sites forming a discrete rectangle (big white dots) with two sides aligned with `boundary' of the strips is plotted. 
Note that the far field displacement, for any of the lattice waveguides, is dominated by the wave modes in the pass band allowed by the corresponding dispersion curves in part (b). 
The lack of symmetry across ${\upxi}=\pm{\frac{1}{2}}\pi$ in the dispersion relation of waves ahead of the unzipped portion, i.e. blue curves, in Figs. \ref{crackSol_tg_P} and \ref{crackSol_tg_P_altinc} is explained in \cite{Bls9}. 
Recall the last paragraph of \S2.2.
It is clear that (with ${z}=e^{-i{\upxi}}$) when the wave number ${\upxi}$ is around $\pi$, the numerator ${\mathscr{N}}$ and denominator ${\mathscr{D}}$ of the kernel ${\sLNtg}$ exhibit a different behavior as compared to that around $0$. 
This is clear from the expression of ${\sLNtg}_{{}}$ in \eqref{Lk_tg_P} as ${z}$ lies on ${\mathbb{T}}$ near $-1$. The dispersion relations for the odd modes in the portion ahead of the unzipped portion and behind are almost overlapping.

Using \eqref{farfield_k_tg}, for incidence from the tubular portion, the reflectance and the transmittance, respectively, are given by
\begin{subequations}
\beqan
\hspace{-.75in}&&{\mathscr{R}}_{{{\mathcal{L}}}\leftarrow{{\mathcal{R}}}}=\hspace{-.05in}-\frac{{\mathtt{V}_g}({\upxi}_{\srad{{\mathsf{a}}}})^{-1}}{|{\sLNtg}_{{}+}({{z}}_{\srad{{\mathsf{a}}}})|^2}
\hspace{-.05in}\sum\limits_{{\mathsf{a}}=1}^{N^{{{\mathcal{R}}}}}
\bigg|\frac{{{a}}_{(\srad{{\mathsf{a}}}){0}}(1+{{z}}_{\srad{{\mathsf{a}}}}^{-1})}{{{a}}_{({{\mathsf{a}}})0}(1+{{z}}_{{\mathsf{a}}}^{-1})}\bigg|^2
\hspace{-.05in}
\frac{{\mathtt{V}_g}({\upxi}_{{{\mathsf{a}}}})}{|{{z}}_{{\mathsf{a}}}-{{z}}_{\srad{{\mathsf{a}}}}|^2} \bigg|\frac{{\mathscr{N}}_+({{z}}_{{\mathsf{a}}})}{{\mathscr{D}}'_+({{z}}_{{\mathsf{a}}})}\bigg|^2, \label{Rk_tg_P}\\
\hspace{-.75in}&&{\mathscr{T}}_{{{\mathcal{L}}}\leftarrow{{\mathcal{R}}}}=\hspace{-.05in}\frac{{\mathtt{V}_g}({\upxi}_{\srad{{\mathsf{a}}}})^{-1}}{|{\sLNtg}_{{}+}({{z}}_{\srad{{\mathsf{a}}}})|^2}
\hspace{-.05in}
\sum\limits_{{\mathsf{b}}=1}^{N^{{{\mathcal{L}}}}}
\bigg|\frac{{{a}}_{(\srad{{\mathsf{a}}}){0}}(1+{{z}}_{\srad{{\mathsf{a}}}}^{-1})}{{{a}}_{({{\mathsf{b}}})0}(1+{{z}}_{{\mathsf{b}}}^{-1})}\bigg|^2
\hspace{-.05in}
\frac{{\mathtt{V}_g}({\upxi}_{{{\mathsf{b}}}})}{|{{z}}_{{\mathsf{b}}}-{{z}}_{\srad{{\mathsf{a}}}}|^2} \bigg|\frac{{\mathscr{D}}_-({{z}}_{{\mathsf{b}}})}{{\mathscr{N}}'_-({{z}}_{{\mathsf{b}}})}\bigg|^2, \label{Tk_tg_P}\\
\hspace{-.75in}&&\quad
\text{where }
{{z}}=e^{-i{\upxi}}. 
\label{zxidefn}
\eeqan\end{subequations}
For incidence from the unzipped the expressions for ${\mathscr{R}}_{{{\mathcal{L}}}\to{{\mathcal{R}}}}$ and ${\mathscr{T}}_{{{\mathcal{L}}}\to{{\mathcal{R}}}}$ are respectively interchanged with ${\mathscr{T}}_{{{\mathcal{L}}}\leftarrow{{\mathcal{R}}}}$ and ${\mathscr{R}}_{{{\mathcal{L}}}\leftarrow{{\mathcal{R}}}}$ and the coefficient appearing in front of the sum contains $-{|{\sLNtg}_{{}-}^{-1}({{z}}_{\srad{{\mathsf{b}}}})|^2}$ in place of ${|{\sLNtg}_{{}+}({{z}}_{\srad{{\mathsf{a}}}})|^2}$.
The expression ${\mathtt{V}_g}({\upxi})$ denotes the group velocity \cite{Brillouin} of the propagating wave with wave number ${\upxi}$ in the appropriate lattice strip.

\begin{figure}[t]
\centering
{{}{\includegraphics[width=.6\linewidth]{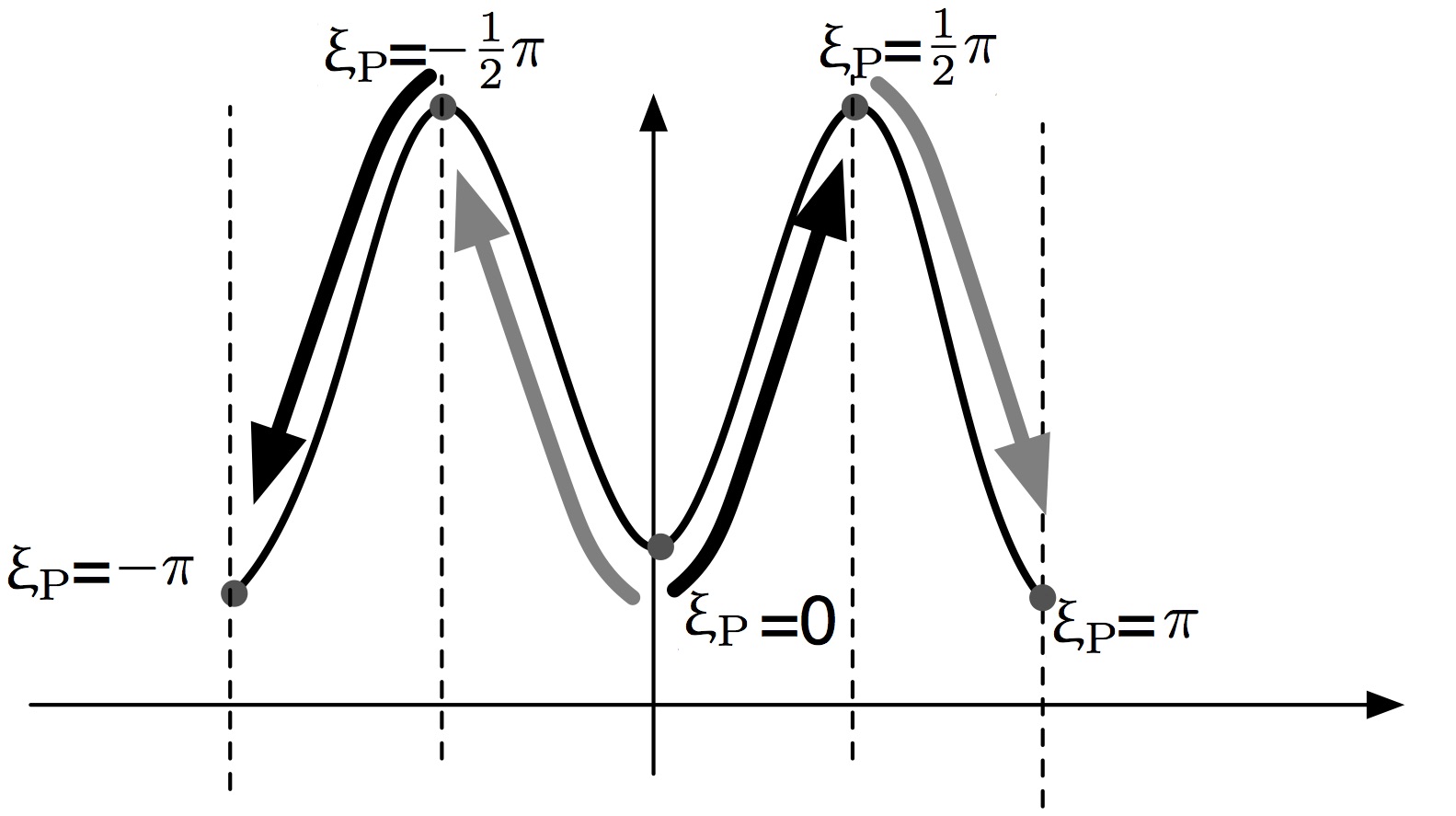}}}
\caption{\footnotesize Choice of ${\upxi}_{\srad{{\mathsf{a}}}}$ (${\upxi}_{\srad{{\mathsf{b}}}}$), so that the group velocity is positive (resp. negative), for wave incident from the tubular (resp. unzipped) portion shown as black (resp. gray) arrow.}
\label{kincchosen}
\end{figure}

Using the periodic boundary condition (on the tubular side), 
${{{a}}_{({{\mathsf{a}}}){0}}=-{{a}}_{({{\mathsf{a}}}){-1}}}=$ $\sqrt{1/{\mathrm{N}}}\sin{\frac{1}{2}}{{\upeta}_{{\kappa}}}$ 
with ${\upeta}_{{\kappa}}={\kappa}\pi/{\mathrm{N}}$ for an appropriate ${\kappa}$
 \cite{Bls9}. 
Similarly, behind the edge of unzipped portion, 
${{{a}}_{({{\mathsf{b}}}){0}}=-{{a}}_{({{\mathsf{b}}}){-1}}}=c(\sin{\mathrm{N}}{\upeta}_{{\mathsf{b}}}+\sin({\mathrm{N}}-1){\upeta}_{{\mathsf{b}}}),$ with $c^{-2}=(1+{\vartheta})({\mathrm{N}}+\frac{\alpha(\alpha{\vartheta}-1)}{1+\alpha^2-2{\vartheta}\alpha}), \alpha=\frac{1+\cos{\upxi}}{1-\cos{\upxi}+2{\vartheta}}$. By reference to \eqref{Lk_tg_P}, ${\mathscr{D}}({{z}})=({z}+{z}^{-1})({\vartheta}+1){{\mathtt{U}}}_{{\mathrm{N}}-1}, {\mathscr{N}}({{z}})={\frac{1}{2}}({z}+{z}^{-1}){{\mathtt{W}}}_{{\mathrm{N}}}-{{\mathtt{W}}}_{{\mathrm{N}}-1}.$ Hence, in the expression \eqref{vm0asym_tg}, (assuming that ${z}\ne\pm1, {\vartheta}\ne-1, {\mathscr{D}}({z})=0$, using \eqref{ChebUVWder}) it is found that 
${\mathscr{D}}'({{z}})=({z}+{z}^{-1})({\vartheta}+1){\vartheta}'({{z}}){\mathrm{N}}({\vartheta}^2-1)^{-1}{\mathtt{T}}_{{\mathrm{N}}}.$ On the other hand, (while observing that ${z}\ne\pm1,$ ${\mathscr{N}}({z})=0$, using \eqref{ChebUVWder}) it is found that
${\mathscr{N}}'({{z}})={\frac{1}{2}}{z}^{-1}({z}-{z}^{-1}){{\mathtt{W}}}_{{\mathrm{N}}}+({\vartheta}-1)^{-1}{\vartheta}'(\cos{\upxi}({\mathrm{N}}+{\frac{1}{2}})(1-\alpha){\mathtt{U}}_{{\mathrm{N}}}-({\mathrm{N}}-{\frac{1}{2}})(\alpha-(2\alpha{\vartheta}-1)){\mathtt{U}}_{{\mathrm{N}}})$ (using ${\mathtt{U}}_{{\mathrm{N}}-1}=\alpha {\mathtt{U}}_{{\mathrm{N}}}, \alpha=(1+\cos{\upxi})/(1-\cos{\upxi}+2{\vartheta})$). 
For the purpose of the manipulations presented below, consider ${\vartheta}({z}, {\upomega})$, i.e. as a function of ${z}$ and ${\upomega}$; same consideration applies to other relevant functions. Thus, treating ${\mathscr{N}}({z}, {\upomega})=0$ as the implicit definition of the dispersion relation ${\upomega}={\Omega}({\upxi})$ ahead of the edge of unzipped portion (since ${\mathscr{N}}(e^{-i{\upxi}}, {\Omega}({\upxi}))=0$)
\begin{subequations}
\begin{eqn}
0&=\od{}{{\upxi}}{\mathscr{N}}(e^{-i{\upxi}}, {\Omega}({\upxi}))|_{e^{-i{\upxi}}={{z}}, {\upomega}={\Omega}({\upxi})}
=-i{z}{\mathscr{N}}'({z}, {\upomega})|_{{{z}}=e^{-i{\upxi}}}+{\mathtt{V}_g}({\upxi})\pd{}{{\upomega}}{\mathscr{N}}({z}, {\upomega})|_{{{z}}=e^{-i{\upxi}}}.
\label{NLdereqn}
\end{eqn}
Similarly, with ${\mathscr{D}}(e^{-i{\upxi}}, {\Omega}({\upxi}))=0$ (corresponding to the dispersion relation in the unzipped portion),
\begin{eqn}
0&=\od{}{{\upxi}}{\mathscr{D}}(e^{-i{\upxi}}, {\Omega}({\upxi}))|_{e^{-i{\upxi}}={{z}}, {\upomega}={\Omega}({\upxi})}
=-i{z}{\mathscr{D}}'({z}, {\upomega})|_{{{z}}=e^{-i{\upxi}}}+{\mathtt{V}_g}({\upxi})\pd{}{{\upomega}}{\mathscr{D}}({z}, {\upomega})|_{{{z}}=e^{-i{\upxi}}}.
\label{DLdereqn}
\end{eqn}
Thus, the following relations are obtained concerning the group velocity of wave modes, ahead and behind the edge of unzipped portion,
\begin{eqn}
\frac{{\mathtt{V}_g}}{{\mathscr{N}}'({z})}=\frac{i{z}}{\pd{}{{\upomega}}{\mathscr{N}}({z}, {\upomega})}|_{{{z}}=e^{-i{\upxi}}}, 
\frac{{\mathtt{V}_g}}{{\mathscr{D}}'({z})}=\frac{i{z}}{\pd{}{{\upomega}}{\mathscr{D}}({z}, {\upomega})}|_{{{z}}=e^{-i{\upxi}}},
\label{NDLder}
\end{eqn} 
respectively.
Using
\eqref{NDLder}, with $\pd{}{{\upomega}}{\mathpzc{Q}}
=-\frac{3}{{2}}\frac{{\upomega}}{\cos{\upxi}}$, as well as \eqref{zxidefn},
\begin{eqn}
\frac{{\mathtt{V}_g}}{{\mathscr{N}}'({z})}
&=\frac{i{z}}{-3{\upomega}}\frac{{2}{\mathpzc{H}}\cos{\upxi}}{\splitfrac{(\cos{\upxi}({\mathrm{N}}+{\frac{1}{2}})(1-\alpha)}{-({\mathrm{N}}-{\frac{1}{2}})(1+\alpha-2\alpha{\vartheta})){\mathtt{U}}_{{\mathrm{N}}}}}, \\
\frac{{\mathtt{V}_g}}{{\mathscr{D}}'({z})}
&=\frac{i{z}{\mathpzc{H}}}{-{3}{{\upomega}}{\mathrm{N}}{\mathtt{T}}_{{\mathrm{N}}}},
\end{eqn}
\end{subequations}
for the respective pieces ahead and behind the edge of unzipped portion.
Then, \eqref{Rk_tg_P} implies
\begin{subequations}
\beqan
\hspace{-.35in}{\mathscr{R}}_{{{\mathcal{L}}}\leftarrow{{\mathcal{R}}}}
&=&-\frac{{\mathtt{V}_g}({\upxi}_{\srad{{\mathsf{a}}}})^{-1}}{|{\sLNtg}_{{}+}({{z}}_{\srad{{\mathsf{a}}}})|^2{\mathrm{N}}}\sum\limits_{{\mathsf{a}}=1}^{N^{{{\mathcal{R}}}}}\frac{\cos^2{\frac{1}{2}}{\upxi}_{\srad{{\mathsf{a}}}}}{\cos^2{\frac{1}{2}}{\upxi}_{{\mathsf{a}}}}\frac{{\mathpzc{H}}({{z}}_{\srad{{\mathsf{a}}}})}{{\mathpzc{H}}({{z}}_{{\mathsf{a}}})}
\frac{1}{{{z}}_{{\mathsf{a}}}-{{z}}_{\srad{{\mathsf{a}}}}}\frac{1}{\overline{{z}}_{{\mathsf{a}}}-\overline{{z}}_{\srad{{\mathsf{a}}}}}
\frac{\overline{{\mathscr{D}}_-({{z}}_{{\mathsf{a}}}){\mathscr{N}}_+({{z}}_{{\mathsf{a}}})}}{{\mathscr{D}}'_+({{z}}_{{\mathsf{a}}}){\mathscr{N}}_-({{z}}_{{\mathsf{a}}})}\frac{i\overline{{z}}_{{\mathsf{a}}}{\mathpzc{H}}{\mathscr{N}}({{z}}_{{\mathsf{a}}})}{{3}{{\upomega}}{\mathtt{T}}_{{\mathrm{N}}}}\notag\\
&=&{\mathtt{C}}_{RT}\sum\limits_{{\mathsf{a}}=1}^{N^{{{\mathcal{R}}}}}\frac{\overline{{\mathscr{D}}_-({{z}}_{{\mathsf{a}}}){\mathscr{N}}_+({{z}}_{{\mathsf{a}}})}}{{\mathscr{D}}'_+({{z}}_{{\mathsf{a}}}){\mathscr{N}}_-({{z}}_{{\mathsf{a}}})}\frac{{{z}}_{\srad{{\mathsf{a}}}}}{({{z}}_{{\mathsf{a}}}-{{z}}_{\srad{{\mathsf{a}}}})^2}, \label{Rk_tg_P_final}\\
{\mathtt{C}}_{RT}&=&\frac{{2}}{3{\upomega}}{i}\frac{{\mathpzc{H}}({{z}}_{\srad{{\mathsf{a}}}})\cos^2{\frac{1}{2}}{\upxi}_{\srad{{\mathsf{a}}}}}{{\mathtt{V}_g}({\upxi}_{\srad{{\mathsf{a}}}})|{\sLNtg}_{{}+}({{z}}_{\srad{{\mathsf{a}}}})|^2{\mathrm{N}}}.
\label{Ck_tg_P}
\eeqan
Note that
$2{{\mathtt{T}}}_{{\mathtt{N}}}={{\mathtt{U}}}_{{\mathtt{N}}}-{{\mathtt{U}}}_{{\mathtt{N}}-2},$ so that when ${{\mathtt{U}}}_{{\mathtt{N}}-1}=0,$ by the relation $2{\vartheta}{{\mathtt{U}}}_{{\mathtt{N}}-1}={{\mathtt{U}}}_{{\mathtt{N}}}+{{\mathtt{U}}}_{{\mathtt{N}}-2},$ it follows that $2{{\mathtt{T}}}_{{\mathtt{N}}}=2{{\mathtt{U}}}_{{\mathtt{N}}}.$ Also
the identity ${{\mathtt{U}}}_{n}^2=1+{{\mathtt{U}}}_{n-1}{{\mathtt{U}}}_{n+1}$ implies ${{\mathtt{U}}}_{{\mathtt{N}}}^2=1$ when ${{\mathtt{U}}}_{{\mathtt{N}}-1}=0,$ so that ${{\mathtt{T}}}_{{\mathtt{N}}}={{\mathtt{U}}}_{{\mathtt{N}}}=\pm1.$ But ${{\mathtt{T}}}_{{\mathtt{N}}}\ne1$ when ${\mathscr{D}}\ne0$ which leaves only one choice. 
Finally, the expected, and elegant, form of the expression \eqref{Rk_tg_P_final} results.

Similarly, (using $(1+\alpha^2-2{\vartheta}\alpha){\mathtt{U}}^2_{{\mathrm{N}}}=1$) it is found that
\beqan
{\mathscr{T}}_{{{\mathcal{L}}}\leftarrow{{\mathcal{R}}}}&=&{2}i\frac{\sin^2{\frac{1}{2}}{{\upeta}_{\srad{{\mathsf{a}}}}}\cos^2{\frac{1}{2}}{\upxi}_{\srad{{\mathsf{a}}}}{\mathtt{V}_g}({\upxi}_{\srad{{\mathsf{a}}}})^{-1}}{3{\upomega}|{\sLNtg}_{{}+}({{z}}_{\srad{{\mathsf{a}}}})|^2{\mathrm{N}}}
\sum\limits_{{\mathsf{b}}=1}^{N^{{{\mathcal{L}}}}}\frac{\overline{{\mathscr{D}}_-({{z}}_{{\mathsf{b}}}){\mathscr{N}}_+({{z}}_{{\mathsf{b}}})}}{{\mathscr{N}}'_-({{z}}_{{\mathsf{b}}}){\mathscr{D}}_+({{z}}_{{\mathsf{b}}})}\frac{{{z}}_{\srad{{\mathsf{a}}}}}{({{z}}_{{\mathsf{b}}}-{{z}}_{\srad{{\mathsf{a}}}})^2}\frac{4{\mathpzc{R}}}{({{\mathtt{U}}}_{{\mathrm{N}}-1}+{{\mathtt{U}}}_{{\mathrm{N}}-2})^2}\notag\\&&
\frac{\cos^2{\upxi}_{{\mathsf{b}}}{\mathtt{U}}_{{\mathrm{N}}-1}}{\cos^2{\frac{1}{2}}{\upxi}_{{\mathsf{b}}}}
\frac{({\mathrm{N}}+\frac{\alpha(\alpha{\vartheta}-1)}{1+\alpha^2-2{\vartheta}\alpha})}{(\cos{\upxi}_{{\mathsf{b}}}({\mathrm{N}}+{\frac{1}{2}})(1-\alpha)-({\mathrm{N}}-{\frac{1}{2}})(1+\alpha-2\alpha{\vartheta})){\mathtt{U}}_{{\mathrm{N}}}}\\
&=&{\mathtt{C}}_{RT}\sum\limits_{{\mathsf{b}}=1}^{N^{{{\mathcal{L}}}}}\frac{\overline{{\mathscr{D}}_-({{z}}_{{\mathsf{b}}}){\mathscr{N}}_+({{z}}_{{\mathsf{b}}})}}{{\mathscr{N}}'_-({{z}}_{{\mathsf{b}}}){\mathscr{D}}_+({{z}}_{{\mathsf{b}}})}\frac{{{z}}_{\srad{{\mathsf{a}}}}}{({{z}}_{{\mathsf{b}}}-{{z}}_{\srad{{\mathsf{a}}}})^2}.
\label{Tk_tg_P_final}
\eeqan\end{subequations}
Indeed, it is easily shown that ${\mathtt{C}}_{RT}$ expressed by \eqref{Ck_tg_P} can be simplified to \eqref{CRexp_sq}, i.e., the same expression as that stated by for the square lattice.
In fact, by inspection it is clear that the relations \eqref{Rk_tg_P_final} and \eqref{Tk_tg_P_final} are same as those for the square lattice structure;
moreover, for the incidence from the unzipped portion, the expressions retains the same form (recall \eqref{CRexp_sq_altinc} and statement preceding it).

\begin{figure}[H]
\centering
{{}{\raisebox{-0.05\height}{\includegraphics[width=.8\linewidth]{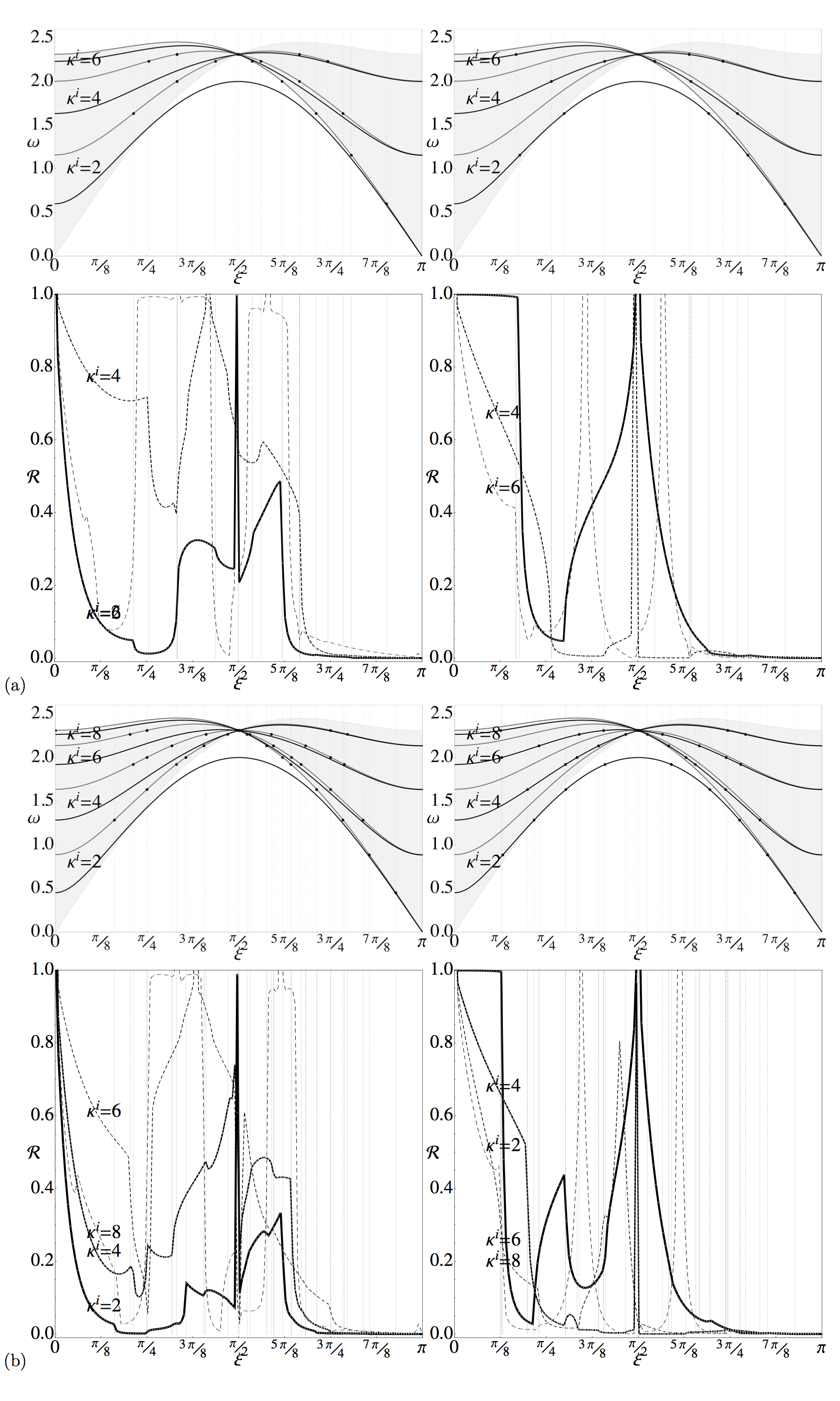}}}}
\caption{\footnotesize Reflectance for the partly unzipped tube with (a) ${\mathtt{N}}=6$ (b) ${\mathtt{N}}=8$. The critical point associated with local maximum of the dispersion curves in the interval $(0, {\frac{1}{2}}\pi)\cup({\frac{1}{2}}\pi, \pi)$ are not shown. The incidence from tubular portion is assumed for the left plot while it is from the unzipped portion for the right plot.
The direction of increase in the variable ${\upxi}_{\srad{{\mathsf{a}}}}$ (resp. ${\upxi}_{\srad{{\mathsf{b}}}}$) is shown in Fig. \ref{kincchosen} with black (resp. gray) curve. 
}
\label{latticestrip_Reflectance_k_tg}
\end{figure}

It is worth a non-trivial note that the final expression for ${\mathscr{R}}$ and ${\mathscr{T}}$ for both types of lattice structures studied in this paper, while accounting for the specific multiplicative Wiener--Hopf factors of the kernel and their zeroes and poles associated with the outgoing wave modes, is {\em identical} to the expression that has been discovered in the context of a general family of bifurcated square lattice strips by \cite{Bls9s}. In particular, with ${{z}}_{{P}}={{z}}_{\srad{{\mathsf{a}}}}\delta_{{{\mathfrak{s}}}, {{\mathcal{R}}}}+{{z}}_{\srad{{\mathsf{b}}}}\delta_{{{\mathfrak{s}}}, {{\mathcal{L}}}},$ in the first two equations,
\beqans
\hspace{-.35in}{\mathscr{R}}\delta_{{{\mathfrak{s}}}, {{\mathcal{R}}}}+{\mathscr{T}}\delta_{{{\mathfrak{s}}}, {{\mathcal{L}}}}={\mathtt{C}}_{RT}\sum\limits_{{\mathsf{a}}=1}^{N^{{{\mathcal{R}}}}}\frac{\overline{{\mathscr{D}}_-({{z}}_{{\mathsf{a}}}){\mathscr{N}}_+({{z}}_{{\mathsf{a}}})}}{{\mathscr{N}}_-({{z}}_{{\mathsf{a}}}){\mathscr{D}}'_+({{z}}_{{\mathsf{a}}})}\frac{{{z}}_{{P}}}{({{z}}_{{\mathsf{a}}}-{{z}}_{{P}})^2},\label{Ref_tg}\\
\hspace{-.35in}{\mathscr{R}}\delta_{{{\mathfrak{s}}}, {{\mathcal{L}}}}+{\mathscr{T}}\delta_{{{\mathfrak{s}}}, {{\mathcal{R}}}}={\mathtt{C}}_{RT}\sum\limits_{{\mathsf{b}}=1}^{N^{{{\mathcal{L}}}}}\frac{\overline{{\mathscr{D}}_-({{z}}_{{\mathsf{b}}}){\mathscr{N}}_+({{z}}_{{\mathsf{b}}})}}{{\mathscr{N}}'_-({{z}}_{{\mathsf{b}}}){\mathscr{D}}_+({{z}}_{{\mathsf{b}}})}\frac{{{z}}_{{P}}}{({{z}}_{{\mathsf{b}}}-{{z}}_{{P}})^2},\label{Trans_tg}\\
\hspace{-.35in}{\mathtt{C}}_{RT}=\frac{{{z}}_{\srad{{\mathsf{a}}}}{\mathscr{N}}_-({{z}}_{\srad{{\mathsf{a}}}}){\mathscr{D}}_+({{z}}_{\srad{{\mathsf{a}}}})}{\overline{{\mathscr{D}}'_-({{z}}_{\srad{{\mathsf{a}}}})}\overline{{\mathscr{N}}_+({{z}}_{\srad{{\mathsf{a}}}})}}\delta_{{{\mathfrak{s}}}, {{\mathcal{R}}}}+\frac{{{z}}_{\srad{{\mathsf{b}}}}{\mathscr{D}}_+({{z}}_{\srad{{\mathsf{b}}}}){\mathscr{N}}_-({{z}}_{\srad{{\mathsf{b}}}})}{\overline{{\mathscr{N}}'_+({{z}}_{\srad{{\mathsf{b}}}})}\overline{{\mathscr{D}}_-({{z}}_{\srad{{\mathsf{b}}}})}}\delta_{{{\mathfrak{s}}}, {{\mathcal{L}}}}.
\label{CRexp_tg}
\eeqans{RTk_tg_full}
The expression \eqref{Ref_tg} for reflectance ${\mathscr{R}}_{{{\mathcal{L}}}\leftarrow{{\mathcal{R}}}}$ and ${\mathscr{R}}_{{{\mathcal{L}}}\to{{\mathcal{R}}}}$ is plotted in Fig. \ref{latticestrip_Reflectance_k_tg} with ${\upxi}_{\srad{{\mathsf{a}}}}$ and ${\upxi}_{\srad{{\mathsf{b}}}}$, respectively, as a variable on horizontal axis.

\begin{figure*}[h]
\centering
{{}{\raisebox{-0.05\height}{\includegraphics[width=\linewidth]{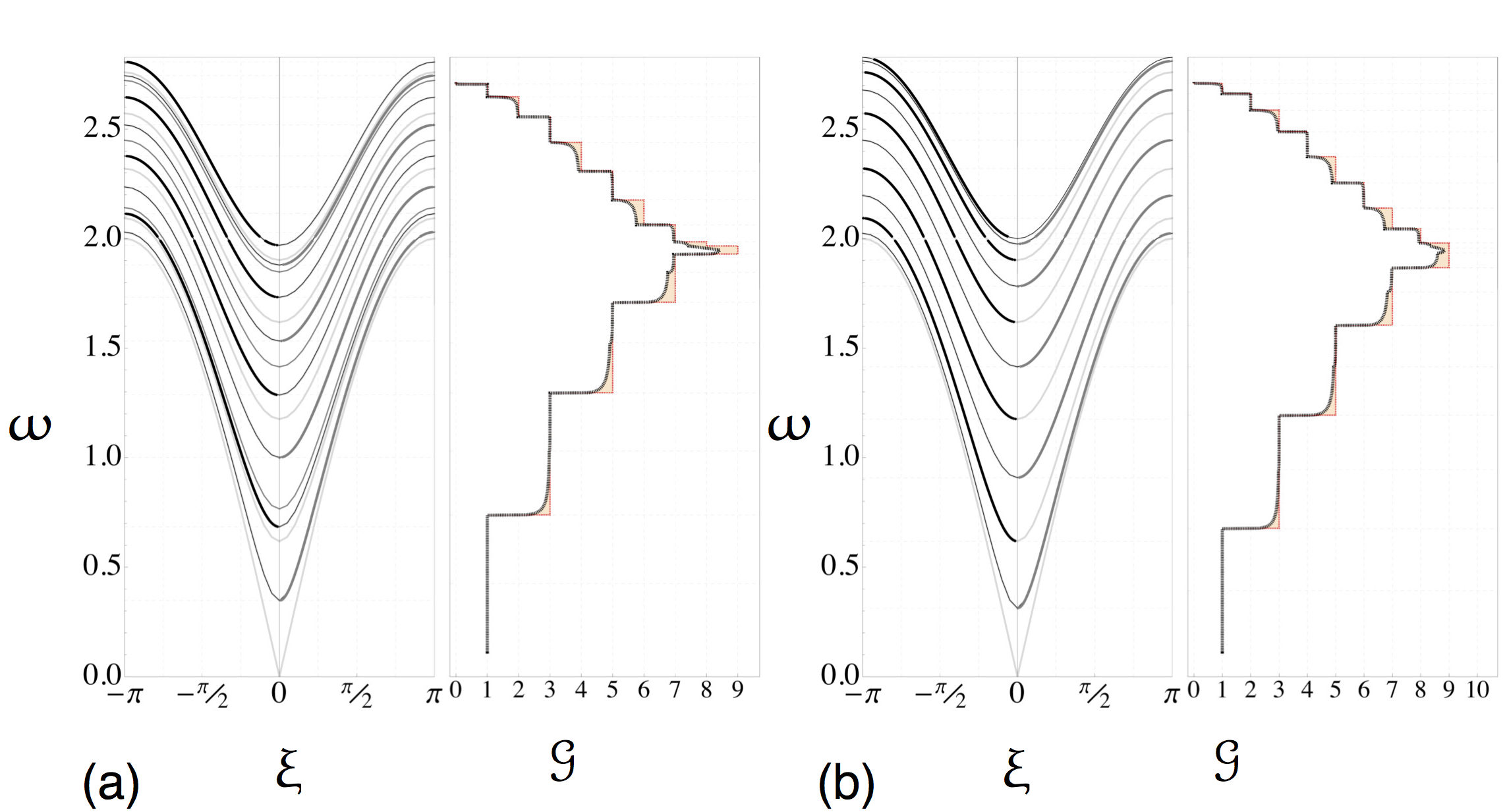}}}}
\caption{Conductance ${\mathscr{G}}$ ($={\mathscr{G}}_{{{{\mathcal{L}}}}\leftarrow{{{\mathcal{R}}}}}={\mathscr{G}}_{{{{\mathcal{L}}}}\to{{{\mathcal{R}}}}}$) for partly unzipped tube of square lattice structure: (a) ${\mathtt{N}}=9,$ (b) ${\mathtt{N}}=10$.}\label{latticestrip_Conductance_sq_P}\end{figure*}
\begin{figure*}[h]
\centering
{{}{\raisebox{-0.05\height}{\includegraphics[width=\linewidth]{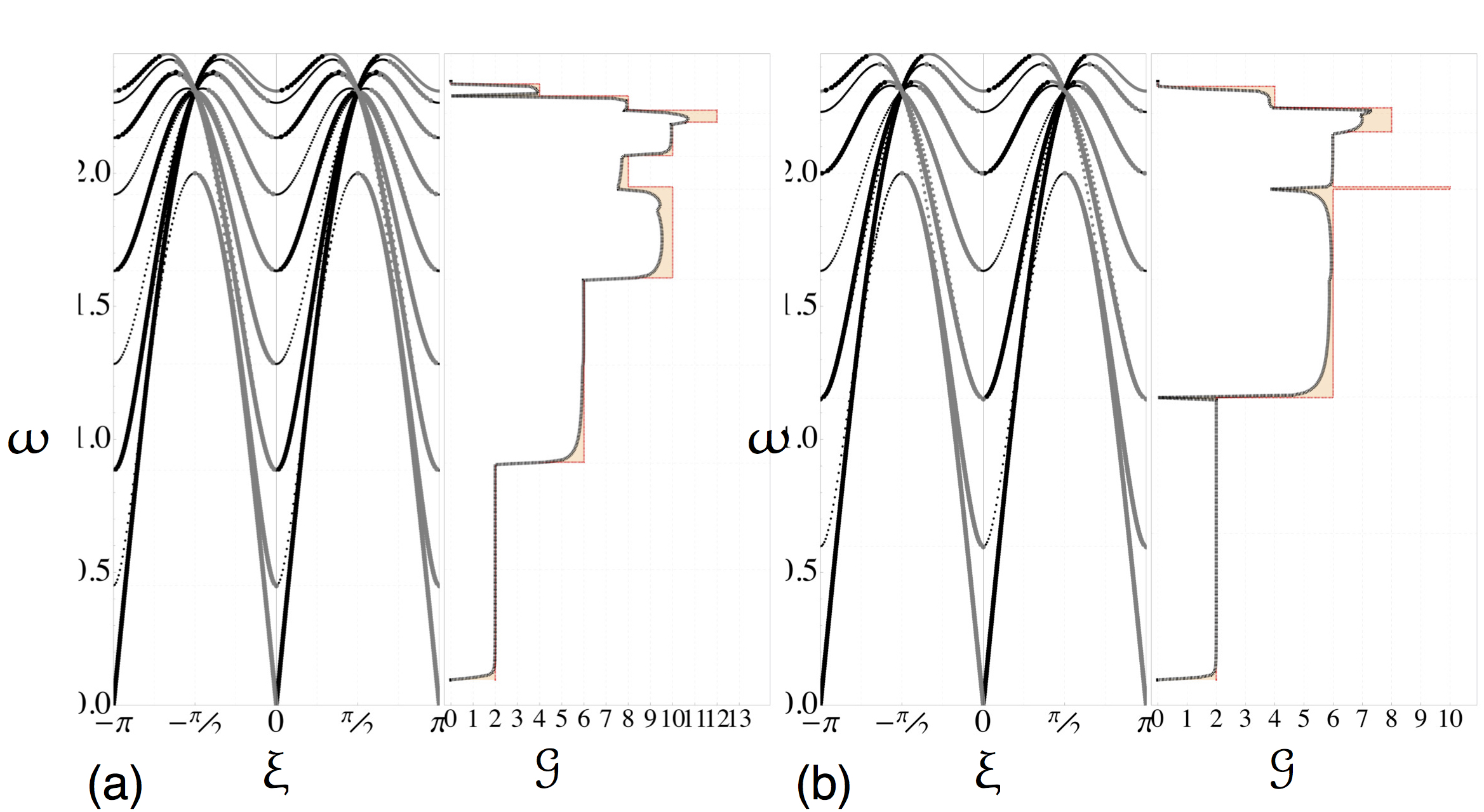}}}}
\caption{Conductance ${\mathscr{G}}$ ($={\mathscr{G}}_{{{{\mathcal{L}}}}\leftarrow{{{\mathcal{R}}}}}={\mathscr{G}}_{{{{\mathcal{L}}}}\to{{{\mathcal{R}}}}}$) for partly unzipped tube of triangular lattice structure: (a) ${\mathtt{N}}=6,$ (b) ${\mathtt{N}}=8$.}\label{latticestrip_Conductance_tg_P}\end{figure*}
\begin{figure*}[h]
\centering
{}{\raisebox{-0.05\height}{\includegraphics[width=\linewidth]{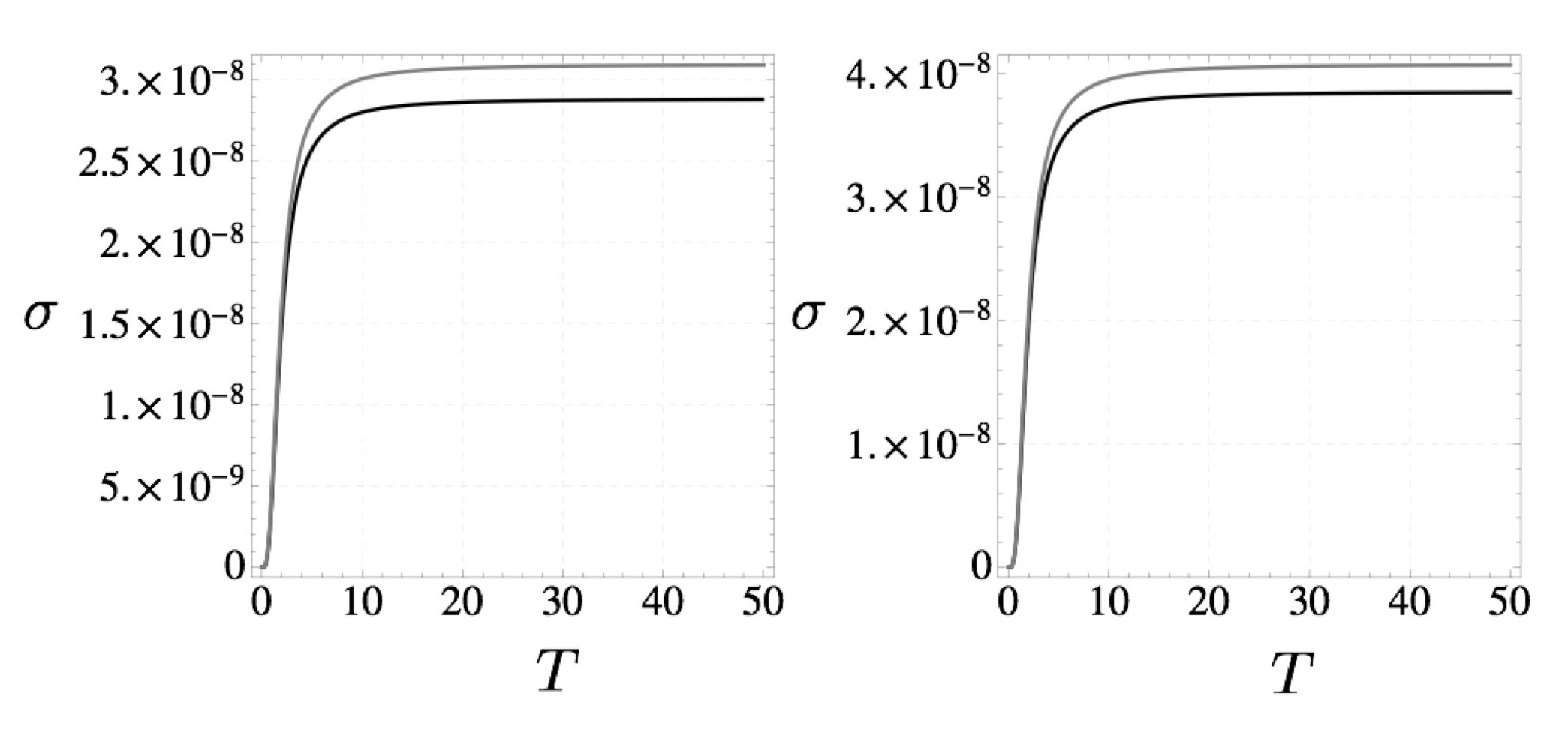}}}
\caption{{Thermal conductance (on the vertical axis) vs absolute temperature (on the horizontal axis) in unzipped square lattice tube (a) ${\mathtt{N}}=9,$ (b) ${\mathtt{N}}=10$. The gray curve represents ballistic conductance.}}
\label{latticestrip_ThermalConductance_sq_P}
\end{figure*}
\begin{figure*}[h]
\centering
{}{\raisebox{-0.05\height}{\includegraphics[width=\linewidth]{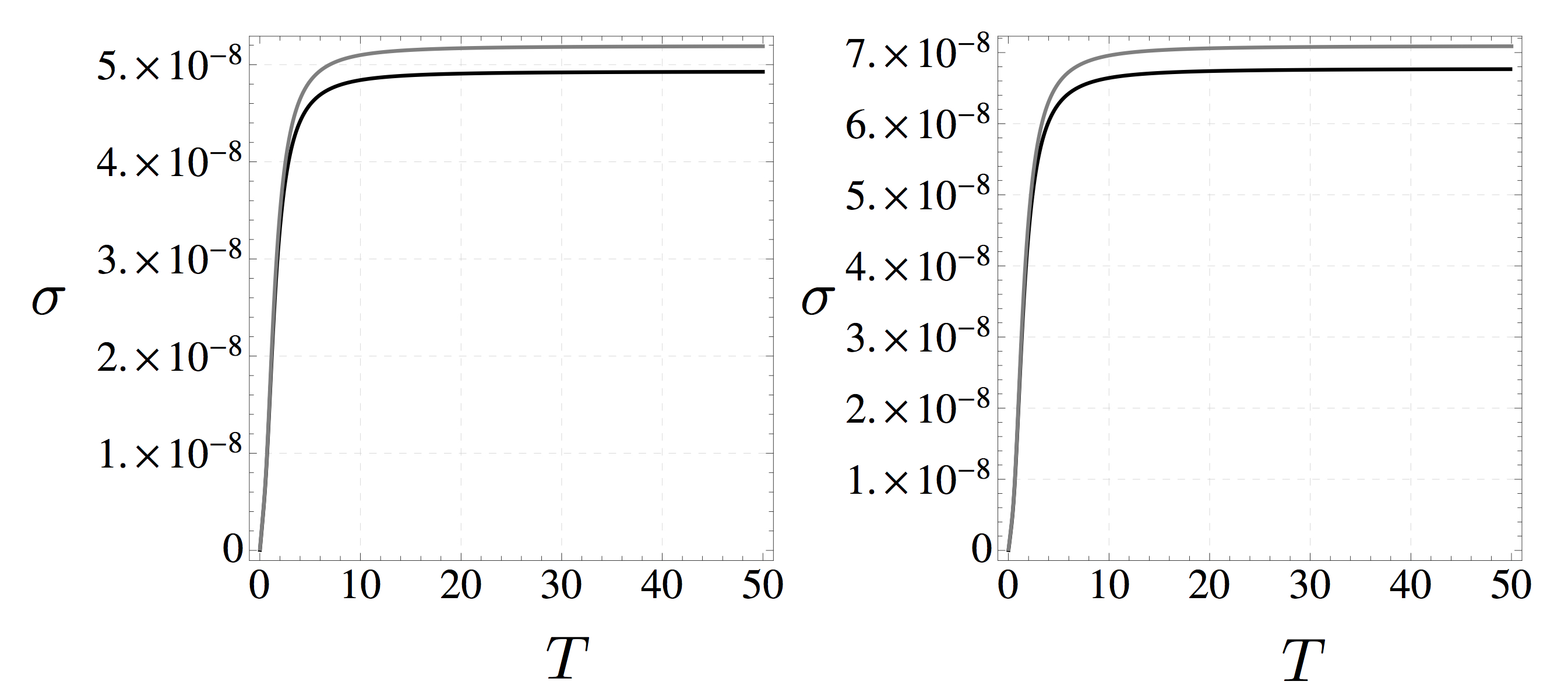}}}
\caption{{Thermal conductance (on the vertical axis) vs absolute temperature (on the horizontal axis) in unzipped triangular lattice tube (a) ${\mathtt{N}}=6,$ (b) ${\mathtt{N}}=8$. The gray curve represents ballistic conductance.}}
\label{latticestrip_ThermalConductance_tg_P}
\end{figure*}
\begin{figure*}[h]
\centering
{}{\raisebox{-0.05\height}{\includegraphics[width=\linewidth]{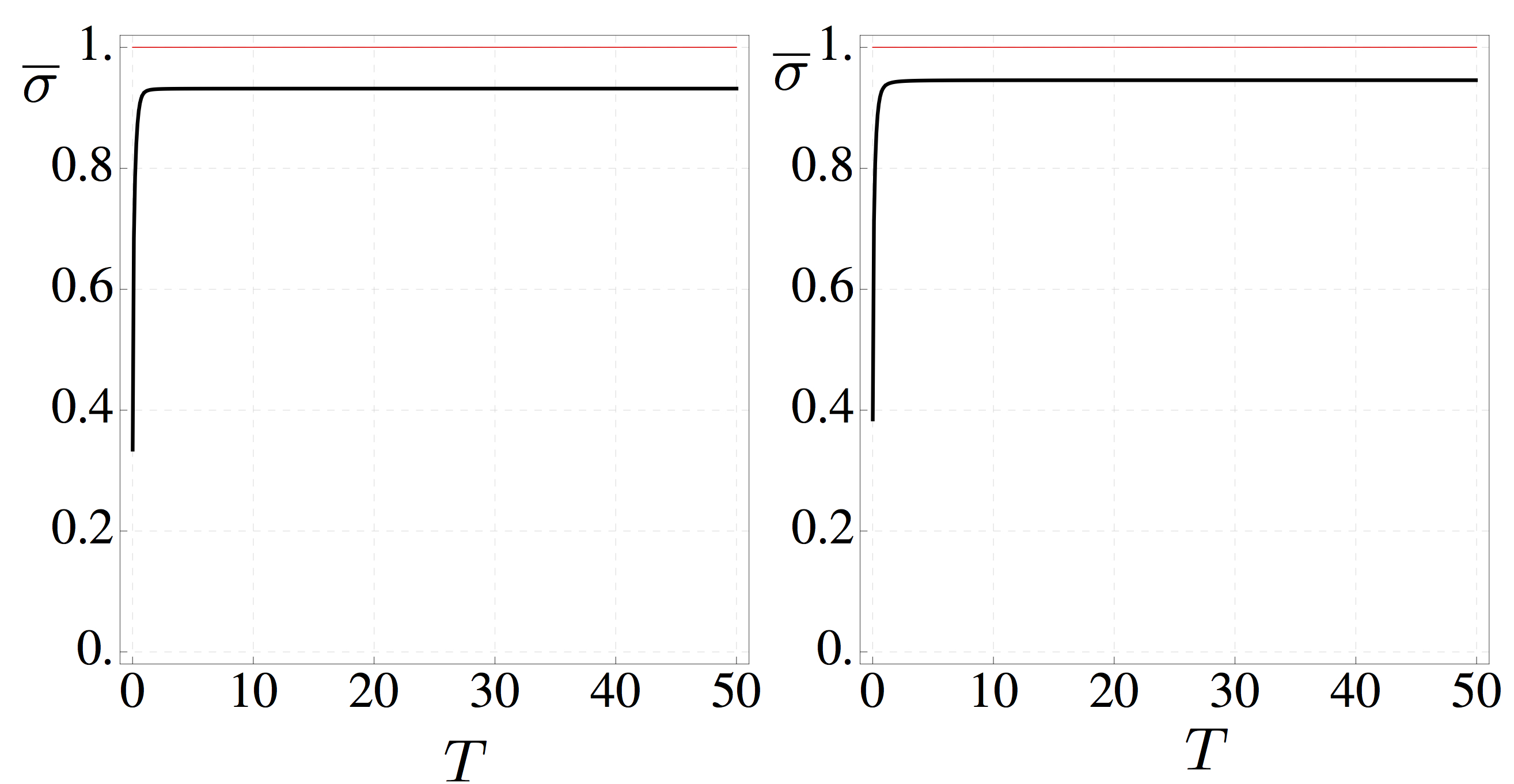}}}
\caption{{Thermal conductance relative to ballistic limit vs absolute temperature (on the horizontal axis) in unzipped square lattice tube (a) ${\mathtt{N}}=9,$ (b) ${\mathtt{N}}=10$.}}
\label{latticestrip_ThermalConductanceRelative_sq_P}
\end{figure*}
\begin{figure*}[h]
\centering
{}{\raisebox{-0.05\height}{\includegraphics[width=\linewidth]{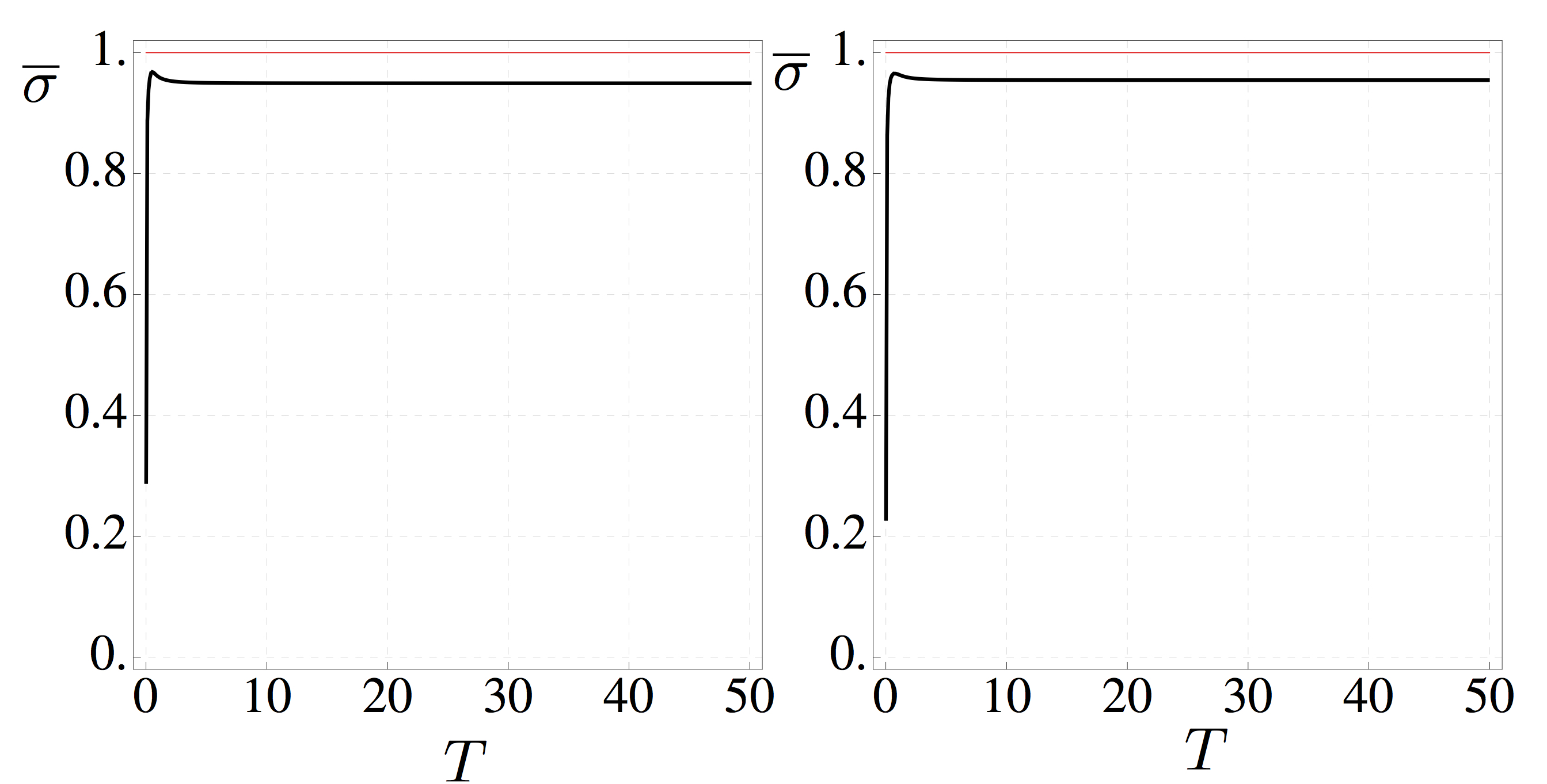}}}
\caption{{Thermal conductance relative to ballistic limit vs absolute temperature (on the horizontal axis) in unzipped triangular lattice tube (a) ${\mathtt{N}}=6,$ (b) ${\mathtt{N}}=8$.}}
\label{latticestrip_ThermalConductanceRelative_tg_P}
\end{figure*}

\section{Conductance}
Finally, the conductance for the transmission from right to left can be expressed at a given frequency ${\upomega}$ as
\begin{eqn}
{\mathscr{G}}_{{{{\mathcal{L}}}}\leftarrow{{{\mathcal{R}}}}}=\sum_{\srad{{\mathsf{a}}}=1}^{N^{{\mathcal{R}}}}{\mathscr{T}}_{{{\mathcal{L}}}\leftarrow{{\mathcal{R}}}}\label{conductance_gen}\end{eqn}
which equals that for transmission from left to right given by ${\mathscr{G}}_{{{{\mathcal{L}}}}\to{{{\mathcal{R}}}}}=\sum_{\srad{{\mathsf{b}}}=1}^{N^{{\mathcal{L}}}}{\mathscr{T}}_{{{\mathcal{L}}}\to{{\mathcal{R}}}}$. Let ${\mathscr{G}}$ denote the common symbol for both ${\mathscr{G}}_{{{{\mathcal{L}}}}\leftarrow{{{\mathcal{R}}}}}$ and ${\mathscr{G}}_{{{{\mathcal{L}}}}\to{{{\mathcal{R}}}}}$. Note that the sum includes the modes with odd as well as even reflection symmetry on square lattice structure (illustrated in Fig. \ref{latticestrip_Conductance_sq_P}) and on triangular lattice structure (illustrated in Fig. \ref{latticestrip_Conductance_tg_P}). 

In the case of square lattice, due to the geometric symmetry, the even modes are transmitted without any scattering across the edge of unzipped portion so that their transmittance is unity. 
On the other hand, for the triangular lattice structure the `even' modes are also scattered in the same manner as the `odd' modes. Notice that this symmetry based classification on the triangular lattice structure is not natural as there is no intrinsic symmetry on triangular lattice structure. The orange region shown in Fig. \ref{latticestrip_Conductance_sq_P} and Fig. \ref{latticestrip_Conductance_tg_P} depicts the `gap' that arises due to the scattering phenomenon at the edge of unzipped portion, so that the conductance is not completely ballistic.

For the kinematically restricted phonon transmission in the thermal transport,
the thermal conductance ${\mathscr{G}}T$ \cite{Rego1998,Angelescu1998,Blencowe1999,WangJS2008} is given by (using $\omega={\mathpzc{h}}earvel{\upomega}/{\mathrm{b}}$ \cite{Bls0}, and $\beta=\hbar {\mathpzc{h}}earvel/ {\mathrm{k}_B} T {\mathrm{b}}$)
\begin{eqn}
{\mathscr{G}}T&=\frac{\hbar}{2\pi}\int_0^\infty {\mathscr{G}}({\upomega})\omega\pd{f(\omega,T)}{T}d\omega
=\frac{3\beta^3{\mathscr{G}}T_Q}{\pi^2}\int_0^\infty {\mathscr{G}}({\upomega})\frac{{\upomega}^2 exp(\beta {\upomega})}{(exp(\beta{\upomega}) -1)^2}d{\upomega},
\label{sigma_eq}
\end{eqn}
where ${\mathscr{G}}$ is presented above. The expression $f(\omega,T) = (exp(\hbar \omega/ {\mathrm{k}_B} T) -1)^{-1}$ is Bose-Einstein distribution function \cite{Rego1998,Angelescu1998} for heat carriers in the terminals. Typically ${\mathscr{G}}T$ is expressed in terms of the unit thermal quantum \cite{Rego1998,Angelescu1998} conductance ${\mathscr{G}}T_Q=\pi^2{\mathrm{k}_B}^2T/(3h)$. 
Note that ${\mathrm{k}_B}$ is the Boltzmann constant, $T$ is the (absolute) temperature, and $\hbar$ is Planck's constant. 
For perfect (ballistic) transmission, the expression \eqref{sigma_eq} becomes the ballistic conductance
${\mathscr{G}}T^{{\text{ball}}}$.
Let
${\overline{{\mathscr{G}}T}}{:=}\frac{{\mathscr{G}}T}{{\mathscr{G}}T^{{\text{ball}}}}$.
As illustrations, Fig. \ref{latticestrip_ThermalConductance_sq_P} and Fig. \ref{latticestrip_ThermalConductance_tg_P} depict the temperature dependence of ${\mathscr{G}}T$, while Fig. \ref{latticestrip_ThermalConductanceRelative_sq_P} and Fig. \ref{latticestrip_ThermalConductanceRelative_tg_P} present the same results in terms of $\overline{{\mathscr{G}}T}$ for both types of lattice structures (described in the Figure captions).

\section{Concluding remarks}
A closed form expression has been provided for the conductance in partly unzipped tubes of square and triangular lattice, and in fact the same form of expression holds. An exact solution of the wave propagation problem has been harnessed for this purpose.
A provision of the reflection and transmission coefficients using the Chebyshev polynomials is also one of the main results of the paper. 
For the approximation of the transmission properties of certain `double' junctions of the same type, which are separated at a distance, the analysis of this paper can be extended using the existing framework of scattering matrices for dealing with a combination of scatterers \cite{Cahay1988}.
The algebraic problem studied in the present paper is also relevant for the analysis of the electronic energy bands in the tight-binding approximation of the many body Schr{\"{o}}dinger equation. After a suitable re-statement for the equivalent problem in electronic, magnetic, and photonic waveguides, with square and triangular lattice structure, the analysis can be pursued in the same manner. 
The analysis of periodic boundary condition, corresponding to a junction formed by partial splitting a tube, in the context of carbon nano-tubes/nano-ribbons is possible (see for example, \cite{Bls5k_tube}). A natural analog to the honeycomb lattice waveguides (graphene ribbons, i.e. zGNRs) related to the work presented in this paper is published elsewhere (see for example, \cite{Bls5c_tube,Bls5c_tube_media}) based on the analysis for the infinite lattice \cite{Bls5}.
The generalization to the coupling with in-plane modes of the lattice strip as well as other degrees of freedom of the tubular portion shall be considered in future elsewhere.

\section*{Acknowledgments}
The partial support of SERB MATRICS grant MTR/2017/000013 and funds from IITK/ME/20090027 is gratefully acknowledged. 

\appendix
\section{Auxiliary derivations}
\subsection{}
\label{sqtg_recall}
The discrete Fourier transform ${\mathtt{u}}_{{\mathtt{y}}}^F: {\mathbb{C}}\to{\mathbb{C}}$ of $\{{\mathtt{u}}_{{\mathtt{x}}, {\mathtt{y}}}\}_{{\mathtt{x}}\in{\mathbb{Z}}}$ (along the ${\mathtt{x}}$ axis) is defined by
${\mathtt{u}}_{{\mathtt{y}}}^F={\mathtt{u}}_{{\mathtt{y}};+}+{\mathtt{u}}_{{\mathtt{y}};-},$
${\mathtt{u}}_{{\mathtt{y}};+}({{z}})=\sum\nolimits_{{\mathtt{x}}=0}^{+\infty} {\mathtt{u}}_{{\mathtt{x}}, {\mathtt{y}}}{{z}}^{-{\mathtt{x}}}, {\mathtt{u}}_{{\mathtt{y}};-}({{z}})=\sum\nolimits_{{\mathtt{x}}=-\infty}^{-1} {\mathtt{u}}_{{\mathtt{x}}, {\mathtt{y}}}{{z}}^{-{\mathtt{x}}}.$
Using the discrete Fourier transform, the discrete Helmholtz equation \eqref{dHelmholtz_sq} can be expressed as
\beqans
{{\mathpzc{Q}}}{\mathtt{u}}_{{\mathtt{y}}}^F-({\mathtt{u}}_{{{\mathtt{y}}}+1}^F+{\mathtt{u}}_{{{\mathtt{y}}}-1}^F)=0\text{ on } {{\mathscr{A}}}_{\mathtt{u}}, \text{where } \label{dHelmholtzF_sq} \\
{{\mathpzc{Q}}}({{z}}){:=}(4-{{z}}-{{z}}^{-1}-{\upomega}^2), {{z}}\in{\mathbb{C}}, \label{q2_sq}
\eeqans{dHelmholtzFfull_sq}
for all ${\mathtt{y}}\in{\mathbb{Z}}$ inside the lattice structure but away from the boundary. 
The relevant definition of ${{\mathscr{A}}}_u$ for the unzipped square lattice tube \cite{Bls9s} is 
\begin{eqn}{{\mathscr{A}}}_u{:=}\{{{z}}\in{\mathbb{C}}: {{\mathit{R}}}_+< |{{z}}|< {{\mathit{R}}}_-\}, {{\mathit{R}}}_\pm=e^{\mp{\upkappa}_2}.\label{annAu_sq}\end{eqn}
In above definition, ${\upkappa}_2$ is the imaginary part of ${\upkappa}_x$.

The square root function, $\sqrt{\cdot}$, has the usual branch cut in the complex plane running from $-\infty$ to $0$.
The general solution of homogeneous equation \eqref{dHelmholtzF_sq} is given by
${\mathtt{u}}^F_{{\mathtt{y}}}={\mathit{c}}_1{\lambda}^{{\mathtt{y}}}+{\mathit{c}}_2{\lambda}^{-{\mathtt{y}}}, {\mathtt{y}}\in{\mathbb{Z}}_0^{{\mathtt{N}}-1},$
where ${\mathit{c}}_{1, 2}$ are arbitrary analytic functions of ${{z}}$ in ${{\mathscr{A}}}$ (to be specified later) and
 \cite{Bls9s}\begin{eqn}
{{\lambda}}{:=}\frac{{\sr}-{{\mathpzc{h}}}}{{\sr}+{{\mathpzc{h}}}}\text{ on }{\mathbb{C}}\setminus{\mathscr{B}}, {{\mathpzc{h}}}{:=}\sqrt{{{\mathpzc{Q}}}-2}, 
{\sr}{:=}\sqrt{{{\mathpzc{Q}}}+2}.
\label{lambdadef_sq}
\end{eqn}
where ${\mathscr{B}}$ denotes the union of branch cuts for ${{\mathpzc{h}}}$ and ${\sr}$ such that $|{{\lambda}}({{z}})|\le1, {{z}}\in{\mathbb{C}}\setminus{\mathscr{B}}.$
Using the definition of Chebyshev polynomial of the second kind, for $ 0<n\in{\mathbb{Z}}$, it follows that
${\lambda}^{-n}-{\lambda}^{n}=({\lambda}^{-1}-{\lambda}){\mathtt{U}}_{n-1}({\frac{1}{2}}({\lambda}+{\lambda}^{-1})).$

Due to their frequent appearance in the rest of the paper, 
it is also useful to define 
\begin{eqn}
\hspace{-.2in}
{{z}}_{{P}}{:=} e^{-i{\upkappa}_x},\delta_{D+}({{z}}){:=}\sum\limits_{n=0}^{+\infty}{{z}}^{-n}, \delta_{D-}({{z}}){:=}\sum\limits_{n=-\infty}^{-1}{{z}}^{-n}. 
\label{zPdef_sq}
\end{eqn}

\subsection{}
\label{appLk_sq_P}
When ${\mathtt{N}}$ is even, i.e. ${\mathtt{N}}=2{\mathrm{N}}$, 
\begin{eqn}
{\sLNsq}_{{}}&=\frac{\prod\nolimits_{j=1}^{2{\mathrm{N}}}({{\mathpzc{h}}}^2+4\sin^2{\frac{1}{2}}\frac{j-1}{2{\mathrm{N}}}\pi)}{\prod\nolimits_{j=1}^{2{\mathrm{N}}}({{\mathpzc{h}}}^2+4\sin^2{\frac{1}{2}}\frac{j}{{\mathrm{N}}}\pi)}
=\frac{\prod\nolimits_{j=1}^{{\mathrm{N}}}({{\mathpzc{h}}}^2+4\sin^2\frac{j-{\frac{1}{2}}}{2{\mathrm{N}}}\pi)}{\prod\nolimits_{j=1}^{{\mathrm{N}}}({{\mathpzc{h}}}^2+4\sin^2{\frac{1}{2}}\frac{j}{{\mathrm{N}}}\pi)},
\end{eqn}
and for ${\mathtt{N}}$ ($=2{\mathrm{N}}+1$) odd,
\begin{eqn}
{\sLNsq}_{{}}&=\frac{\prod\nolimits_{j=1}^{2{\mathrm{N}}+1}({{\mathpzc{h}}}^2+4\sin^2{\frac{1}{2}}\frac{j-1}{2{\mathrm{N}}+1}\pi)}{\prod\nolimits_{j=1}^{2{\mathrm{N}}+1}({{\mathpzc{h}}}^2+4\sin^2{\frac{1}{2}}\frac{2j}{2{\mathrm{N}}+1}\pi)}
=\frac{\prod\nolimits_{j=1}^{{\mathrm{N}}}({{\mathpzc{h}}}^2+4\sin^2\frac{j-{\frac{1}{2}}}{2{\mathrm{N}}+1}\pi)}{\prod\nolimits_{j=1}^{{\mathrm{N}}}({{\mathpzc{h}}}^2+4\sin^2\frac{j}{2{\mathrm{N}}+1}\pi)}.
\end{eqn}

\section{Wiener--Hopf factorization:}
\label{appWHfac}
Following \cite{Bls9s}, assuming that $|{{z}}_{{F}}|<1, {{z}}_{{F}}\in{\mathbb{C}}$, let
${{F}}({{z}}; {{z}}_{{F}}){:=}{{z}}_{{F}}^{-1}(1-{{z}}_{{F}}{{z}})(1-{{z}}_{{F}}{{z}}^{-1}),$
and ${{F}}_\pm({{z}}; {{z}}_{{F}})={{z}}_{{F}}^{-{\frac{1}{2}}}(1-{{z}}_{{F}}{{z}}^{\mp1}).$ The Wiener--Hopf kernel \eqref{Lk_sq_P} can be expressed as $\sLNsq_{{}}={\mathscr{N}}({{z}})/{\mathscr{D}}({{z}})=\sLNsq_{{}+}\sLNsq_{{}-}$, 
where ${\sLNsq}_{{}\pm}$ are given by
\begin{eqn}
\sLNsq_{{}\pm}&=\frac{{\mathscr{N}}_\pm({{z}})}{{\mathscr{D}}_\pm({{z}})}=\frac{\prod\nolimits_{j=1}^{{\mathtt{N}}}{{F}}_\pm({{z}}; {{z}}_{{F}_{{{\mathcal{R}}} j}})}{\prod\nolimits_{j=1}^{{\mathtt{N}}}{{F}}_\pm({{z}}; {{z}}_{{F}_{{{\mathcal{L}}} j}})}.
\label{Lkpmexp}
\end{eqn}
{In view of these definitions, let}
\begin{eqn}
{{\mathscr{A}}}_{{}}{:=}{{\mathscr{A}}}_u\cap\{{{z}}\in{\mathbb{C}} : {{\mathit{R}}}_{L_{{}}}< |{{z}}|< {{\mathit{R}}}_{L_{{}}}^{-1}\}, 
{\mathit{R}}_{L_{{}}}{:=}\max(\{|{{z}}_{{F}_{{{\mathcal{R}}} j}}|\}_{j=1}^{{\mathtt{N}}}\cup\{|{{z}}_{{F}_{{{\mathcal{L}}} j}}|\}_{j=1}^{{\mathtt{N}}}). 
\label{annAALk_sq}
\end{eqn} 
The function ${{\sLNsq}_{{}}}_{+}$ (resp. ${{\sLNsq}_{{}}}_{-}$) is analytic, without any zeros, in the exterior (resp. interior) of a disk centered at $0$ in ${\mathbb{C}}$ with radius ${{\mathit{R}}}_{L_{{}}}$ (resp. ${{\mathit{R}}}_{L_{{}}}^{-1}$). 
Also ${{\sLNsq}_{{}}}_{\pm}({{z}})={{\sLNsq}_{{}}}_{\mp}({{z}}^{-1})$; $\pm$ signs concur in \eqref{Lkpmexp}. 

The details for triangular lattice waveguide are analogous to those presented above for square lattice waveguide (based on \cite{Bls9s}).
The kernel ${\sLNtg}_{{}}$ \eqref{Lk_tg_P} is also a ratio of a polynomial ${\mathscr{N}}$ (of ${z}$) in the numerator and another polynomial ${\mathscr{D}}$ (of ${z}$) in the denominator. 
To express succinctly the multiplicative factors for the Wiener--Hopf kernels \eqref{Lk_tg_P},
a notational device developed by \cite{Bls9s} is employed. 
Thus $\sLNtg_{{}}={\mathscr{N}}({{z}})/{\mathscr{D}}({{z}})$, where
\begin{eqn}
{\mathscr{N}}({{z}})=\prod\nolimits_{j=1}^{2{\mathrm{N}}}{{F}}({{z}}; {{z}}_{{F}_{{\mathscr{N}} j}})
={\mathscr{N}}_+({{z}}){\mathscr{N}}_-({{z}}), 
{\mathscr{D}}({{z}})=\prod\nolimits_{j=1}^{2{\mathrm{N}}}{{F}}({{z}}; {{z}}_{{F}_{{\mathscr{D}} j}})
={\mathscr{D}}_+({{z}}){\mathscr{D}}_-({{z}}).
\label{NDk_tg_P}
\end{eqn}
The factor of two appears in the product limits because \eqref{chebx_sq} (with ${\mathpzc{Q}}$ defined by \eqref{q2_tg}) involves ${z}^2$ and ${z}^{-2}$. 
For example, in case of the periodic boundary, i.e. ${\mathfrak{T}\hspace{-.4ex}}{\mathbin{\substack{{\circledcirc}\\{\circ}}}}$, the expressions of ${\mathscr{N}}$ and ${\mathscr{D}}$ \eqref{Lk_tg_P} can be found to be
\begin{eqn}
{\mathscr{N}}({z})&=\prod\limits_{j=1}^{{\mathrm{N}}}(6-({z}^2+{z}^{-2})-\frac{3}{2}{\upomega}^2-2({z}+{z}^{-1})\cos\frac{j}{{{\mathrm{N}}}+1}\pi)
-2\prod\limits_{j=1}^{{{\mathrm{N}}}-1}(6-({z}^2+{z}^{-2})-\frac{3}{2}{\upomega}^2-2({z}+{z}^{-1})\cos\frac{j}{{\mathrm{N}}}\pi)\\
{\mathscr{D}}({z})&=\prod\limits_{j=1}^{{\mathrm{N}}}(6-({z}^2+{z}^{-2})-\frac{3}{2}{\upomega}^2-2({z}+{z}^{-1})\cos\frac{2j}{2{{\mathrm{N}}}+1}\pi),
\label{ND_tg_P}
\end{eqn}
which indicates that ${\mathscr{N}}$ and ${\mathscr{D}}$ are polynomials of degree $2{\mathrm{N}}$ in ${z}+{z}^{-1}$.
The Wiener--Hopf factors of ${\mathscr{N}}$ and ${\mathscr{D}}$ are, respectively, given by
${\mathscr{N}}_\pm({{z}})=\prod\nolimits_{j=1}^{2{\mathrm{N}}}{{F}}_\pm({{z}}; {{z}}_{{F}_{{\mathscr{N}} j}}),$ ${\mathscr{D}}_\pm({{z}})=\prod\nolimits_{j=1}^{2{\mathrm{N}}}{{F}}_\pm({{z}}; {{z}}_{{F}_{{\mathscr{D}} j}}).$
Explicit expressions for the factors ${\sLNtg}_{{}\pm}$ are \begin{eqn}
\sLNtg_{{}\pm}({{z}})&=\frac{{\mathscr{N}}_\pm({{z}})}{{\mathscr{D}}_\pm({{z}})}=\frac{\prod\nolimits_{j=1}^{2{\mathrm{N}}}{{F}}_\pm({{z}}; {{z}}_{{F}_{{\mathscr{N}} j}})}{\prod\nolimits_{j=1}^{2{\mathrm{N}}}{{F}}_\pm({{z}}; {{z}}_{{F}_{{\mathscr{D}} j}})}.
\label{Lkpmexp}
\end{eqn}
{In view of above definitions, let}
\begin{eqn}
{{\mathscr{A}}}_{{}}{:=}{{\mathscr{A}}}_u\cap\{{{z}}\in{\mathbb{C}} : {{\mathit{R}}}_{L_{{}}}< |{{z}}|< {{\mathit{R}}}_{L_{{}}}^{-1}\}, 
{\mathit{R}}_{L_{{}}}{:=}\max\big(\{|{{z}}_{{F}_{{\mathscr{N}} j}}|\}_{j=1}^{2{\mathrm{N}}}\cup\{|{{z}}_{{F}_{{\mathscr{D}} j}}|\}_{j=1}^{2{\mathrm{N}}}\big).
\label{annAALk_tg}
\end{eqn} 
In \eqref{Lkpmexp}, $\pm$ signs concur and it has been implicitly assumed that $\sLNtg_{{}\pm}({{z}})=\sLNtg_{{}\mp}({{z}}^{-1})$. 
The function ${\sLNtg}_{{}+}$ (resp. ${\sLNtg}_{{}-}$) is analytic, in fact it has neither poles nor zeros, in the exterior (resp. interior) of a disk centered at $0$ in ${\mathbb{C}}$ with radius ${{\mathit{R}}}_{L_{{}}}$ (resp. ${{\mathit{R}}}_{L_{{}}}^{-1}$); thus, $1/{{\sLNtg}_{{}+}}$ (resp. $1/{{\sLNtg}_{{}-}}$) is analytic in the same region as ${\sLNtg}_{{}+}$ (resp. ${\sLNtg}_{{}-}$). 

\section{Auxiliary derivations}
\label{app_aux}

\subsection{}
\label{Liouville_argument}
{In particular, 
$J({{z}}){:=} {{\mathpzc{L}}_{{}}}^{-1}_{+}({{z}}){{\mathtt{v}}}_{+}({{z}})-{{\mathpzc{C}}}_+({{z}})=-{{\mathpzc{L}}_{{}}}_{-}({{z}}){{\mathtt{v}}}_-({{z}})+{{\mathpzc{C}}}_-({{z}})=0, {{z}}\in{{\mathscr{A}}},$
The function $J({{z}})$ is analytic at ${{z}}\in{\mathbb{C}}$ with $|{{z}}|>\max\{{{\mathit{R}}}_+, {{\mathit{R}}}_{{\mathpzc{L}}_{{}}}\}$, {\em and } also at ${{z}}\in{\mathbb{C}}$ with $|{{z}}| <\min\{{{\mathit{R}}}_-, {{\mathit{R}}}_{{\mathpzc{L}}_{{}}}^{-1}\}$, i.e., in the whole of the complex plane ${\mathbb{C}}$, since the two regions overlap in the annulus ${{\mathscr{A}}}_{{}}$, defined in \eqref{annAALk_sq}. Using \eqref{Lkpmexp}, 
and \eqref{CpmK_sq}, as ${{z}}\to0,$ ${{\mathpzc{L}}_{{}}}_{-}({{z}})\to {\mathit{c}}_1$, ${\mathtt{u}}_-({{z}})\to0,$ and ${{\mathpzc{C}}}_-({{z}})\to0$, while as ${{z}}\to\infty,$ ${{\mathpzc{L}}_{{}}}_{+}({{z}})\to {\mathit{c}}_2$, ${\mathtt{u}}_+({{z}})\to{\mathit{c}}_3$ and ${{\mathpzc{C}}}_+({{z}})\to{\mathit{c}}_4$, for some constants ${\mathit{c}}_1, \dotsc, {\mathit{c}}_4$. Since $J({{z}})$ is bounded on the complex plane and tends to zero as ${{z}}$ tends to $0$, it follows that $J\equiv0$.}

\subsection{}
Recall that the incident wave is ${{\mathrm{A}}}{{a}}_{({{{\kappa}}^{i}}){{\mathtt{y}}}}e^{i{\upkappa}_x {{\mathtt{x}}}}$ on the physical sub-lattice, while the incident wave is $-{{\mathrm{A}}}{{a}}_{({{{\kappa}}^{i}}){{\mathtt{N}}-{\mathtt{y}}-1}}e^{i{\upkappa}_x {{\mathtt{x}}}}$ on the `{replicated}' sub-lattice. In particular, at ${\mathtt{y}}=0$ and even ${\mathtt{x}}$, ${{\mathrm{A}}}{{a}}_{({{{\kappa}}^{i}}){0}}e^{i{\upkappa}_x {{\mathtt{x}}}}$ holds on the physical sub-lattice while for odd ${\mathtt{x}}$, $-{{\mathrm{A}}}{{a}}_{({{{\kappa}}^{i}}){{\mathtt{N}}-1}}e^{i{\upkappa}_x {{\mathtt{x}}}}$ holds on the `{replicated}' sub-lattice.
As a consequence of \eqref{evenoddsymm}, ${{a}}_{({{{\kappa}}^{i}}){0}}=\pm{{a}}_{({{{\kappa}}^{i}}){{\mathtt{N}}-1}}.$
For the incidence from the tubular portion, using \eqref{y0_tg_P}, hence,
\begin{eqn}
q^F({{z}})&=\sum\limits_{{{\mathtt{x}}}=-\infty}^{-1}{{z}}^{-{{\mathtt{x}}}}({\mathtt{u}}^{i}_{{{\mathtt{x}}}, 0}-{\mathtt{u}}^{i}_{{{\mathtt{x}}}-1, -1})
+\sum\limits_{{{\mathtt{x}}}=-\infty}^{-2}{{z}}^{-{{\mathtt{x}}}}({\mathtt{u}}^{i}_{{{\mathtt{x}}}, 0}-{\mathtt{u}}^{i}_{{{\mathtt{x}}}+1, -1})\\
&={{\mathrm{A}}}\sum\limits_{{\mathtt{x}} \text{even},{{\mathtt{x}}}=-\infty}^{-1}{{z}}^{-{{\mathtt{x}}}}({{a}}_{({{{\kappa}}^{i}}){0}}{{z}}_{{P}}^{{\mathtt{x}}}-{{a}}_{({{{\kappa}}^{i}}){{\mathtt{N}}-1}}{{z}}_{{P}}^{{\mathtt{x}}-1})
+{{\mathrm{A}}}\sum\limits_{{\mathtt{x}} \text{even},{{\mathtt{x}}}=-\infty}^{-2}{{z}}^{-{{\mathtt{x}}}}({{a}}_{({{{\kappa}}^{i}}){0}}{{z}}_{{P}}^{{\mathtt{x}}}-{{a}}_{({{{\kappa}}^{i}}){{\mathtt{N}}-1}}{{z}}_{{P}}^{{\mathtt{x}}+1})\\
&+{{\mathrm{A}}}\sum\limits_{{\mathtt{x}} \text{odd},{{\mathtt{x}}}=-\infty}^{-1}{{z}}^{-{{\mathtt{x}}}}(-{{a}}_{({{{\kappa}}^{i}}){{\mathtt{N}}-1}}{{z}}_{{P}}^{{\mathtt{x}}}+{{a}}_{({{{\kappa}}^{i}}){0}}{{z}}_{{P}}^{{\mathtt{x}}-1})
+{{\mathrm{A}}}\sum\limits_{{\mathtt{x}} \text{odd},{{\mathtt{x}}}=-\infty}^{-2}{{z}}^{-{{\mathtt{x}}}}(-{{a}}_{({{{\kappa}}^{i}}){{\mathtt{N}}-1}}{{z}}_{{P}}^{{\mathtt{x}}}+{{a}}_{({{{\kappa}}^{i}}){0}}{{z}}_{{P}}^{{\mathtt{x}}+1})\\
&={{\mathrm{A}}}{{a}}_{({{{\kappa}}^{i}}){0}}(\mp{{z}}_{{P}}^{-1}+1)(1+{{z}})\delta_{D-}(\mp{{z}}{{z}}_{{P}}^{-1}).
\label{qF_plusinc}
\end{eqn}
Similarly, for incidence from the unzipped portion, \begin{eqn}
-q^F({{z}})&=\sum\limits_{{{\mathtt{x}}}=0}^{+\infty} {{z}}^{-{{\mathtt{x}}}}({\mathtt{u}}^{i}_{{{\mathtt{x}}}, 0}-{\mathtt{u}}^{i}_{{{\mathtt{x}}}-1, -1})+\sum\limits_{{{\mathtt{x}}}=-1}^{+\infty} {{z}}^{-{{\mathtt{x}}}}({\mathtt{u}}^{i}_{{{\mathtt{x}}}, 0}-{\mathtt{u}}^{i}_{{{\mathtt{x}}}+1, -1})\\
&={{\mathrm{A}}}{{a}}_{({{{\kappa}}^{i}}){0}}(1+{{z}})(1\pm{{z}}_{{P}}^{-1})\delta_{D+}(\pm{{z}}{{z}}_{{P}}^{-1}).
\label{qF_minusinc}
\end{eqn}

\subsection{}
The simplification of $(1+{{z}}^{-1})({\mathtt{u}}_+({{z}})+{\mathtt{u}}_-({{z}}))$, stated in \eqref{uzfull_tg}, is detailed as follows 
\begin{eqn}
&(1+{{z}}^{-1}){\mathtt{u}}^F({{z}})=
-{\frac{1}{2}}{{{\mathrm{A}}}{{a}}_{({{{\kappa}}^{i}}){0}}}{{z}}_{{P}}^{-1}\bigg({\mathpzc{L}}_{{}+}({{z}}){{z}}(\frac{1}{{\mathpzc{L}}_{{}+}({{z}})}-\overline{l}_{{}+0})+1-({{z}}+1)+\frac{1}{{\mathpzc{L}}_{{}-}({{z}})}(\overline{l}_{{}+0}{{z}})\bigg)
(\delta_{D+}({{z}}{{z}}_{{P}}^{-1})-\delta_{D+}(-{{z}}{{z}}_{{P}}^{-1}))\\&-{\frac{1}{2}}{{{\mathrm{A}}}{{a}}_{({{{\kappa}}^{i}}){0}}}{{z}}_{{P}}^{-1}\bigg({{z}}_{{P}}({{\mathpzc{L}}_{{}-}({{z}}_{{P}})}-\overline{l}_{{}+0})+{{\mathpzc{L}}_{{}-}({{z}}_{{P}})}\bigg)
\delta_{D+}({{z}}{{z}}_{{P}}^{-1})(\frac{1}{{\mathpzc{L}}_{{}-}({{z}})}-{\mathpzc{L}}_{{}+}({{z}}))\\&+{\frac{1}{2}}{{{\mathrm{A}}}{{a}}_{({{{\kappa}}^{i}}){0}}}{{z}}_{{P}}^{-1}\bigg((-{{z}}_{{P}})({{\mathpzc{L}}_{{}-}(-{{z}}_{{P}})}-\overline{l}_{{}+0})+{{\mathpzc{L}}_{{}-}(-{{z}}_{{P}})}\bigg)
\delta_{D+}(-{{z}}{{z}}_{{P}}^{-1})(\frac{1}{{\mathpzc{L}}_{{}-}({{z}})}-{\mathpzc{L}}_{{}+}({{z}})),\label{app_aux_usim_1}\end{eqn} 
which can be be expanded further and simplified to obtain
\begin{eqn}
&(1+{{z}}^{-1}){\mathtt{u}}^F({{z}})={\frac{1}{2}}{{{\mathrm{A}}}{{a}}_{({{{\kappa}}^{i}}){0}}}{{z}}_{{P}}^{-1}\big((\overline{l}_{{}+0}{{z}}+\frac{{z}}{{z}-{{z}}_{{P}}}\frac{{{z}}_{{P}}}{{\mathpzc{L}}_{{}-}^{-1}({{z}}_{{P}})})
+\frac{{z}}{{z}-{{z}}_{{P}}}{{\mathpzc{L}}_{{}-}({{z}}_{{P}})}\big)(\frac{1}{{\mathpzc{L}}_{{}-}}-{\mathpzc{L}}_{{}+})\\&-{\frac{1}{2}}{{{\mathrm{A}}}{{a}}_{({{{\kappa}}^{i}}){0}}}{{z}}_{{P}}^{-1}\big((\overline{l}_{{}+0}{{z}}+\frac{{z}}{{z}+{{z}}_{{P}}}\frac{-{{z}}_{{P}}}{{\mathpzc{L}}_{{}-}^{-1}(-{{z}}_{{P}})})
+\frac{{z}}{{z}+{{z}}_{{P}}}{{\mathpzc{L}}_{{}-}(-{{z}}_{{P}})}\big)(\frac{1}{{\mathpzc{L}}_{{}-}}-{\mathpzc{L}}_{{}+})\\&
{\frac{1}{2}}{{{\mathrm{A}}}{{a}}_{({{{\kappa}}^{i}}){0}}}{{z}}_{{P}}^{-1}\big((\frac{{z}}{{z}-{{z}}_{{P}}}{{\mathpzc{L}}_{{}-}({{z}}_{{P}})}+\frac{{z}}{{z}+{{z}}_{{P}}}{{\mathpzc{L}}_{{}-}(-{{z}}_{{P}})}){{z}}_{{P}}
+(\frac{{z}}{{z}-{{z}}_{{P}}}{{\mathpzc{L}}_{{}-}({{z}}_{{P}})}-\frac{{z}}{{z}+{{z}}_{{P}}}{{\mathpzc{L}}_{{}-}(-{{z}}_{{P}})})\big)(\frac{1}{{\mathpzc{L}}_{{}-}}-{\mathpzc{L}}_{{}+})\\&={\frac{1}{2}}{{{\mathrm{A}}}{{a}}_{({{{\kappa}}^{i}}){0}}}{z}\big((\frac{1}{{z}-{{z}}_{{P}}}{{\mathpzc{L}}_{{}-}({{z}}_{{P}})}+\frac{1}{{z}+{{z}}_{{P}}}{{\mathpzc{L}}_{{}-}(-{{z}}_{{P}})})
+(\frac{1}{{z}-{{z}}_{{P}}}{{\mathpzc{L}}_{{}-}({{z}}_{{P}})}-\frac{1}{{z}+{{z}}_{{P}}}{{\mathpzc{L}}_{{}-}(-{{z}}_{{P}})}){{z}}_{{P}}^{-1}\big)(\frac{1}{{\mathpzc{L}}_{{}-}}-{\mathpzc{L}}_{{}+})\\
&={\frac{1}{2}}{{{\mathrm{A}}}{{a}}_{({{{\kappa}}^{i}}){0}}}{z}\big(\frac{{{z}}_{{P}}^{-1}+1}{{z}-{{z}}_{{P}}}{{\mathpzc{L}}_{{}-}({{z}}_{{P}})}+\frac{1-{{z}}_{{P}}^{-1}}{{z}+{{z}}_{{P}}}{{\mathpzc{L}}_{{}-}(-{{z}}_{{P}})}\big)
(\frac{1}{{\mathpzc{L}}_{{}-}({{z}})}-{\mathpzc{L}}_{{}+}({{z}})).
\label{app_aux_usim_2}
\end{eqn}

\section{Chebyshev polynomials}
\label{appCheb}
The Chebyshev polynomials, following Appendix A of \cite{Bls9}, are:
first kind ${\mathtt{T}}_n=
{\frac{1}{2}}(({\vartheta}+\sqrt{{\vartheta}^2-1})^n+({\vartheta}-\sqrt{{\vartheta}^2-1})^n),$ 
second kind $
{\mathtt{U}}_n={\frac{1}{2}}(({\vartheta}+\sqrt{{\vartheta}^2-1})^n-({\vartheta}-\sqrt{{\vartheta}^2-1})^n)/\sqrt{{\vartheta}^2-1},$ 
third kind $
{\mathtt{V}}_n=
{\frac{1}{2}}(w^{n+1}+w^{-n})/(w+1),$ 
and fourth kind $
{\mathtt{W}}_n=
{\frac{1}{2}}(w^{n+1}-w^{-n})/(w-1), {\vartheta}={\frac{1}{2}}(w+w^{-1}).$ 
Several identities involving the Chebyshev polynomials find applications in the paper. For example, the following are standard relations 
\beqans
\hspace{-.3in}{\mathtt{T}}'_n=n{\mathtt{U}}_{n-1}, {\mathtt{U}}'_{n}=(n+1)({\vartheta}^2-1)^{-1}{\mathtt{T}}_{n+1}, 
{\mathtt{V}}'_n=(n+{\frac{1}{2}})({\vartheta}+1)^{-1}{\mathtt{W}}_n, {\mathtt{W}}'_n=(n+{\frac{1}{2}})({\vartheta}-1)^{-1}{\mathtt{V}}_n. 
\label{ChebUVWder}
\eeqans{Chebrels}

\section{Scattering matrix}

Ignore the superscript ${t}$ on ${\mathtt{u}}^{{t}}$. Using the analysis presented in the paper for the incidence from the tubular portion, 
ahead and behind the edge of unzipped portion, respectively,
\begin{eqn}
{\mathtt{u}}_{\srad{{\mathsf{a}}}}\sim{\mathtt{u}}_{{\srad{{\mathsf{a}}}}}+\sum\limits_{{{\mathsf{a}}}=1}^{N^{{{\mathcal{R}}}}}{{\tau}}^{{{\mathcal{R}}}\srad{{{\mathcal{R}}}}}_{{\mathsf{a}}{{\srad{{\mathsf{a}}}}}}\sqrt{\frac{|{{\mathtt{V}_g}}_{{\srad{{\mathsf{a}}}}}|}{|{{\mathtt{V}_g}}_{{\mathsf{a}}}|}}{\mathtt{u}}_{{\mathsf{a}}}, 
\text{ and }
{\mathtt{u}}_{\srad{{\mathsf{a}}}}\sim\sum\limits_{{{\mathsf{b}}}=1}^{N^{{{\mathcal{L}}}}}{{\tau}}^{{{\mathcal{L}}}\srad{{{\mathcal{R}}}}}_{{\mathsf{b}}{\srad{{\mathsf{a}}}}}\sqrt{\frac{|{{\mathtt{V}_g}}_{\srad{{\mathsf{a}}}}|}{|{{\mathtt{V}_g}}_{{\mathsf{b}}}|}}{\mathtt{u}}_{{\mathsf{b}}}. 
\end{eqn}
Also for the incidence from the unzipped portion, ahead and behind the edge of unzipped portion, respectively, 
\begin{eqn}
{\mathtt{u}}_{\srad{{\mathsf{b}}}}\sim{\mathtt{u}}_{\srad{{\mathsf{b}}}}+\sum\limits_{{\mathsf{b}}=1}^{N^{{{\mathcal{L}}}}}{{\tau}}^{{{\mathcal{L}}}\srad{{{\mathcal{L}}}}}_{{\mathsf{b}}{\srad{{\mathsf{b}}}}}\sqrt{\frac{|{{\mathtt{V}_g}}_{{\srad{{\mathsf{b}}}}}|}{|{{\mathtt{V}_g}}_{{\mathsf{b}}}|}}{\mathtt{u}}_{{\mathsf{b}}}, 
{\mathtt{u}}_{\srad{{\mathsf{b}}}}\sim\sum\limits_{{\mathsf{a}}=1}^{N^{{{\mathcal{R}}}}}{{\tau}}^{{{\mathcal{R}}}\srad{{{\mathcal{L}}}}}_{{\mathsf{a}}{\srad{{\mathsf{b}}}}}\sqrt{\frac{|{{\mathtt{V}_g}}_{\srad{{\mathsf{b}}}}|}{|{{\mathtt{V}_g}}_{{\mathsf{a}}}|}}{\mathtt{u}}_{{\mathsf{a}}}.
\label{scaamp_sq_altinc}
\end{eqn}
The general expression for the incident waves is 
\begin{eqn}
{\mathtt{u}}=\widetilde{{\mathsf{I}}}_{{{\mathcal{R}}}}{\mathtt{u}}_{\srad{{\mathsf{a}}}}+\widetilde{{\mathsf{I}}}_{{{\mathcal{L}}}}{\mathtt{u}}_{\srad{{\mathsf{b}}}}.
\end{eqn}
The asymptotic form of solution to equation of motion is thus
\begin{eqn}
{\mathtt{u}}\to\begin{cases}
\widetilde{{\mathsf{I}}}_{{{\mathcal{R}}}}{\mathtt{u}}_{{\srad{{\mathsf{a}}}}}+\widetilde{{\mathsf{I}}}_{{{\mathcal{R}}}}\sum\limits_{{{\mathsf{a}}}=1}^{N^{{{\mathcal{R}}}}}{{\tau}}^{{{\mathcal{R}}}\srad{{{\mathcal{R}}}}}_{{\mathsf{a}}{{\srad{{\mathsf{a}}}}}}\sqrt{\frac{|{{\mathtt{V}_g}}_{{\srad{{\mathsf{a}}}}}|}{|{{\mathtt{V}_g}}_{{\mathsf{a}}}|}}{\mathtt{u}}_{{\mathsf{a}}}+\widetilde{{\mathsf{I}}}_{{{\mathcal{L}}}}\sum\limits_{{\mathsf{a}}=1}^{N^{{{\mathcal{R}}}}}{{\tau}}^{{{\mathcal{R}}}\srad{{{\mathcal{L}}}}}_{{\mathsf{a}}{\srad{{\mathsf{b}}}}}\sqrt{\frac{|{{\mathtt{V}_g}}_{\srad{{\mathsf{b}}}}|}{|{{\mathtt{V}_g}}_{{\mathsf{a}}}|}}{\mathtt{u}}_{{\mathsf{a}}}
=\widetilde{{\mathsf{I}}}_{{{\mathcal{R}}}}{\mathtt{u}}_{{\srad{{\mathsf{a}}}}}+(\widetilde{{\mathsf{I}}}_{{{\mathcal{R}}}}{{\tau}}^{{{\mathcal{R}}}\srad{{{\mathcal{R}}}}}_{{\mathsf{a}}{{\srad{{\mathsf{a}}}}}}\sqrt{\frac{|{{\mathtt{V}_g}}_{{\srad{{\mathsf{a}}}}}|}{|{{\mathtt{V}_g}}_{{\mathsf{a}}}|}}+\widetilde{{\mathsf{I}}}_{{{\mathcal{L}}}}{{\tau}}^{{{\mathcal{R}}}\srad{{{\mathcal{L}}}}}_{{\mathsf{a}}{\srad{{\mathsf{b}}}}}\sqrt{\frac{|{{\mathtt{V}_g}}_{\srad{{\mathsf{b}}}}|}{|{{\mathtt{V}_g}}_{{\mathsf{a}}}|}}, {\mathtt{x}}\to+\infty,\\
\widetilde{{\mathsf{I}}}_{{{\mathcal{L}}}}{\mathtt{u}}_{\srad{{\mathsf{b}}}}+\widetilde{{\mathsf{I}}}_{{{\mathcal{R}}}}\sum\limits_{{{\mathsf{b}}}=1}^{N^{{{\mathcal{L}}}}}{{\tau}}^{{{\mathcal{L}}}\srad{{{\mathcal{R}}}}}_{{\mathsf{b}}{\srad{{\mathsf{a}}}}}\sqrt{\frac{|{{\mathtt{V}_g}}_{\srad{{\mathsf{a}}}}|}{|{{\mathtt{V}_g}}_{{\mathsf{b}}}|}}{\mathtt{u}}_{{\mathsf{b}}}+\widetilde{{\mathsf{I}}}_{{{\mathcal{L}}}}\sum\limits_{{\mathsf{b}}=1}^{N^{{{\mathcal{L}}}}}{{\tau}}^{{{\mathcal{L}}}\srad{{{\mathcal{L}}}}}_{{\mathsf{b}}{\srad{{\mathsf{b}}}}}\sqrt{\frac{|{{\mathtt{V}_g}}_{{\srad{{\mathsf{b}}}}}|}{|{{\mathtt{V}_g}}_{{\mathsf{b}}}|}}{\mathtt{u}}_{{\mathsf{b}}}
=\widetilde{{\mathsf{I}}}_{{{\mathcal{L}}}}{\mathtt{u}}_{\srad{{\mathsf{b}}}}+(\widetilde{{\mathsf{I}}}_{{{\mathcal{R}}}}{{\tau}}^{{{\mathcal{L}}}\srad{{{\mathcal{R}}}}}_{{\mathsf{b}}{\srad{{\mathsf{a}}}}}\sqrt{\frac{|{{\mathtt{V}_g}}_{\srad{{\mathsf{a}}}}|}{|{{\mathtt{V}_g}}_{{\mathsf{b}}}|}}+\widetilde{{\mathsf{I}}}_{{{\mathcal{L}}}}{{\tau}}^{{{\mathcal{L}}}\srad{{{\mathcal{L}}}}}_{{\mathsf{b}}{\srad{{\mathsf{b}}}}}\sqrt{\frac{|{{\mathtt{V}_g}}_{{\srad{{\mathsf{b}}}}}|}{|{{\mathtt{V}_g}}_{{\mathsf{b}}}|}}, {\mathtt{x}}\to-\infty.
\end{cases}
\end{eqn}
Suppose \cite{Nazarovbook2009}
\begin{eqn}
{{\mathsf{O}}}_{{{\mathcal{R}}}}&\equiv \sqrt{|{{\mathtt{V}_g}}_{{\mathsf{a}}}|}{\widetilde{{\mathsf{O}}}}_{{{\mathcal{R}}}}={\widetilde{{\mathsf{I}}}}_{{{\mathcal{R}}}}{{\tau}}^{{{\mathcal{R}}}\srad{{{\mathcal{R}}}}}_{{\mathsf{a}}\srad{{\mathsf{a}}}}\sqrt{{{\mathtt{V}_g}}_{\srad{{\mathsf{a}}}}}+{\widetilde{{\mathsf{I}}}}_{{{\mathcal{L}}}}{{\tau}}^{{{\mathcal{R}}}\srad{{{\mathcal{L}}}}}_{{\mathsf{a}}\srad{{\mathsf{b}}}}\sqrt{{{\mathtt{V}_g}}_{\srad{{\mathsf{b}}}}}
\equiv {{\mathsf{I}}}_{{{\mathcal{R}}}}{{\tau}}^{{{\mathcal{R}}}\srad{{{\mathcal{R}}}}}_{{\mathsf{a}}\srad{{\mathsf{a}}}}+{{\mathsf{I}}}_{{{\mathcal{L}}}}{{\tau}}^{{{\mathcal{R}}}\srad{{{\mathcal{L}}}}}_{{\mathsf{a}}\srad{{\mathsf{b}}}},
\end{eqn}
\text{and }
\begin{eqn}
{{\mathsf{O}}}_{{{\mathcal{L}}}}&\equiv \sqrt{|{{\mathtt{V}_g}}_{{\mathsf{b}}}|}{\widetilde{{\mathsf{O}}}}_{{{\mathcal{L}}}}={\widetilde{{\mathsf{I}}}}_{{{\mathcal{R}}}}{{\tau}}^{{{\mathcal{L}}}\srad{{{\mathcal{R}}}}}_{{\mathsf{b}}\srad{{\mathsf{a}}}}\sqrt{{{\mathtt{V}_g}}_{\srad{{\mathsf{a}}}}}+{\widetilde{{\mathsf{I}}}}_{{{\mathcal{L}}}}{{\tau}}^{{{\mathcal{L}}}\srad{{{\mathcal{L}}}}}_{{\mathsf{b}}\srad{{\mathsf{b}}}}\sqrt{{{\mathtt{V}_g}}_{\srad{{\mathsf{b}}}}}
\equiv {{\mathsf{I}}}_{{{\mathcal{R}}}}{{\tau}}^{{{\mathcal{L}}}\srad{{{\mathcal{R}}}}}_{{\mathsf{b}}\srad{{\mathsf{a}}}}+{{\mathsf{I}}}_{{{\mathcal{L}}}}{{\tau}}^{{{\mathcal{L}}}\srad{{{\mathcal{L}}}}}_{{\mathsf{b}}\srad{{\mathsf{b}}}},
\end{eqn}
where ${{\mathsf{O}}}_{{{\mathcal{R}}},{{\mathcal{L}}}} \equiv \sqrt{|{{\mathtt{V}_g}}_{{\mathsf{a}},{\mathsf{b}},{\mathsf{b}}}|}{\widetilde{{\mathsf{O}}}}_{{{\mathcal{R}}},{{\mathcal{L}}}}$ and ${{\mathsf{I}}}_{{{\mathcal{R}}},{{\mathcal{L}}}}\equiv \sqrt{|{{\mathtt{V}_g}}_{\srad{{\mathsf{a}}},\srad{{\mathsf{b}}}}|}{\widetilde{{\mathsf{I}}}}_{{{\mathcal{R}}},{{\mathcal{L}}}}$ are the flux amplitudes.
Above defines a linear relation between the flux amplitudes of outgoing and incoming waves which can be written in a matrix form. The matrix $\mathbf{S}$, that relates the outgoing flux amplitudes ${{\mathsf{O}}}_{{{\mathcal{R}}},{{\mathcal{L}}}}$ to the incoming flux amplitudes ${{\mathsf{I}}}_{{{\mathcal{R}}},{{\mathcal{L}}}}$, is called the $\mathbf{S}$ matrix \cite{Nazarovbook2009}.
The $\mathbf{S}$ matrix is a $(N^{{{\mathcal{R}}}}+N^{{{\mathcal{L}}}}) \times (N^{{{\mathcal{R}}}}+N^{{{\mathcal{L}}}})$ matrix.
With
$\boldsymbol{{\mathsf{O}}}=[{{\mathsf{O}}}^1_{{{\mathcal{R}}}}~
\cdots~
{{\mathsf{O}}}^{N^{{{\mathcal{R}}}}}_{{{\mathcal{R}}}}~
{{\mathsf{O}}}^1_{{{\mathcal{L}}}}~
\cdots~
{{\mathsf{O}}}^{N^{{{\mathcal{L}}}}}_{{{\mathcal{L}}}}]^T,$ $\boldsymbol{{\mathsf{I}}}=[{{\mathsf{I}}}^1_{{{\mathcal{R}}}}~
\cdots~
{{\mathsf{I}}}^{N^{{{\mathcal{R}}}}}_{{{\mathcal{R}}}}~
{{\mathsf{I}}}^1_{{{\mathcal{L}}}}~
\cdots~
{{\mathsf{I}}}^{N^{{{\mathcal{L}}}}}_{{{\mathcal{L}}}}]^T$
\begin{eqn}
\boldsymbol{{\mathsf{O}}}=\mathbf{S}\boldsymbol{{\mathsf{I}}},
\text{where }
\mathbf{S}&=\begin{bmatrix}
{{\tau}}^{{{\mathcal{R}}}\srad{{{\mathcal{R}}}}}_{N^{{{\mathcal{R}}}}\times N^{{{\mathcal{R}}}}}&{{\tau}}^{{{\mathcal{R}}}\srad{{{\mathcal{L}}}}}_{N^{{{\mathcal{R}}}}\times N^{{{\mathcal{L}}}}}\\
{{\tau}}^{{{\mathcal{L}}}\srad{{{\mathcal{R}}}}}_{N^{{{\mathcal{L}}}}\times N^{{{\mathcal{R}}}}}&{{\tau}}^{{{\mathcal{L}}}\srad{{{\mathcal{L}}}}}_{N^{{{\mathcal{L}}}}\times N^{{{\mathcal{L}}}}}\\
\end{bmatrix},
\end{eqn}
${{\tau}}^{{a}\srad{{b}}}_{N^{{a}}\times N^{{b}}}=\left[
\begin{array}{cccccc}
{{\tau}}^{{a}\srad{b}}_{11}&\dotsc&{{\tau}}^{a\srad{b}}_{1N^{a}}\\
\vdots&\ddots\\
{{\tau}}^{a\srad{b}}_{N^{a}1}&\dotsc&{{\tau}}^{a\srad{b}}_{N^{a}N^{b}}
\end{array}
\right],$ for all four choices of $a, b\in\{{{\mathcal{R}}}, {{\mathcal{L}}}\}$.
As a necessary consequence of the unitarity of $\mathbf{S}$ it follows that
$\sum\nolimits_{{{\mathsf{a}}}=1}^{N^{{{\mathcal{R}}}}}|{{\tau}}^{{{\mathcal{R}}}\srad{{{\mathcal{R}}}}}_{{\mathsf{a}}\srad{{\mathsf{a}}}}|^2+\sum\nolimits_{{{\mathsf{b}}}=1}^{N^{{{\mathcal{L}}}}}|{{\tau}}^{{{\mathcal{L}}}\srad{{{\mathcal{R}}}}}_{{\mathsf{b}}\srad{{\mathsf{a}}}}|^2=1,$
$\sum\nolimits_{{{\mathsf{a}}}=1}^{N^{{{\mathcal{R}}}}}|{{\tau}}^{{{\mathcal{R}}}\srad{{{\mathcal{L}}}}}_{{\mathsf{a}}\srad{{\mathsf{b}}}}|^2+\sum\nolimits_{{{\mathsf{b}}}=1}^{N^{{{\mathcal{L}}}}}|{{\tau}}^{{{\mathcal{L}}}\srad{{{\mathcal{L}}}}}_{{\mathsf{b}}\srad{{\mathsf{b}}}}|^2=1,$
for arbitrary $\srad{{\mathsf{a}}}$ and $\srad{{\mathsf{b}}}$, respectively. 
The latter has been established as the zero lemma by \cite{Bls9s} for the incidence from one direction (the other case is analogous).
Indeed, in the conservative case (${\upomega}_2=0$), the principle of conservation of energy implies that 
all of the incident energy is scattered completely while getting subdivided in the reflected waves and transmitted waves. 
In view of \eqref{energyflux_inc}, the statement ${\mathscr{R}}+{\mathscr{T}}=1$ \cite{Bls9s} can be re-written as
${\mathscr{E}^{i}}{\mathscr{R}}+{\mathscr{E}^{i}}{\mathscr{T}}={\mathscr{E}^{i}}.$
Finally, the conductance for transmission from left to right can be expressed
in terms of the elements of the $\mathbf{S}$ matrix, at a given frequency ${\upomega}$, as
\begin{eqn}
{\mathscr{G}}_{{{{\mathcal{L}}}}\to{{{\mathcal{R}}}}}
&=\sum\limits_{\srad{{\mathsf{b}}}=1}^{N^{{{\mathcal{L}}}}}\sum\limits_{{\mathsf{a}}=1}^{N^{{{\mathcal{R}}}}}{{\tau}}^{{{\mathcal{R}}}\srad{{{\mathcal{L}}}}}_{{\mathsf{a}}\srad{{\mathsf{b}}}}{{\tau}}^{{{\mathcal{R}}}\srad{{{\mathcal{L}}}}\ast}_{{\mathsf{a}}\srad{{\mathsf{b}}}}
=\text{ tr }{{\tau}}_{N^{{{\mathcal{R}}}}\times N^{{{\mathcal{L}}}}}^{{{\mathcal{R}}}\srad{{{\mathcal{L}}}}}{{\tau}}_{N^{{{\mathcal{R}}}}\times N^{{{\mathcal{L}}}}}^{{{\mathcal{R}}}\srad{{{\mathcal{L}}}}\dagger}
=\text{ tr }{{\tau}}_{N^{{{\mathcal{R}}}}\times N^{{{\mathcal{L}}}}}^{{{\mathcal{R}}}\srad{{{\mathcal{L}}}}\dagger}{{\tau}}_{N^{{{\mathcal{R}}}}\times N^{{{\mathcal{L}}}}}^{{{\mathcal{R}}}\srad{{{\mathcal{L}}}}},
\label{VentrabookEq3_210}
\end{eqn}
which is also equal conductance ${\mathscr{G}}_{{{{\mathcal{L}}}}\leftarrow{{{\mathcal{R}}}}}$ from right to left. 

\renewcommand*{\bibfont}{\footnotesize}
\printbibliography

\end{document}